\documentclass{CUP-JNL-DTM-mod}

\usepackage{graphicx}
\usepackage{multicol,multirow}
\usepackage{amsmath,amssymb,amsfonts}
\usepackage{mathrsfs}
\usepackage{amsthm}
\usepackage{rotating}
\usepackage{appendix}
\usepackage{overpic}
\usepackage[numbers]{natbib}
\usepackage{ifpdf}
\usepackage[T1]{fontenc}
\usepackage{newtxtext}
\usepackage{newtxmath}
\usepackage{textcomp}
\usepackage{xcolor}
\usepackage{lipsum}
\usepackage[colorlinks,allcolors=blue]{hyperref}

\usepackage[latin,english]{babel}
\usepackage[utf8]{inputenc}
\usepackage{eufrak}
\usepackage{geometry}
\usepackage{bbm}
\usepackage{amsbsy}
\usepackage{makeidx}
\usepackage{mathtools}
\usepackage{mathrsfs, dsfont}
\usepackage{cases}
\usepackage{braket}
\usepackage{dirtytalk}
\usepackage{epstopdf}
\usepackage{todonotes}
\usepackage{fancyhdr}
\usepackage{subcaption}

\allowdisplaybreaks

\newcommand{\Real}{\mathbb{R}}
\newcommand{\Complex}{\mathbb{C}}
\def\Re{\mathop{\rm Re}\nolimits}
\def\Im{\mathop{\rm Im}\nolimits}

\usepackage{todonotes}

\newtheorem{RHP}{RHP}
\newtheorem{remark}{Remark}
\newtheorem{lemma}{Lemma}
\newtheorem{proposition}{Proposition}

\theoremstyle{definition}

\jname{Journal of Nonlinear Waves}
\articletype{RESEARCH ARTICLE}
\jyear{~}

\usepackage{geometry}

\begin{document}

\begin{Frontmatter}

\title[Numerical IST]{Numerical inverse scattering transform for the defocusing nonlinear Schr\"odinger equation with  box-type initial conditions on a nonzero background}

\author[1]{A. Gkogkou}
\author[2]{B. Prinari}
\author[3]{T. Trogdon}

\authormark{A. Gkogkou \textit{et al}.}

\address[1]{\orgdiv{Department of Mathematics}, \orgname{Tulane University}, \orgaddress{\city{New Orleans}, \postcode{70118-5698}, \state{LA},  \country{USA}}}

\address[2]{\orgdiv{Department of Mathematics}, \orgname{University at Buffalo}, \orgaddress{\city{Buffalo}, \postcode{14260-2900}, \state{NY},  \country{USA}} \email{bprinari@buffalo.edu}}

\address[3]{\orgdiv{Department of Applied Mathematics}, \orgname{University of Washington}, \orgaddress{\city{Seattle}, \postcode{98195-3925}, \state{WA},  \country{USA}}}

\keywords{Integrable systems, Riemann-Hilbert problems, nonlinear steepest descent}

\keywords[MSC Codes]{\codes[Primary]{35Q55}; \codes[Secondary]{37K15, 35Q15, 45E05}}

\abstract{We present a method to solve numerically the Cauchy problem for the defocusing nonlinear Schr\"{o}dinger (NLS) equation with a box-type initial condition (IC) having a nontrivial background of amplitude $q_o>0$ as $x\to \pm \infty$ by implementing numerically the corresponding Inverse Scattering Transform (IST). The Riemann--Hilbert problem associated to the inverse transform is solved numerically by means of appropriate contour deformations in the complex plane following the numerical implementation of the  Deift-Zhou nonlinear steepest descent method. 
In this work, the box parameters are chosen so that there is no discrete spectrum (i.e., no solitons).  The numerical method is demonstrated to be accurate within the two asymptotic regimes corresponding to two different regions of the $(x,t)$-plane depending on whether $|x/(2t)| < q_o$ or $|x/(2t)| > q_o$, as $t \to \infty$.}
\end{Frontmatter}

\section{Introduction}
\label{s:intro}

In this work, we present a method to numerically solve the Cauchy problem for the defocusing nonlinear Schr\"{o}dinger (NLS) equation:
\begin{gather}
\label{eq:dNLS}
i q_{t} + q_{xx} + 2(q_o^2 - |q|^2)q=0,
\end{gather}
on a nonzero background, and specifically with the following constant nonzero boundary conditions at infinity
\begin{gather}\label{eq:NZBCs}
\lim_{x \to \pm \infty} q(x,t) = q_{\pm} = q_o e^{i \theta_{\pm}}, \qquad q_o>0,\ \theta_\pm\in \Real.
\end{gather}
Note that due to the phase invariance of the NLS equation \eqref{eq:dNLS}, in the following we will set 
$$
\theta_{+} = - \theta_{-} = \theta
$$
without loss of generality. Note also that the additional linear term in \eqref{eq:dNLS}, compared to the standard form of the equation when $q_o=0$, can be removed by a gauge transformation
$q\to e^{2iq_o^2t}q$, and has simply the effect of ensuring that the boundary conditions \eqref{eq:NZBCs} are time-independent.

As it is well-known, the NLS equation is a completely integrable system, which implies that it admits a Lax pair (an overdetermined system of ordinary differential equations (ODEs) whose compatibility condition is equivalent to the integrable partial differential equation (PDE) expressed by Eq.~\eqref{eq:dNLS}), and its Cauchy problem can be solved by means of the Inverse Scattering Transform (IST), a nonlinear analog of the Fourier transform. Similarly to the case of linear PDEs, the solution of the
Cauchy problem by IST proceeds in three steps: (i) Direct problem - the transformation of the initial data from the
original ``physical'' variables $(q(x, 0))$, to the transformed ``scattering'' variables $\mathcal{S}(k,0)$
(namely, reflection coefficient, discrete eigenvalues and norming constants); (ii) Time
dependence - the evolution of the scattering data, i.e., finding $\mathcal{S}(k, t)$; (iii) Inverse problem - the recovery of
the evolved solution $q(x, t)$ from the evolved solution in the transformed variables $\mathcal{S}(k, t)$. 
The eigenfunctions of the first operator in the Lax pair, referred to as the ``scattering problem'', play a crucial role in the theory:
the direct problem involves integral equations for the eigenfunctions, through which the scattering data are
defined; the inverse problem can be formulated in terms of a Riemann-Hilbert problem (RHP) for the
eigenfunctions, from which one then reconstructs the solution of the nonlinear PDE. Furthermore, the inverse problem
reveals that the solitons are the portion of the solution associated with the discrete spectrum (discrete eigenvalues
and related norming constants).

Although the IST has been around for over 60 years, its numerical implementation is a relatively recent effort.
The numerical IST has been implemented in the literature for a number of nonlinear integrable equations: Korteweg-de Vries and modified Korteweg-de Vries equations \cite{TOD2012,OT2012}, focusing and defocusing NLS equations \cite{TO2013}, and the Toda lattice \cite{BT2017}, in all cases under the assumption that the initial condition (IC) is rapidly decaying as $|x|\to \infty$ (typically, in Schwartz class). For non-decaying ICs for the Korteweg-de Vries equation see \cite{BT2020,BNT2023}.
As the above works already showed, the main advantage of the numerical IST over direct numerical simulations is that the computational cost to approximate the solution at given values of $x,t$ can be independent of $x$ and $t$, while the computational cost using time-stepping methods grows rapidly in time, thus making traditional numerical methods inefficient to capture the solution for large times. This is particularly problematic for the NLS equations, both focusing and defocusing, due to the presence of an oscillatory dispersive tail.

In the present work, we will consider for the numerical implementation of the IST a piecewise constant initial condition of box-type, i.e.,
\begin{gather}\label{eq:IC}
q(x,0) = \begin{cases}
q_{-} = q_o e^{-i \theta}, & x< -L\\
q_{c} = h e^{i \alpha}, & -L < x< L\\
q_{+} = q_o e^{i \theta}, & x>L\\
\end{cases}
\end{gather}
where $h, q_o$ and $L$ are arbitrary non-negative parameters and $\theta$ and $\alpha$ are arbitrary real phases. Furthermore, we will restrict ourselves to values of the box parameters for which no discrete eigenvalues (i.e., no solitons) are present. The rationale for restricting to a piecewise constant IC such as above is the following: to begin with, for these types of ICs one can explicitly compute the scattering data (following, e.g., \cite{BP2014}), which avoids the need for a numerical implementation of the direct problem. Furthermore, for the inverse problem, the chosen IC yields analyticity of the reflection coefficient, which in turn avoids the need for a $\bar{\partial}$-problem when the associated Riemann--Hilbert problem is deformed. Moreover, having at our disposal the explicit expression of the reflection coefficient allows us to perform consistency checks and better control the convergence of the numerical integrations. Finally, discontinuous ICs are also notoriously much more difficult to handle from the point of view of direct numerical simulations, thus showcasing the advantages of a numerical IST over traditional time-stepping methods \cite{FornbergFlyer,TrogdonLiu}.
On the other hand, the discontinuous IC results in slower asymptotic decay of the reflection coefficient, which will require a more careful handling of the numerical integrations involved in the solution of the inverse problem.

\begin{figure}[tbp]
    \centering
    \begin{overpic}[width=.9\linewidth,unit=1mm]{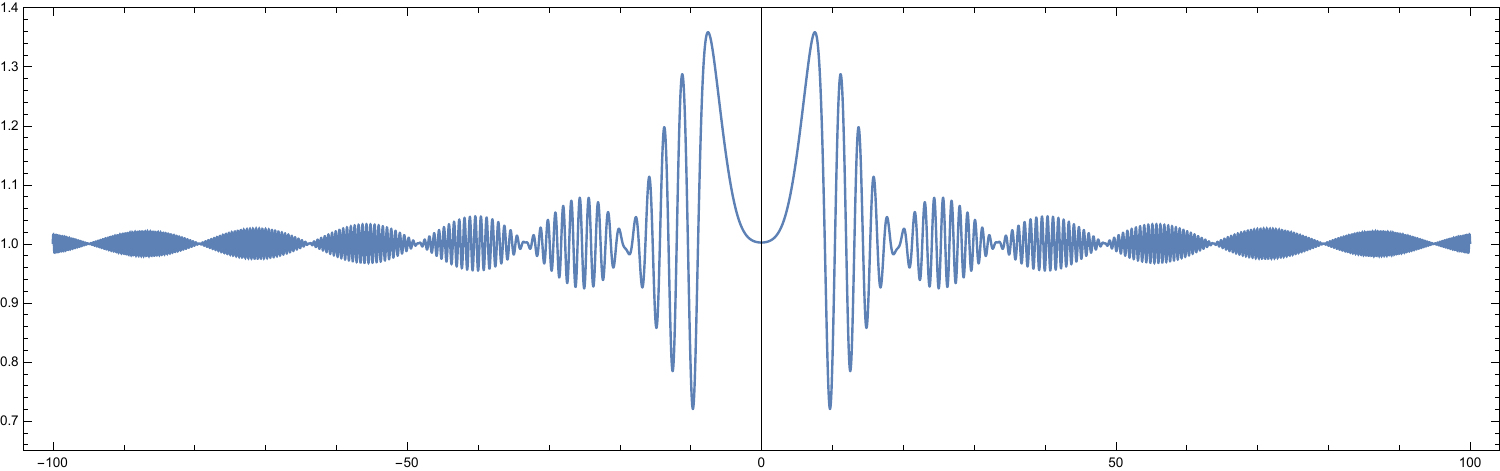}
    \put(-4,10){\rotatebox{90}{$|q(x,2.5)|^2$}}
    \put(50,-1.5){$x$}
    \end{overpic}\vspace{.1in}
    \caption{The squared modulus of the solution of the NLS equation with IC \eqref{eq:IC} at $t = 2.5$ with box parameters as in \eqref{eq:values}.}
    \label{fig:mod_t_2_5}
\end{figure}

To elaborate on the numerical challenges, we note that the most common approach to numerical approximations of solutions of whole-like PDE problems is using a periodic approximation with a large period.  This allows one to use efficient Fourier methods and the technology of exponential integrators \cite{TrefethenKassam,Klein}.  But for discontinuous data one is necessarily combating classical Gibbs phenomenon while simultaneously trying to resolve the nonlinear Gibbs phenomenon that is present in the true solution \cite{DiFrancoMcLaughlin}.  To compare the numerical IST approach with classical approaches, we incorporated the artificially damping/filtering of \cite{TrogdonLiu} and still needed a period of length 800 and over a million Fourier modes to achieve an approximation accurate to with an error $10^{-3}$ at $t =1$.  The runtime of such a simulation on a laptop is likely to be on the order of hours, in contrast to the method we propose here that requires a matter of seconds to evaluate the solution at a point in the $(x,t)$ plane (see, for example, the plot in Fig.~\ref{fig:mod_t_2_5}).

From a theoretical point of view, one of the challenges of dealing with a constant, nonzero background (here, $q_o>0$) in the IST is the fact that the asymptotic eigenvalues and eigenfunctions have branching in the spectral plane, originating from $\lambda=\sqrt{k^2-q_o^2}$. The IST can be carried out by introducing an appropriate branch cut in the complex $k$-plane, and this is also the framework adopted in \cite{BM2019} for their ``robust'' IST for the focusing NLS equation. For our numerical implementation, we will take advantage of the formulation of the IST in terms of the uniformization variable $z =k+\lambda$, for which $k = (z + q_o^2/z)/2$ and $\lambda=(z-q_o^2/z)/2$, and which allows the direct and inverse problems to be formulated in the entire complex $z$-plane and avoid branching altogether. The uniform variable was first introduced by Faddeev and Takhtajan in \cite{FT1987}, and since then it has been successfully utilized in a large number of papers dealing with the IST of scalar, vector and matrix NLS equations on a nontrivial, symmetric background. This is also the approach used in \cite{CJ2016} and \cite{Fan2022}, where the Deift-Zhou nonlinear steepest descent method was employed to compute the long-time asymptotic behavior of solutions of the defocusing NLS.
According to the results in \cite{CJ2016,Fan2022}, there are two different asymptotic regions, which require different contour deformations in the associated oscillatory RHP for the eigenfunctions, depending on whether $|\xi| = |x/(2t)| < 1$ (the so-called ``solitonic region'', where the phase function in the RHP has no real stationary points), and $|\xi| > 1$ (the ``solitonless'' region, where the phase function exhibits 2 real stationary points). The long-time asymptotics problem in the solitonic region has been studied in \cite{CJ2016}, while \cite{Fan2022} investigated the long-time asymptotics in the solitonless region.

The numerical IST scheme for the solution of the inverse problem has two major components: the first is the use of a Chebyshev collocation method for solving RHPs (as in the case of rapidly decaying ICs dealt with in \cite{TOD2012,OT2012,TO2013}, see \cite{SO2012,TO2015} for an overview of the framework), and the second is contour deformation in the complex plane, which mimics the corresponding deformations used for estimating the long-time asymptotics in \cite{CJ2016,Fan2022}.
Once the proper contour deformations are implemented numerically, one has to handle the numerical computation of three different Cauchy-type integrals, featuring: (i) integrands with slow decay at infinity (due to the discontinuous IC); (ii) integrands with rapid oscillations at $z=0$ (introduced by the uniform variable, since $k\to \infty$ on one of the two sheets of the Riemann surface is mapped onto $z=0$);
(iii) integrands with a logarithmic singularity at $z=\pm q_o$ (these singularities depend on the symmetries of the reflection coefficient, and are present also with a smooth IC, see \cite{CJ2016}). For the first two integrals, we can simply judiciously truncate the integration domains to achieve good accuracy while still keeping the computational costs at bay. As to the last integrals, we generalize the contour deformation used to address the analogous problem for the Toda lattice \cite{BT2017}, and reduce each integral to one that can be analytically computed.

The plan of the paper is as follows. In Sec.~\ref{s:IST} we give a succinct overview of the IST for the defocusing NLS equation
formulated in terms of the uniform variable $z$ for a general IC with symmetric nonzero boundary conditions \eqref{eq:NZBCs} as $|x|\to \infty$. 
We will assume $q(x,t)$ to be decaying to the boundary conditions sufficiently rapidly, specifically in analogy with the function class considered in \cite{CJ2016}.
In Sec.~\ref{s:RHP}, we first discuss the jump matrix factorizations and contour deformations that will be utilized for the numerical solution of the RHP,  and then show one can remove the singularity of the RHP at $z=0$.
The required contour deformations are different in the solitonic and solitonless regions,  and they are discussed in Sec.~\ref{s:solitonregion} and Sec.~\ref{s:solitonlessregion}, respectively. Sec.~\ref{s:numerics} provides the details of the numerical implementation of the RHP and its solution for a box-type IC of the form \eqref{eq:IC} with box parameters chosen to ensure absence of discrete eigenvalues/solitons, and hence of poles in the RHP,
via the routine RHSolve (a {\tt Mathematica} implementation of the code will be available as electronic supplementary material). In turn, the numerical solution of the RHP at any given $(x,t)$ is then used to obtain the approximation for $q(x,t)$, and some plots illustrating the solution at various times are provided. The accuracy of the numerical code is verified by substituting the computed $q(x,t)$ into the defocusing NLS equation using a second-order finite difference scheme in both time and space. Finally, Sec.~\ref{s:conclusion} is devoted to some concluding remarks.

The generalization of this work to an arbitrary smooth IC decaying sufficiently rapidly to a symmetric nonzero background as $|x|\to \infty$ will be the subject of future investigation. We anticipate that the numerical implementation of the direct problem, required for a general IC, will not pose any particular challenge, as it can be done similarly to the case of rapidly decaying ICs.  For data with some degree of exponential decay, the deformations outlined here will likely apply for the inverse problem, indicating that indeed, the methodology presented here paves the way for these future implementations.

\section{Overview of the IST}\label{sec:background}
\label{s:IST}

In this section we provide a review of the IST for the defocusing NLS equation on a nonzero symmetric background. 

\subsection{Integrability and Jost eigenfunctions}

Eq.~\eqref{eq:dNLS} is an integrable system and admits the following Lax pair
\begin{subequations}
\begin{gather}
\Phi_{x} = X \Phi, \quad \Phi_{t} = T \Phi,
\end{gather}
\begin{equation}
X(x,t;k) = - ik \sigma_3 + Q, \quad T(x,t;k) = -2ik^2 \sigma_3 + i \sigma_3 (Q_x - Q^2 + q_o^2 I_2) + 2k Q,
\end{equation}
\begin{equation}
Q(x,t) = \begin{pmatrix}
0 & q(x,t)\\
q^*(x,t) & 0
\end{pmatrix}, \quad \sigma_1 = \begin{pmatrix}
0 & 1\\
1 & 0
\end{pmatrix}, \quad \sigma_2 = \begin{pmatrix}
0 & -i\\
i & 0
\end{pmatrix}, \quad \sigma_3 = \begin{pmatrix}
1 & 0\\
0 & -1
\end{pmatrix}.
\end{equation}
\end{subequations}
Here and in the following, $^*$ denotes complex conjugation, and $\sigma_j$ for $j=1,2,3$ are the usual Pauli matrices.
For the purpose of reviewing the IST on a nonzero background, we will assume the IC to be in the analog functional class as in \cite{CJ2016}. Specifically, we introduce
\begin{equation}
\label{e:qtilde}
\tilde{q}(x)=q_o\left[ \cos \theta +i \sin \theta \tanh x \right]    
\end{equation}
to generalize $\tanh x$ in \cite{CJ2016} to a smooth function which approaches, at exponential rates, the boundary conditions \eqref{eq:NZBCs}. 
We consider the weighted spaces 
$L^{p,s}(\Real)$ with norm defined as
\begin{equation}
||f||_{L^{p,s}(\Real)}:=\left(
\int_\Real \langle x \rangle^{2s}|f(x)|^p\, dx \right)^{1/p},
\qquad
\langle x\rangle:=\sqrt{1+x^2}
\end{equation}
(note that $2s=q$, compared to \cite{CJ2016}), as well as the Sobolev spaces
\begin{equation}
H^\ell(\Real)=\left\{ 
f:\Real \to \Complex:
f^{(k)}\in L^2(\Real)\ \text{for } k=0,\dots,\ell
\right\}
\end{equation}
and
\begin{equation}
H^{1,\ell}(\Real):=L^{2,\ell}(\Real)\cap H^\ell(\Real),
\end{equation}
where 
the $L^2$-Sobolev bijectivity of the IST for the focusing NLS equation was established in the case of rapidly decaying ICs \cite{Zhou1998}.
Then, combining results from \cite{Gallo2004,Gallo2008,DPVV2013,CJ2016}, we will assume the IC such that $q(x,0)-\tilde{q}(x)\in H^{1,\ell}(\Real)$ for $\ell=1,3/2,2$ (these are the same as the spaces $\Sigma_m$ in \cite{CJ2016} with $m=2,3,4$, respectively), and highlight the corresponding implications on the IST as it becomes relevant.

The Jost eigenfuctions $\Phi_\pm$ are defined as simultaneous solutions of the two linear equations of the Lax pair with boundary conditions
\begin{subequations}\label{e:4n}
\begin{gather}
\Phi_{\pm}(x,t;k) \sim \left[ Y_{\pm}(k) + o(1)\right] e^{i \Omega(x,t;k) \sigma_3}, \quad x \to \pm \infty,\label{e:4na}\\
\Omega(x,k;t) = - \lambda x - 2 k \lambda t, \quad \lambda^2 = k^2-q_o^2,
\label{e:4nb}\\
Y_{\pm}(k) = I_2 - \frac{i}{k+\lambda} \sigma_3 Q_{\pm}, \quad Q_{\pm} = \begin{pmatrix}
0 & q_{\pm}\\
q_{\pm}^* & 0
\end{pmatrix}. \label{e:4nc}
\end{gather}
\end{subequations}
It is worth remarking that $i\lambda(k)$ with $\lambda(k)$ defined in \eqref{e:4nb} is the eigenvalue parameter for the constant coefficient matrices $X_\pm=X-Q_\pm$, where $Q_\pm=\lim_{x\to \pm \infty}Q(x,t)$ (and $Y_\pm$ are the associated matrices of eigenvectors). 
Then \eqref{e:4na} determines the solutions when
$\lambda(k)\in \Real$ (purely oscillatory behavior), and this corresponds to $k^2 \ge q_o^2$, showing that
the continuous spectrum for the problem in the $k$-plane is $\Real\setminus(-q_o,q_o)$, i.e., there is a gap.

Note that in the expression for $\Omega$ above, the second term has the opposite sign compared to \cite{BP2014,CJ2016,Fan2022}. The correct sign is the one in \eqref{e:4nb}, but the sign error in \cite{BP2014,CJ2016,Fan2022} does not affect in any significant way the results in those papers, as it ultimately only switches the regions with $t>0$ and those with $t<0$. For convenience, we introduce the uniformization variable $z$ defined by the map
\begin{gather}
z=k+\lambda
\end{gather}
which is inverted via the relations
\begin{gather}\label{eq:eq25n}
k(z) = \frac{1}{2} \left(z+q_o^2/z \right), \quad \lambda(z)=\frac{1}{2}\left(z-q_o^2/z \right).
\end{gather}
As mentioned in Sec.~1, the map $k\to z$ in the context of the IST of the defocusing NLS with nonzero background was first introduced by Faddeev and Takhtajan in \cite{FT1987}. The interested reader can find a discussion on the $2$-to-$1$ nature of the map and other relevant details in \cite{PAB2006}.

The asymptotic behavior of the eigenfunctions can then be written as:
\begin{subequations}\label{e:5}
\begin{gather}
\Phi_{\pm}(x,t;z) \sim \left[ Y_{\pm}(z) + o(1)\right] e^{i \Omega(x,t;z) \sigma_3
} , \quad x \to \pm \infty,\label{e:5a}\\
Y_{\pm}(z) = \begin{pmatrix}
1 & -\frac{i}{z} q_{\pm}\\
\frac{i}{z}q_{\pm}^* & 1
\end{pmatrix} = \begin{pmatrix}
1 & -\frac{i}{z} q_o e^{\pm i \theta}\\
\frac{i}{z} q_o e^{\mp i \theta} & 1
\end{pmatrix}, \label{e:5b} \\
\Omega(x,t;z) = -\frac{1}{2}(z-q_o^2/z)x-\frac{1}{2}(z^2-q_o^4/z^2)t.
\end{gather}
\end{subequations}
Note that in terms of $z$ the phase function $\Omega$ has singularities at $z=0$ and $z=\infty$. As usual, it is more convenient to introduce modified eigenfunctions
\begin{gather}\label{eq:eq210}
M_{\pm}(x,t;z) = \Phi_{\pm}(x,t;z) e^{-i \Omega(x,t;z) \sigma_3}
\end{gather}
which satisfy the boundary conditions
\begin{gather}
\label{e:Mpm_x_as}
M_{\pm}(x,t;z) \sim Y_{\pm}(z) + o(1), \quad x \to \pm \infty.
\end{gather}

If the IC $q(x,0)$ is such that $q(x,0)-\tilde{q}(x)\in H^{1,1}(\Real)$, then standard Neumann series arguments on the Volterra integral equations for $M_\pm$ can be used to show that the modified eigenfunctions $M_{-,1}(x,t;z)$, $M_{+,2}(x,t;z)$ are analytic in the upper half-plane of $z$ and continuous up to $\Real\setminus\left\{0\right\}$, while $M_{-,2}(x,t;z)$, $M_{+,1}(x,t;z)$ are analytic in the lower half-plane of $z$ and continuous up to $\Real\setminus\left\{0\right\}$ (see \cite{DPVV2013,CJ2016}). Here and in the following, the additional subscript $j=1,2$ in $M_\pm$ denotes the corresponding column of the matrix function. 
Furthermore, the following lemma holds.
\begin{lemma}
The modified eigenfunctions $M_{\pm}$ satisfy the symmetries
\begin{gather}\label{eq:syms212}
M_{\pm}^*(z^*) = \sigma_1 \, M_{\pm}(z) \, \sigma_1, \quad M_{\pm} ( q_o^2/z ) = \frac{z}{q_o} \, M_{\pm}(z) \, \sigma_2 \, e^{\mp i \theta \sigma_3}
\end{gather}
for any $q_o >0$ and $\theta \in \mathbb{R}$, where each column on both sides of the equations is considered for $z\in \Complex^+\cup \Real \setminus\{0\}$ or $z\in \Complex^-\cup \Real \setminus\{0\}$ depending on the respective domain of analyticity.
\end{lemma}
\begin{proof}
Using the symmetries of the Lax pair
\begin{subequations}
\begin{gather}
X^*(x,t;z^*) = \sigma_1 \, X(x,t;z) \, \sigma_1, \quad X (x,t;q_o^2/z) = X(x,t;z) \\
T^*(x,t;z^*) = \sigma_1 \, T(x,t;z) \, \sigma_1, \quad T(x,t;q_o^2/z) = T(x,t;z)
\end{gather}
\end{subequations}
and the asymptotics of $\Phi_{\pm}$ as $x \to \pm \infty$, one can show that $\Phi_{\pm}$ satisfy the symmetries
\begin{gather}\label{eq:eq214n}
\Phi_{\pm}^*(z^*) = \sigma_1 \, \Phi_{\pm}(z) \, \sigma_1, \quad \Phi_{\pm} ( q_o^2/z ) = \frac{z}{q_o} \, \Phi_{\pm}(z) \, \sigma_2 \, e^{\mp i \theta \sigma_3}.
\end{gather}
Symmetries \eqref{eq:syms212} follow directly from equation \eqref{eq:eq210} and the symmetries of the function $\Omega$, namely
\begin{gather}
\label{e:18}
\Omega^*(x,t;z^*) = \Omega(x,t;z), \quad \Omega (x,t;q_o^2/z) = - \Omega(x,t;z).
\end{gather}
For instance, to prove that the second of \eqref{eq:eq214n} holds, one needs to show that the two matrix functions satisfy the same differential equations and have the same asymptotic behavior as $x\to \pm \infty$. We have:
\begin{align*}
\partial_x \Phi_{\pm} ( q_o^2/z ) &= \frac{z}{q_o} \, \partial_x \Phi_{\pm}(z) \, \sigma_2 \, e^{\mp i \theta \sigma_3}= \frac{z}{q_o} X(x,t;z) \, \Phi_{\pm}(z) \, \sigma_2 \, e^{\mp i \theta \sigma_3}\\
&= \frac{z}{q_o} X(x,t;q_o^2/z) \, \Phi_{\pm}(z) \, \sigma_2 \, e^{\mp i \theta \sigma_3}\\
&=\frac{z}{q_o} \, \frac{q_o}{z} X(x,t;q_o^2/z) \, \Phi_{\pm}(q_o^2/z) \, e^{\pm i \theta \sigma_3}\, \sigma_2 \, \sigma_2 \, e^{\mp i \theta \sigma_3} \\
&=X(x,t;q_o^2/z) \, \Phi_{\pm}(q_o^2/z)
\end{align*}
and similarly one can show the same for the time-dependence problem. One can also check that the asymptotic behavior coincides, as well. Indeed, one has:
\begin{equation*}
\Phi_{\pm}(z) \sim Y_{\pm}(z) e^{i \Omega(z) \sigma_3}, \qquad \Phi_{\pm}(q_o^2/z) \sim Y_{\pm}(q_o^2/z) e^{i \Omega(q_o^2/z) \sigma_3} \quad \text{as }x \to \pm \infty.
\end{equation*}
Moreover, assuming that
\begin{equation*}
\Phi_{\pm}(q_o^2/z) = \frac{z}{q_o} \Phi_{\pm}(z) \sigma_2 e^{\mp i \theta \sigma_3} \sim \frac{z}{q_o}  \left[  Y_{\pm}(z) e^{i \Omega(z) \sigma_3}  \right] \sigma_2 e^{\mp i \theta \sigma_3},  \quad \text{as }x \to \pm \infty
\end{equation*}
then one must show that
\begin{equation*}\label{eq:eq6}
Y_{\pm}(q_o^2/z) e^{i \Omega(q_o^2/z) \sigma_3} \equiv \frac{z}{q_o}  \left[  Y_{\pm}(z) e^{i \Omega(z) \sigma_3}  \right] \sigma_2 e^{\mp i \theta \sigma_3},
\end{equation*}
which can be easily verified using the definition of $Y_{\pm}$ together with the second of \eqref{e:18}.

It is worth mentioning here that the symmetries \eqref{eq:syms212} can be established for $z\in \Real$, where all the columns are simultaneously defined, and then extended to $\Complex^\pm$ column-wise.
\end{proof}

\subsection{Direct problem: scattering data}\label{e:eqscdata}

The Jost eigenfunctions  $\Phi_{\pm}$ are two fundamental solutions of the Lax pair for any $z\in \Real\setminus\left\{\pm q_o\right\}$, since
\begin{gather}\label{eq:determinant}
\det \Phi_{\pm}(x,t;z) \equiv \det Y_\pm (z)= 1 - q_o^2 z^{-2}.
\end{gather}
Therefore, there exists a $2 \times 2$ scattering matrix $S(z)$ (independent of $x,t$) such that
\begin{gather}
\label{e:S}
\Phi_{-}(x,t;z) = \Phi_{+}(x,t;z) S(z), \quad S(z) = \begin{pmatrix}
a(z) & \bar{b}(z)\\
b(z) &  \bar{a}(z)
\end{pmatrix}, \quad z \in \mathbb{R} \setminus \{ \pm q_o\}.
\end{gather}
The time independence of the scattering matrix $S(z)$ is a consequence of having defined the Jost eigenfunctions as simultaneous solutions of the Lax pair, not just of the scattering problem.
After replacing the modified eigenfunctions, the last equation becomes
\begin{subequations}\label{e:14n}
\begin{gather}
M_{-,1}(x,t;z)/a(z) = M_{+,1}(x,t;z) + \rho(z) M_{+,2}(x,t;z) e^{-2i \Omega(x,t;z)},\label{e:14na}\\
M_{-,2}(x,t;z)/a^*(z) = M_{+,2}(x,t;z) + \bar{\rho}(z) M_{+,1}(x,t;z) e^{2i \Omega(x,t;z)},\label{e:14nb}
\end{gather}
\end{subequations}
where we introduced the reflection coefficients 
\begin{gather}
\label{e:rho}
\rho(z) = b(z)/a(z), \qquad \bar{\rho}(z)= \bar{b}(z)/\bar{a}(z).
\end{gather}
It is then easy to verify that the entries of the scattering matrix $S(z)$ are such that 
$$
\bar{a}(z)=a^*(z^*), \qquad \bar{b}(z)=b^*(z^*),
$$
where the symmetry holds for the values of $z$ for which $S(z)$ is defined. In particular, this implies that for all $z\in \Real$ one has $\bar{\rho}(z)=\rho^*(z)$, and 
\begin{equation}
\label{e:1st_symm_barrho}
\bar{\rho}(z)=\rho^*(z^*)
\end{equation}
when the reflection coefficient $\rho(z)$ can be analytically continued off the real $z$-axis, which will be the case for the box-type initial conditions considered later on.

Eq.~\eqref{eq:determinant}, combined with \eqref{e:S}, yields the following expressions for the scattering coefficients:
\begin{gather}\label{eq:sc data}
a(z) = \frac{1}{1 - q_o^2 z^{-2}} \mathrm{det} \left( M_{-,1}(z), M_{+,2}(z) \right), \quad
b(z)=\frac{1}{1 - q_o^2 z^{-2}} \mathrm{det} \left( M_{+,1}(z), M_{-,1}(z) \right)
\end{gather}
where we have omitted the $(x,t)$ dependence in the right-hand side for brevity. The above expressions show that even if the Jost eigenfunctions are continuous at $z=\pm q_o$, in general the scattering coefficients are singular at those points.

Discrete eigenvalues are the zeros of the scattering coefficient $a(z)$ in $\Complex^+$ and their complex conjugates (as zeros of $a^*(z^*)$) in $\Complex^-$. Such zeros are known to be simple, and located on the circle $\mathcal{C}=\left\{z\in \Complex: |z|=q_o\right\}$ due to the self-adjointess of the scattering problem \cite{FT1987}. Note that the only possible zeros of $a(z)$ embedded in the continuous spectrum, i.e., on the real $z$-axis, are $\pm q_o$, since it was shown in \cite{FT1987} that $a(k)\ne 0$ for any $k\in \Real \setminus [-q_o,q_o]$, which corresponds $z\in \Real \setminus \mathbb{C}$. Moreover, in \cite{DPVV2013} it is shown that if $q(x,0)-q_\pm \in L^{1,4}(\Real^\pm)$, then $a(z)$ does not vanish at $\pm q_o$ (and hence the number of discrete eigenvalues is necessarily finite, since the only possible accumulation points are $\pm q_o$).
Finally, Eqs.~\eqref{eq:sc data} imply that at each pair of discrete eigenvalues $z_j,z_j^*$:
\begin{gather}
\label{e:prop_cost}
M_{-,1}(x,t;z_j) = b_j \, M_{+,2}(x,t;z_j) e^{-2i \Omega(x,t;z_j)}, \quad M_{-,2}(x,t;z_j^*) = b_j^* \, M_{+,1}(x,t;z_j^*) e^{2i \Omega(x,t;z_j^*)}
\end{gather}
for some $b_j\in \Complex$.
Additional smoothness and decay properties of the reflection coefficient are established in \cite{CJ2016} under suitable assumptions on the IC.
Specifically: 
\begin{itemize}
\item If $q(x,0)-\tilde{q}(x)\in H^{1,1}(\Real)$, then $\rho \in L^2(\Real)$ and $||\log (1-|\rho|^2)||_{L^p}<\infty$ for all $p\ge 1$;
\item If $q(x,0)-\tilde{q}(x)\in H^{1,3/2}(\Real)$, then $\rho\in H^1(\Real)$ and moreover the discrete eigenvalues are necessarily finite in number.
\end{itemize}
Even though it will not be relevant for the present work, we mention that in \cite{CJ2016} it was also shown that requiring $q(x,0)-\tilde{q}(x)\in H^{1,2}(\Real)$ allows to obtain bounds of the $\bar{\partial}$-derivatives of the reflection coefficient which were there used for the long-time estimates in the inverse problem.

For completeness, we include the symmetries of the scattering data in the following lemma, whose proof follows straightforwardly from the symmetries \eqref{eq:eq214n} of the Jost eigenfunctions.
\begin{lemma}\label{lem:lemma2}
The scattering matrix $S(z)$ satisfies
\begin{gather}\label{eq:eq222nn}
S (q_o^2/z ) = e^{i \theta \sigma_3} \, \sigma_2 \, S(z) \, \sigma_2 \, e^{i \theta \sigma_3}, \qquad \forall z\in \Real\setminus\left\{\pm q_o\right\}.
\end{gather}
Consequently, one has
\begin{subequations}\label{eq:eq221}
\begin{gather}
a ( q_o^2/z ) = e^{2 i \theta} a^*(z^*), \quad z\in \Complex^+ \\
b ( q_o^2/z ) = - b^*(z), \quad \rho (q_o^2/z ) = - e^{-2i \theta} \, \rho^*(z), \quad z\in \Real\setminus\left\{\pm q_o\right\}. \label{eq:eq221b}
\end{gather}   
\end{subequations}
Moreover, the proportionality constants $b_j$ defined in \eqref{e:prop_cost} satisfy
\begin{equation}
b_j^*=-b_j.
\end{equation}
\end{lemma}

We emphasize that with a piecewise IC, the direct problem of the IST can be solved exactly. Specifically, as shown in \cite{BP2014}, for any piecewise IC \eqref{eq:IC}, one can obtain explicit formulas for the scattering coefficients by expressing, at the boundary of each region, the (fundamental) solution on the left as a linear combination of the (fundamental) solution on the right, which yields:
\begin{subequations}
\label{e:ab}
\begin{gather}
a(z) = \frac{1}{\lambda(z) \mu(z)}  e^{i(2 L \lambda(z) + \theta)} \Bigg\{ \mu(z) \cos(2 L \mu(z)) [ \lambda(z) \cos \theta - i k(z) \sin \theta ] \nonumber \\
+ i \sin(2L \mu(z)) \Big[ h q_o \cos \alpha - k(z) ( k(z) \cos \theta - i \lambda(z) \sin \theta ) \Big] \Bigg\}\\
 b(z) = \frac{1}{\lambda(z) \mu(z)} \Bigg\{ -q_o \mu(z) \sin \theta \cos(2 L \mu(z)) \nonumber \\+ \Big[ h k(z) \cos \alpha - k(z) q_o \cos \theta -i h \lambda(z) \sin \alpha \Big] \sin(2L \mu(z)) \Bigg\},
\end{gather}
\end{subequations}
where
\begin{equation}
\label{e:mu}
k(z)=\frac{1}{2}(z+q_o^2/z), \quad \lambda(z)=\sqrt{k(z)^2-q_o^2}\equiv \frac{1}{2}(z-q_o^2/z),\quad \mu(z)=\sqrt{k(z)^2-h^2}\,.
\end{equation}
It is important to point out that although $\mu(z)$ is defined in terms of a square root, there is no branching in the IST.
Indeed, $\mu$ does not enter in the definition of the Jost eigenfunctions, and \eqref{e:ab} show that only even functions of $\mu$ appear in the scattering coefficients, which are therefore independent of the choice of sign of the square root.

\subsection{Asymptotic behavior of eigenfunctions and scattering data}

To formulate the inverse problem properly, one needs to determine the asymptotic behavior of eigenfunctions and scattering data as $z \to 0$, $z \to \infty$ and as $z\to\pm q_o$. 
Specifically, the asymptotic behavior of the eigenfunctions is obtained from integration by parts on appropriate Volterra integral equations for the eigenfunction (cf., e.g., \cite{DPVV2013}), and it is given by: 
\begin{subequations}\label{e:n100}
\begin{gather}
M_{\pm}(x,t;z) = \begin{pmatrix}
1 + i I_0(x,t)/z & - i q(x,t)/z\\
i q^*(x,t)/z & 1 - i I_0(x,t)/z
\end{pmatrix}+O(z^{-2}) , \quad z \to \infty,\\
M_{\pm}(x,t;z) = \begin{pmatrix}
q(x,t)/q_{\pm} & I_0(x,t)/q_{\pm}^* - i q_{\pm}/z\\
I_0(x,t)/q_{\pm} + i q_{\pm}^*/z & q^*(x,t)/q_{\pm}^*
\end{pmatrix}+O(z^2), \quad z \to 0,
\end{gather}
\end{subequations}
where
\begin{gather}
I_0(x,t)= \int_{x}^{+ \infty} (|q(x',t)|^2 - q_o^2) dx'.
\end{gather}
\begin{proposition}The asymptotic behavior of the scattering coefficients as $z\to \infty$ and as $z\to 0$ is given by:
\begin{subequations}
\begin{gather}
a(z)=1 + \mathcal{O}(z^{-1}), \quad b(z) = \mathcal{O}(z^{-1}), \quad \rho(z) = \mathcal{O}(z^{-1}) \quad z \to \infty,\\
a(z) = q_{+}/q_{-} + \mathcal{O}(z), \quad b(z) = \mathcal{O}(z), \quad \rho(z) = \mathcal{O}(z), \quad z \to 0.
\end{gather}
\end{subequations}
\end{proposition}
\begin{proof}
The proof of these results follows by combining Eqs. \eqref{eq:sc data} and \eqref{e:n100}. 
Stronger regularity of the potential (e.g., assuming $q(x,0)\in C^j(\Real)$ for $j\ge 0$) provides faster decay of $b(z)$ and hence of $\rho(z)$ both as $z\to \infty$ and as $z\to 0$ (details can be found in \cite{DPVV2013}). 
\end{proof}

For ICs in $\tilde{q}(x)+H^{1,1}(\Real)$ the modified eigenfuctions are continuous at $z=\pm q_o$, and Eqs. \eqref{eq:sc data} show that generically the scattering data have simple poles at $\pm q_0$:
\begin{gather}\label{eq:eq228nn}
a(z) = \frac{\alpha_{\pm}}{z \mp q_o} + \mathcal{O}(1), \quad
b(z) = \frac{\beta_\pm}{z \mp q_o} + \mathcal{O}(1), \qquad \beta_\pm=\mp ie^{-i\theta} \alpha_\pm,
\end{gather}
where $\alpha_{\pm} = \mathrm{det} \big( M_{-,1}(x,t;\pm q_o), M_{+,2}(x,t;\pm q_o)\big)$ [and $\alpha_\pm\ne 0$ unless the columns of $M_{\pm}$ become linearly dependent at either $z= q_o$ or $z=-q_o$ or both, in which case the points $\pm q_o$ are called ``virtual levels''].
Note that \cite{DPVV2013} provides a sufficient condition on the decay of the potential that guarantees $a(\pm q_o) \ne 0$ (even if one between $\alpha_+$ and $\alpha_-$ or both vanish), namely that
$q(x,0)-q_\pm \in L^{1,4}(\Real^\pm)$, and this will be the case for the ICs considered in this work. Importantly, the reflection coefficient is always bounded at $z=\pm q_o$,
and in particular
\begin{equation}\label{eq:eq222}
\rho(\pm q_o) = \mp i e^{-i \theta}.
\end{equation}
The following result can also be established.
\begin{proposition}\label{eq:prop3}
If $\rho(z), \bar{\rho}(z)\in C^1(\Real)$, the quantity $1-\rho(z) \bar{\rho}(z)$ vanishes quadratically as $z\to \pm q_o$, i.e., 
\begin{equation}
1-\rho(z) \bar{\rho}(z) = O((z\mp q_o)^2), \qquad z\to \pm q_o.
\end{equation}
\end{proposition}
\begin{proof}
Clearly, $1-\rho(\pm q_o) \bar{\rho}(\pm q_o)=0$ because of Eq.~\eqref{eq:eq222}. Differentiating both sides of the second equation in~\eqref{eq:eq221b} with respect to $z$ and evaluating at $z=\pm q_o$ we find:
$$
\frac{d\rho(z)}{dz} \Bigg|_{z=\pm q_o} = e^{-2i\theta}\frac{d\rho^*(z)}{dz}\Bigg|_{z=\pm q_o}.
$$
Using the property $\bar{\rho}(z)=\rho^*(z)$ for $z \in \mathbb{R}$, and the above symmetry for $d\rho/dz$ at $z=\pm q_o$, we get:
$$
\frac{d}{dz}(1-\rho(z) \bar{\rho}(z)) \Bigg|_{z=\pm q_o} = \frac{d\rho^*(z)}{dz}\Bigg|_{z=\pm q_o}
\left( e^{-2i\theta}\rho^*(\pm q_o)+\rho(\pm q_o)\right)\equiv 0
$$
where the term in brackets vanishes owing to \eqref{eq:eq222}.  
\end{proof}
In Appendix~A, we give the explicit expressions of the asymptotics of the scattering coefficients \eqref{e:ab} as $z\to \infty$, $z\to 0$, and $z\to \pm q_o$ in the case of a box IC.

\subsection{Inverse problem: Riemann-Hilbert formulation}

Within the IST framework, the inverse problem amounts to reconstructing the Jost eigenfunctions in terms of the scattering data, from which one can then recover $q(x,t)$ for any $t>0$.
For this purpose, we introduce:
\begin{gather}\label{eq:function m}
m(z) \coloneqq m(x,t;z) = \begin{cases}
\Big( M_{-,1}(x,t;z)/a(z), \quad M_{+,2}(x,t;z)\Big), & z \in \mathbb{C}^{+}\\[5pt]
\Big( M_{+,1}(x,t;z), \quad M_{-,2}(x,t;z)/a^*(z^*)\Big), & z \in \mathbb{C}^{-}.
\end{cases}
\end{gather}
For a generic IC, the problem might support solitons that arise from the zeros of $a(z)$ in $\Complex^+$, and their complex conjugates, zeros of $a^*(z^*)$ in $\Complex^-$. 
Let $N\in \mathbb{N}$ denote the number of solitons (necessarily finite for IC in $\tilde{q}(x)+H^{1,3/2}(\Real)$, as mentioned above),
which implies that $m(z)$ has $N$ poles in $\mathbb{C}^{+}$ with their complex conjugates appearing as poles in $\mathbb{C}^{-}$. The properties established for the eigenfunctions and the scattering coefficient $a(z)$ in the direct problem then yield the following RHP for $m(z)$.  

\begin{RHP}[RHP for $m$ with solitons]\label{eq:RHP1} Find a $2 \times 2$ matrix-valued function $m(z)$ such that:
\begin{itemize}
    \item[1.] (a) $m(z)$ is sectionally meromorphic for $z\in \mathbb{C} \setminus \mathbb{R}$, and (b) for any $\epsilon > 0$, the functions $m(z)|_{z\in\mathbb C^\pm \setminus S_\epsilon}$ extend to continuous functions on the closure of their domains of definition, where
    \begin{align*}
    S_\epsilon = \{ |z| \leq \epsilon\} \cup \bigcup_{j=1}^N\{ |z - z_j| \leq \epsilon \}_{j=1}^N \cup \bigcup_{j=1}^N\{ |z - z^*_j| \leq \epsilon \}_{j=1}^N.
    \end{align*}
    \item[2.] The boundary values $m^{\pm}(x,t;z) = \lim_{\substack{\zeta \to z \\ \zeta \in \mathbb{C}^{\pm}}} m(x,t;\zeta)$,  $z \in \mathbb{R} \setminus \{ 0 \}$, satisfy the following jump relation across the real axis:
\begin{equation*}
m^{+}(z) = m^{-}(z) G (x,t;z), \quad G (x,t;z) = \begin{pmatrix}
1 - \rho(z)\bar{\rho}(z)& - \bar{\rho}(z) e^{2i \Omega(x,t;z)}\\
\rho(z) e^{-2i \Omega(x,t;z)} & 1
\end{pmatrix},
\end{equation*}
\begin{equation}\label{eq:eq2.27b}
\Omega(x,t;z) = - \lambda(z) x -2 k(z) \lambda(z) t.
\end{equation}
\item[3.] $m(z)$ has simple poles at $\{z_j\}_{j=1}^{N}$ in $\mathbb{C}^{+}$ and $\{z_j^{*}\}_{j=1}^{N}$ in $\mathbb{C}^{-}$ with $|z_j|=q_o$ for all $j=1,\cdots,N$, and with residue conditions:
\begin{subequations}
\label{e:res}
\begin{gather}
\mathrm{Res}_{z=z_j} m(z) = \lim_{z\to z_j} m(z) \begin{pmatrix}
0 & 0\\
C_{j} e^{-2i \Omega(x,t;z_j)} & 0
\end{pmatrix}
\end{gather}
\begin{gather}
\mathrm{Res}_{z=z_j^*} m(z) = \lim_{z\to z_j^*} m(z) \begin{pmatrix}
0 & C^*_{j} e^{2i \Omega(x,t;z_j^*)}\\
0 & 0
\end{pmatrix}
\end{gather}
\end{subequations}
where $\left\{C_j\right\}_{j=1}^N$ are the norming constants [which, for $m(z)$ in \eqref{eq:function m},  are defined in terms of the proportionality constants $b_j$ as $C_j = b_j/a'(z_j)$] and 
\begin{gather}
C_j^*=-e^{2i\theta}C_j.
\end{gather}
Note that the symmetries are as in Lemma \ref{lem:lemma2}.
\item[4.] $m(z)$ has the following asymptotic behavior as $z\to \infty$ and as $z\to 0$:
\begin{subequations}\label{eq:eq3.6}
\begin{equation}\label{eq:eq3.6a}
m(z) = I_2 + \mathcal{O}(1/z), \quad z \to \infty,
\end{equation}
\begin{equation}\label{eq:eq3.6b}
m(z) = \frac{q_o}{z} \sigma_2 e^{-i \theta \sigma_3}
+ \mathcal{O}(1), \quad z \to 0, \quad z \in \mathbb{C}^{\pm}.
\end{equation}
\end{subequations}
\item[5.] $m(z)$ satisfies the symmetries: 
\begin{equation}\label{e:equation42}
m(z) = \sigma_1 m^*(z^*) \sigma_1, \quad m ( q_o^2/z) = \frac{z}{q_o} \, m(z) \, \sigma_2 \, e^{-i \theta \sigma_3}.\end{equation}
\end{itemize}
\end{RHP}
For brevity, in the statement of future RHPs, we refer to 1(b) as the condition that $m(z)$ takes continuous boundary values away from the set $\{0\} \cup \{z_j\}_{j=1}^N \cup \{z_j^*\}_{j=1}^N$. Note also that whenever the jump matrix is evaluated on the real $z$-axis, one can use $\bar{\rho}(z)=\rho^*(z)$ and $1+\rho(z)\bar{\rho}(z)=1+|\rho(z)|^2$. However, since in the following we will have to introduce deformations off the real axis, it is preferable to keep both $\rho(z)$ and $\bar{\rho}(z)$, related to each other via symmetry \eqref{e:1st_symm_barrho}. Finally, note that the solution of the above RHP, as well as all the following ones, depends parametrically on $x,t$, but this parametric dependence is usually omitted for brevity unless otherwise specified.

\begin{remark}\label{rem:remark1}
The symmetries \eqref{e:equation42} reflect our use of a slightly different normalization for the Jost eigenfunctions as $x \to \pm \infty$ compared to \cite{CJ2016}. Moreover, we consider an IC with arbitrary amplitude $q_o$ and asymptotic phase difference $\theta_{+} - \theta_{-}\equiv 2\theta $, whereas in \cite{CJ2016}, $q_o=1$ and $\theta=\pi$. While the generalization to arbitrary $q_o\ne 1$ is not essential, since the amplitude can always be rescaled out, being able to include an arbitrary asymptotic phase difference can be relevant for certain applications.
\end{remark}

\begin{proposition}\label{eq:prop1new}
Suppose $m(z)$ is a solution to  RHP \ref{eq:RHP1}. Then the columns $m^+_{1}(z)$ and $m^-_{2}(z)$ have zeros at $z= \pm q_o$ of at least first order.
\end{proposition}
\begin{proof}
Let $m$ be a solution of RHP \ref{eq:RHP1}. Evaluating the jump condition of $m$ at $z= \pm q_o$ we get
\begin{equation*}
m^{+}(x,t;-q_o) = m^{-}(x,t;-q_o) \begin{pmatrix}
0 & i e^{i \theta}\\
i e^{-i \theta} & 1
\end{pmatrix}, \quad m^{+}(x,t;q_o) = m^{-}(x,t;q_o) \begin{pmatrix}
0 & -i e^{i \theta}\\
-i e^{-i \theta} & 1
\end{pmatrix}
\end{equation*}
because $\rho(\pm q_o) = \mp i e^{-i \theta}$ (cf. \eqref{eq:eq222}). The first column of the above equations gives
\begin{gather}\label{eq:437new}
m^+_{1}(x,t;-q_o) = i e^{-i \theta} \, m^-_{2}(x,t;-q_o), \quad m^+_{1}(x,t;q_o) = - i e^{-i \theta} \, m^-_{2}(x,t;q_o).
\end{gather}
Moreover, symmetries \eqref{e:equation42} at $z= \pm q_o$ yield
\begin{gather}
m^+_{1}(x,t;-q_o) = -i e^{-i \theta} \, m^-_{2}(x,t;-q_o), \quad m^+_{1}(x,t;q_o) = i e^{-i \theta} \, m^-_{2}(x,t;q_o),
\end{gather}
which together with Eq.~\eqref{eq:437new} gives 
\begin{gather}
m^+_{1}(x,t;-q_o) = m^-_{2}(x,t;-q_o) =0, \quad m^+_{1}(x,t;q_o) = m^-_{2}(x,t;q_o) =0.
\end{gather}
The analyticity of the jump matrix then implies that $m^+_{1}(z)$ and $m^-_{2}(z)$ have analytic extensions in a neighborhood of $z = \pm q_o$, and therefore the zeros have to be at least of first order.
\end{proof}

\begin{lemma}\label{e:lemma3n}
Any solution $m(z)$ of RHP \ref{eq:RHP1} is such that $\det m(z) = 1 - q_o^2 z^{-2}$.
\end{lemma}
\begin{proof}
The proof of this lemma follows the proof of Lemma 5.3 in \cite{CJ2016}, after re-adjusting the normalization and the symmetries appropriately.
\end{proof}

\begin{remark}\label{eq:remark2}
It is important for future reference to establish the behavior of the columns of $m(x,t;z)$ as $z\to \pm q_o$ from either half-plane. First, note that if $m(x,t;z)$ is defined as in \eqref{eq:function m}, then one can verify that at $t=0$:
\begin{subequations}
\label{e:mt0pmqo}
\begin{gather}
m_1^+(x,0; z)\sim \frac{z\mp q_o}{\alpha_\pm}
\begin{pmatrix}
1 \\ \pm i e^{i\theta} 
\end{pmatrix}, \quad
m_2^+(x,0; z)\sim 
\begin{pmatrix}
\mp i e^{i\theta} \\ 1 
\end{pmatrix}
\qquad z\to \pm q_o, \ z\in \Complex^+, \\
m_1^-(x,0; z)\sim
\begin{pmatrix}
1 \\
\pm ie^{-i\theta} 
\end{pmatrix},
\quad m_2^-(x,0; z)\sim \frac{z\mp q_o}{\alpha_\pm^*}
\begin{pmatrix}
\mp ie^{-i\theta} \\ 1 
\end{pmatrix}\qquad z\to \pm q_o, \ z\in \Complex^-,
\end{gather}
\end{subequations}
where $\alpha_\pm$ are defined in \eqref{eq:eq228nn} (and in the case of box-type ICs they can be explicitly computed, see Appendix~\ref{e:appA}). The above equations show that the relevant columns of $m^\pm(z)$ have simple zeros as $z\to \pm q_o$, at least at $t=0$. Since $q_o,\alpha_\pm,\theta$ are time-independent, we assume that the above behavior holds for all $t\ge 0$, similarly to how one would handle pole singularities corresponding to the discrete eigenvalues. The fact that $m_1^+(x,t;z)$ and $m_2^-(x,t;z)$ have {\textbf{simple}} zeros as $z\to \pm q_o$ for any $t \geq 0$, as it happens at $t=0$ according to Eqs.~\eqref{e:mt0pmqo}, is consistent with Lemma~\ref{e:lemma3n}, which holds for all $t\ge 0$, since a double zero would contradict this lemma. This is important for the deformed RHPs, which will be defined later on, to establish the precise behavior of the respective solutions as $z\to \pm q_o$.
\end{remark}

\begin{lemma}\label{l:RHP1}
If a solution to RHP \ref{eq:RHP1} exists, then it is unique.
\end{lemma}
\begin{proof}

We apply Liouville's theorem to the matrix $m \, \breve{m}^{-1}$, where $m$ and $\breve{m}$ are two arbitrary solutions to RHP \ref{eq:RHP1}. In this case, one of the things to check is whether $m \, \breve{m}^{-1}$ is singular at $z=0$ and/or at the poles from the discrete spectrum, since each of the solutions $m$ and $\breve{m}$ is singular there. Additionally, one must check the points $z=\pm q_o$ where the determinant of both $m$ and $\breve{m}$ vanishes according to Lemma~3. We write:
\begin{equation}\label{eq:eqratio}
m(z) \, \breve{m}^{-1}(z) = \frac{1}{\det \breve{m}(z)} \, m(z) \sigma_2 \breve{m}^{T}(z) \sigma_2
\end{equation}
and like in the proof of Lemma 5.3 in \cite{CJ2016}, we can check from condition 4. of RHP 1, that $m \, \breve{m}^{-1}$ is bounded at $z=0$. Moreover, one can check from the residue conditions 2. of RHP \ref{eq:RHP1} that $m \, \breve{m}^{-1}$ has no poles at $z_j$ and $z_j^*$, for $j=1,2,\cdots N$. These cases were already addressed in the proof of Lemma 5.3 in \cite{CJ2016} in a similar manner. In addition to that proof, note that 
\begin{equation}\label{eq:detnew}
\det \breve{m}(z) = \mathcal{O}(z \mp q_o), \quad z\to \pm q_o,
\end{equation}
which follows directly from Lemma~\ref{e:lemma3n}. Using Eqs.~\eqref{eq:eqratio}-\eqref{eq:detnew}, Proposition~\ref{eq:prop1new} and the fact that any solution to RHP \ref{eq:RHP1} takes continuous boundary values away from the set $\{0\} \cup \{z_j\}_{j=1}^N \cup \{z_j^*\}_{j=1}^N$, then we can show that $m \, \breve{m}^{-1}$ is bounded at $z=\pm q_o$, leaving no singularities at these points. Uniqueness of solution of RHP \ref{eq:RHP1} then follows from the large-$z$ asymptotics of $m \, \breve{m}^{-1}$.
\end{proof}

\begin{remark}
   Note that in RHP~1 the behavior at zero can be inferred from the second symmetry in \eqref{e:equation42}.
\end{remark}

Finally, using the asymptotic property of $M_{\pm}(z)$ at infinity and definition \eqref{eq:function m}, one then reconstructs the potential via the following relation
\begin{equation}\label{eq:recformula}
q(x,t) = \lim_{{\substack{z \to \infty \\ z\in \Complex^+}}}  \Big( i z \, m_{12} (x,t,z) \Big)
\end{equation}
where the subscript denotes the $(1,2)$ entry of the matrix function $m$. 

Our aim is to solve RHP \ref{eq:RHP1} numerically and use relation \eqref{eq:recformula} to recover $q(x,t)$ at arbitrary values of $x\in \Real$, $t>0$. However, RHP \ref{eq:RHP1} presents two numerical challenges. The first arises from the presence of rapidly oscillatory exponential terms in the off-diagonal entries of the jump matrix, which we address by deforming the contour away from the real axis, turning the oscillations into exponential decay. The second challenge arises from Eq.~\eqref{eq:eq3.6b}, which indicates that the matrix $m(z)$ has a singularity at $z=0$. The following sections detail our approach to overcoming these challenges.

\begin{remark}
We emphasize that the order in which we address the aforementioned challenges is crucial. Specifically, we first perform the necessary contour deformation (opening lenses), and, only afterwards, remove the singularity at $z=0$. This allows us to better track and control the singularities introduced by each transformation, and, more importantly, to determine whether any new singularities at the points $z=\pm q_o$ are introduced during the deformation away from the real axis. Reversing this order—removing the singularity first and then opening lenses—makes it significantly harder to justify and manage the resulting analytic structure. This is because removing the singularity at $z=0$ breaks the system's second symmetry \eqref{e:equation42}, which encodes important information about how the respective transformation behaves at $z=\pm q_o$.
\end{remark}

\begin{remark}
Our calculations in Appendix~\ref{e:appA} show that for a box-type IC $\rho(z)$ only decays like $1/z$ as $z\to \infty$, and this means that, without deformation, one has to use Riemann-Hilbert theory for jump matrices in $L^2(\mathbb R) \cap L^\infty(\mathbb R)$. However, provided that $\rho(z) e^{2 t i \Theta(\xi,z)}$ has a principal value integral, this theory can be made effective.  Importantly, after deformation, our work shows that it is possible to find an accurate numerical solution to RHP \ref{eq:RHP1} for any $t>0$ despite this challenge and, in turn, compute the solution $q(x,t)$ for the defocusing NLS with a box-type IC. We believe this IST approach can also be effective for other ICs for which the reflection coefficient only decays as $1/z$ as $z\to \infty$.
\end{remark}

\section{RHP analysis and contour deformation}
\label{s:RHP}

This section focuses on two key issues: the deformation of contours according to the principles of the nonlinear steepest descent method, and the development of an appropriate transformation to remove the singularity of the ``deformed'' RHP at $z=0$. The goal of the contour deformations is ultimately to convert the oscillatory RHP in Eq.~\eqref{eq:eq2.27b} for $z \in \mathbb{R}$ into RHPs with jump matrices that are exponentially decaying for $t\to \infty$. This is the natural approach when one is interested in estimating the long-time asymptotic behavior of the RHP solution and of the corresponding potential $q(x,t)$, but it is also crucial for the numerical solution of the RHP at finite $(x,t)$.

We give an overview of the process and technicalities in the so-called solitonic region, see Sec.~\ref{s:solitonregion}.  We deform the RHP for $m$ (RHP~\ref{eq:RHP100}) to a new RHP for $\hat m$ (RHP~\ref{eq:RHP3}) by passing through an intermediate unknown $\check m$ and solve an auxiliary RHP for $\delta$  (RHP~\ref{eq:RHP4}) which accounts for the remaining jump across (parts of) the real axis.  This involves the classical lensing procedure and careful tracking of singularities.  For $t >0$ this process results in an RHP that has its jump matrix concentrated along steepest descent contours emanating from the origin,  with the added complication of a pole-like singularity at the origin.  In a key development, we introduce a new unknown $\widetilde{m}$ (RHP~\ref{eq:RHP4a}) that is not singular at the origin.
We then establish that if RHP~\ref{eq:RHP4a} has a unique solution, we can build $\hat m$ using $\widetilde{m}$ by means of left multiplication by the matrix $E(z)$ in \eqref{e:def_mtilde}.  The RHP for $\widetilde{m}$ is numerically tractable using standard methods.  The solitonless regions (Sec.~\ref{s:solitonlessregion}) are handled similarly, with additional deformations required due the presence of stationary phase points on the real axis.

\subsection{Analysis of the phase function}

\label{s:phasefunction}
Recall that our aim is to solve RHP \ref{eq:RHP1} numerically, and use it to compute the potential $q(x,t)$. As shown in \cite{CJ2016,Fan2022}, the long-time asymptotic analysis identifies three regions in the space-time domain of $x$ and $t$, depending on the value of the parameter $\xi = x/(2t)$, with different characteristics: the solitonic region when $|\xi|<q_o$, the solitonless region when $|\xi|>q_o$, and the collisionless shock region when $|\xi|=q_o$. While the analysis in \cite{CJ2016,Fan2022} is carried out for the specific case of $q_o=1$, the results can be generalized to any arbitrary value of $q_o$. Therefore, we express the jump matrix $G$ in terms of $\xi$ as follows:
\begin{gather}
G(\xi,z) = \begin{pmatrix}
1-\rho(z)\bar{\rho}(z) & -\bar{\rho}(z) e^{-2t \, i \Theta(\xi,z)}\\
\rho(z) e^{2t \, i \Theta(\xi,z)} & 1
\end{pmatrix}, \quad \Theta(\xi,z) = 2 \left( \lambda(z) \xi + k(z) \lambda(z)\right).
\end{gather}
Note that $G$ involves the exponentials $e^{\pm 2 t \, i \Theta(\xi,z)}$ which are oscillatory for $z \in \mathbb{R}$ where the jump of the RHP is defined. Thus, solving RHP \ref{eq:RHP1} numerically involves implementing contour deformations away from the real axis following the principles of nonlinear steepest descent, similar to those used for estimating the long-time asymptotics (see \cite{DZ1993}, \cite{DVZ1997} and \cite{DZV1998} for more details). We want to emphasize that the deformations are valid for arbitrary $t$,
and they have the advantage that, for large $t$, the  exponential decay off the real axis makes the numerical scheme more efficient, by reducing the number of terms that need to be computed. Therefore, we implement the deformations based on the sign chart of $\Re(i \Theta)$ to achieve exponential decay for large $t$ of all the jumps off the real axis. 

First, we compute the stationary phase points of the function $\Theta$. We have:
\begin{align}
\Theta'(z) & = \xi \left( 1+q_o^2 z^{-2} \right)+\left( z+q_o^4 z^{-3}\right)\\ \notag & = z^{-1}\Big( \xi(z+q_o^2 z^{-1}) + (z+q_o^2 z^{-1})^2-2q_o^2 \Big) \\ \notag & \equiv z^{-1} l(s),
\end{align}
where
\begin{gather}
l(s) = s^2 + \xi s - 2q_o^2, \quad s=z+q_o^2 z^{-1},
\end{gather}
and also
\begin{gather}
\Theta''(z) = -z^{-2} l(s)+z^{-1}l'(s)s'(z) \equiv -z^{-2} l(s) + z^{-1} \left( 1-\frac{q_o^2}{z^2}\right)l'(s).
\end{gather}
Simple analysis shows that real zeros of $\Theta'(z)$ exist if and only if $|\xi| \geq q_o$. In particular, we find that the (real) zeros of $\Theta'$, when $|\xi|\geq q_o$, are given by:
\begin{subequations}\label{eq:zeros}
\begin{gather}
z_k(\xi) = \frac{1}{2} \left| \nu(\xi) + (-1)^{k} \sqrt{\nu^2(\xi) - 4 q_o^2} \right|, \quad k=1,2, \quad \xi \leq -q_o,\\
\hat{z}_k(\xi) = -\frac{1}{2} \left| \nu(\xi) + (-1)^{k} \sqrt{\nu^2(\xi) - 4 q_o^2} \right|, \quad k=1,2, \quad \xi \geq q_o, \\
\nu(\xi) = \frac{1}{2}\left( |\xi|+\sqrt{8 q_o^2 + \xi^2}\right).
\end{gather}
\end{subequations}
It is easy to check that 
\begin{subequations}\label{eq:eq57new}
\begin{gather}
0<z_1(\xi)<q_o<z_2(\xi), \quad \xi<-q_o,\\
\hat{z}_2(\xi)<-q_o<\hat{z}_1(\xi)<0, \quad \xi>q_o.
\end{gather}
\end{subequations}
Furthermore, when $\xi= \pm q_o$ the zeros coincide and collapse to the same value $z_{k}(-q_o)= q_o$ and  $\hat{z}_{k}(q_o)= -q_o$, for $k=1,2$. Based on these observations, we can summarize the following:
\begin{enumerate}
\item[•] There is no stationary phase point on $\mathbb{R}$ when $|\xi|<q_o$. 
\item[•] There is one stationary phase point on $\mathbb{R}$ when $|\xi|=q_o$. 
\item[•] There are two stationary phase points on $\mathbb{R}$ when $|\xi|>q_o$. 
\end{enumerate}

Finally, a direct calculation yields
\begin{gather}
\Re (i \Theta(\xi,z)) = - \xi \, \Im z \left( 1 + \frac{q_o^2}{
|z|^2}\right) - \Re z \, \Im z \left( 1 + \frac{q_o^4}{|z|^4} \right).
\end{gather}

\begin{figure}[t!]
\centering
\begin{subfigure}{0.25\textwidth}
    \includegraphics[width=\textwidth]{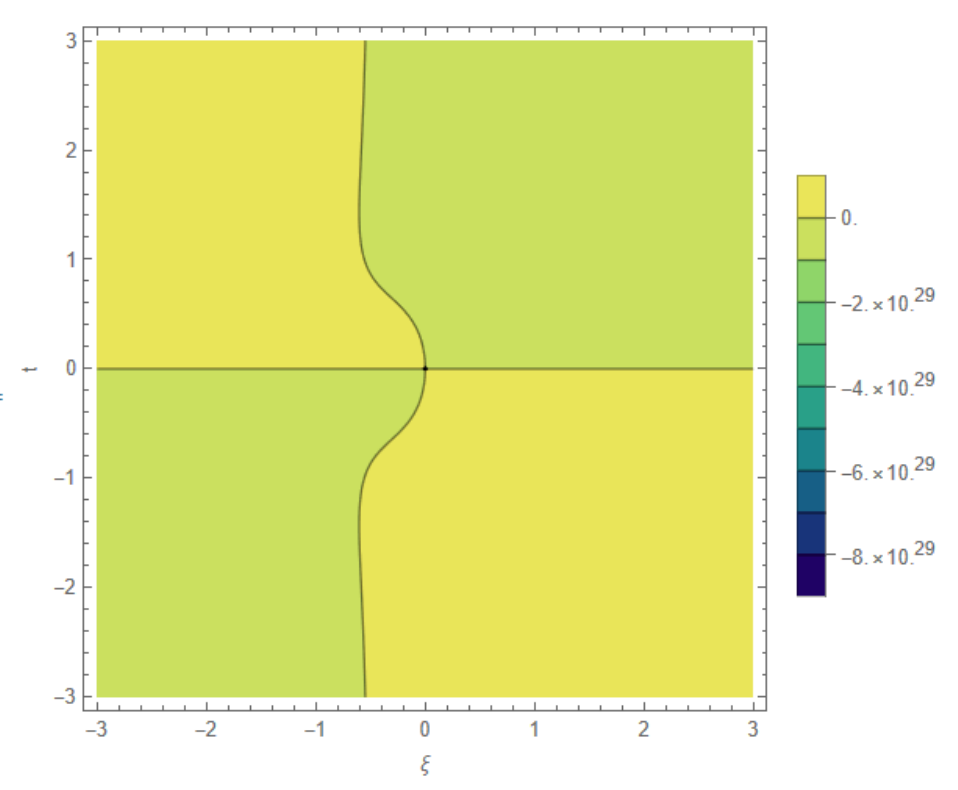}
    \caption{$0 < \xi < 1$}
    \label{fig:first}
\end{subfigure}
\label{fig:figures}
\hfill
\begin{subfigure}{0.25\textwidth}
    \includegraphics[width=\textwidth]{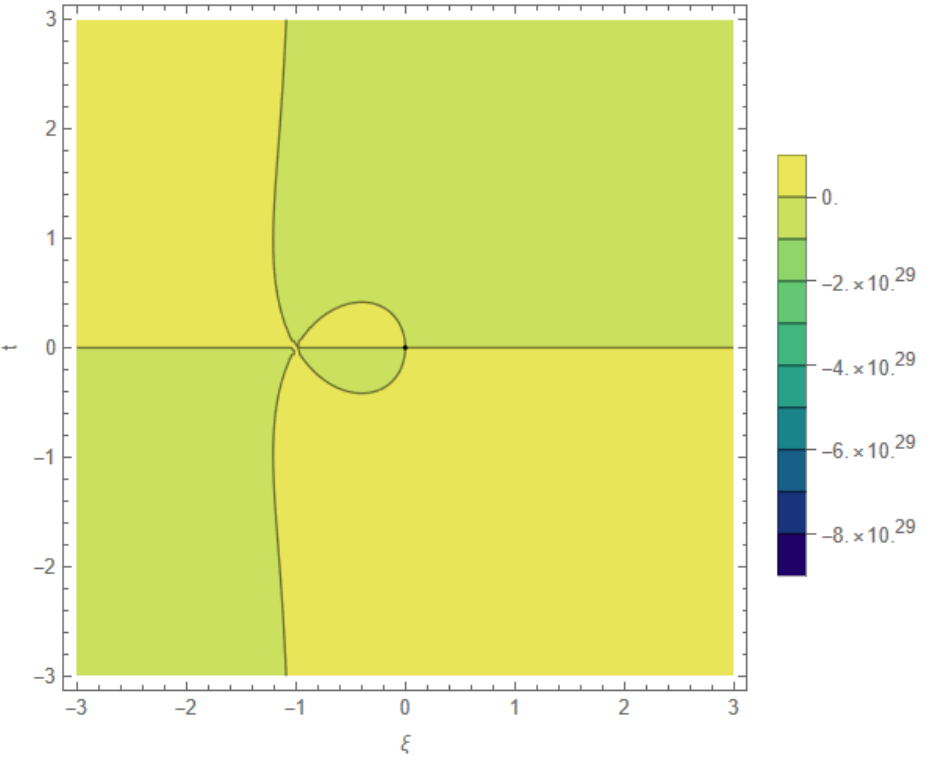}
    \caption{$\xi =1$}
    \label{fig:second1}
\end{subfigure}
\hfill
\begin{subfigure}{0.25\textwidth}
    \includegraphics[width=\textwidth]{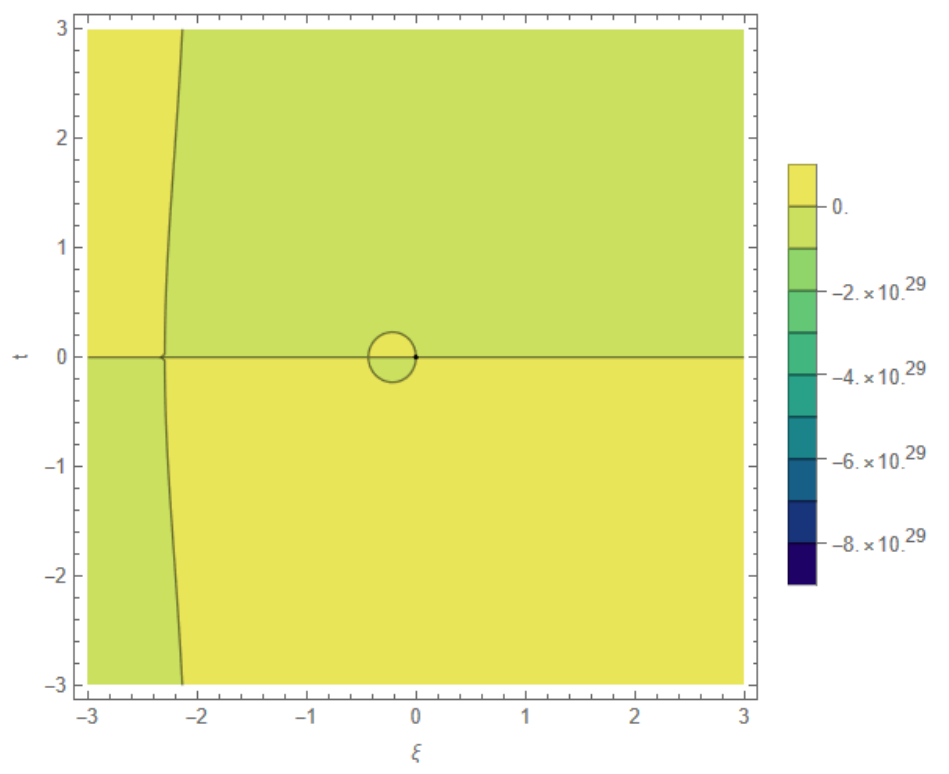}
    \caption{$\xi >1$}
    \label{fig:second2}
\end{subfigure}
\hfill
\begin{subfigure}{0.25\textwidth}
    \includegraphics[width=\textwidth]{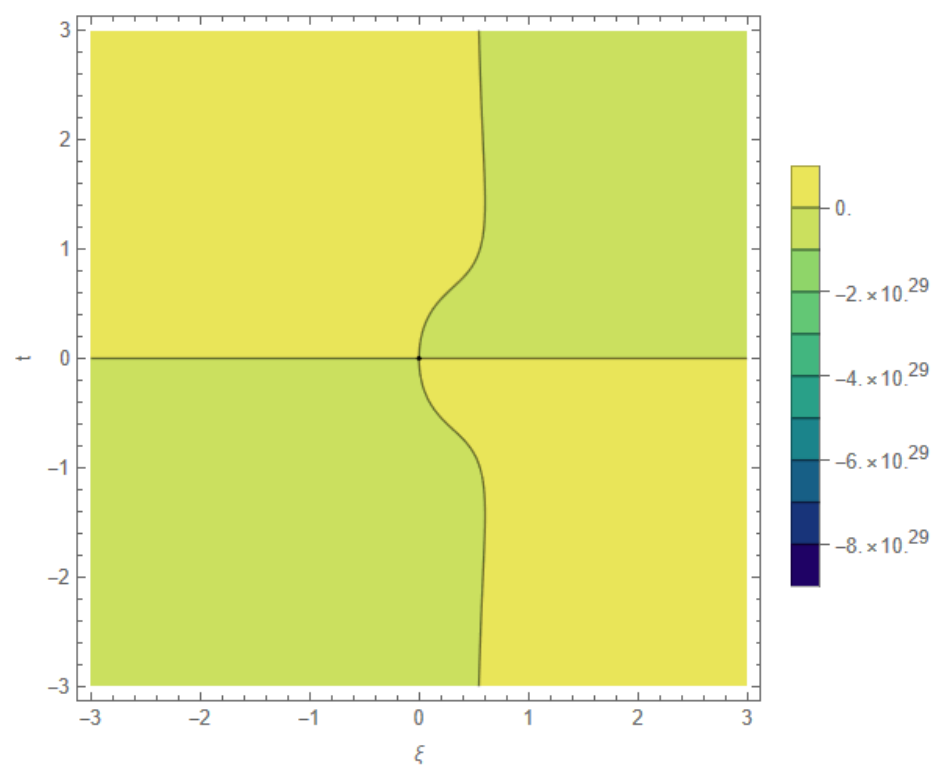}
    \caption{$-1<\xi<0$}
    \label{fig:second3}
\end{subfigure}
\hfill
\begin{subfigure}{0.25\textwidth}
    \includegraphics[width=\textwidth]{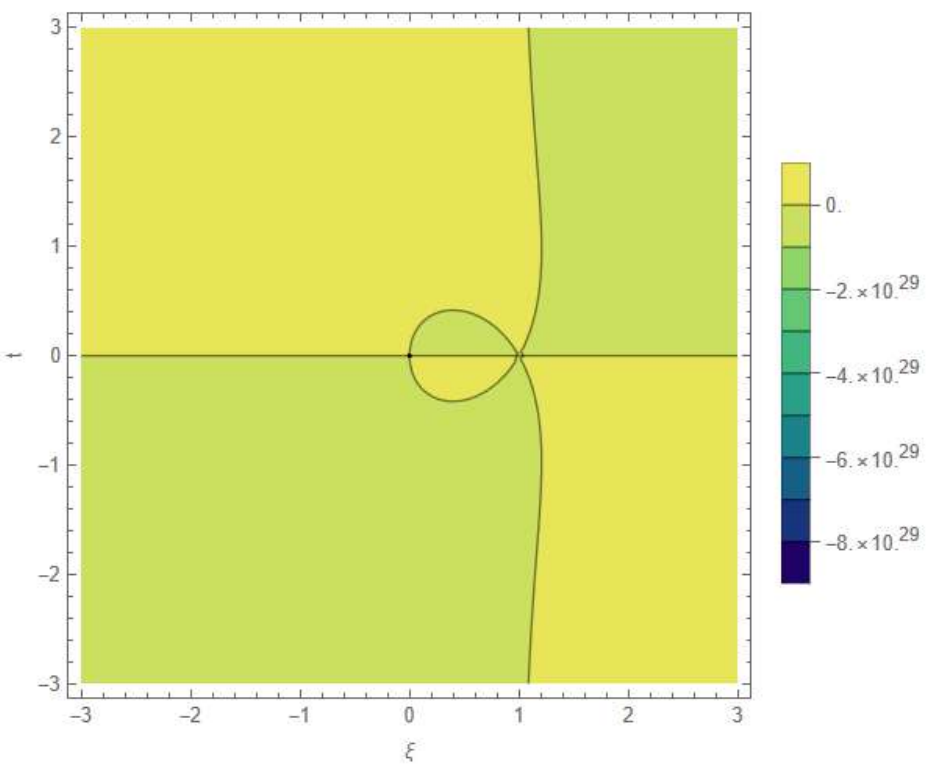} 
    \caption{$\xi=-1$}
    \label{fig:second4}
\end{subfigure}
\hfill
\begin{subfigure}{0.25\textwidth}
    \includegraphics[width=\textwidth]{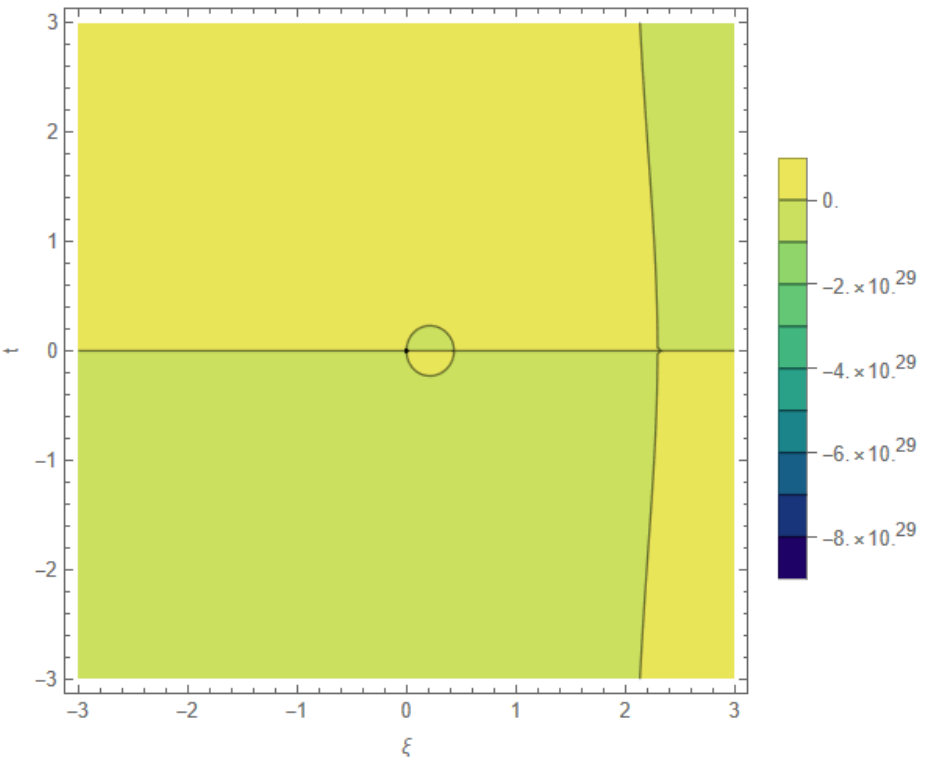}
    \caption{$\xi<-1$}
    \label{fig:second5}
\end{subfigure}
\caption{The sign chart of $\Re(i \Theta)$, when $|\xi|<1$, $|\xi|>1$ and $|\xi|=1$ for the specific case $q_o=1$. The yellow region corresponds to $\Re(i \Theta) >0$, and the green region corresponds to $\Re(i \Theta) <0$.}

\label{e:figgure2}
\end{figure}

Fig.~\ref{e:figgure2} shows the sign chart of $\Re (i \Theta)$ in the three regions for $q_o=1$, along with the location of the stationary phase points in each region. Similar figures can be obtained for arbitrary $q_o$, but the range of $x$-values must be scaled appropriately to account for the location of the stationary phase points.
\begin{remark}
It is clear from Fig.~\ref{e:figgure2} that the required contour deformations will be different in the three regions, and they also depend on whether solitons are present or not. For the remainder of this paper, we focus on a box-type IC that does not support solitons, and we solve the associated RHP numerically in the solitonic and solitonless regions. The case of solitons is beyond the scope of this study, and will be addressed in a future work. Additionally,  the detailed study of the deformations in the collisionless shock region remains an open problem also as far as the long-time asymptotics is concerned. As we will show in Sec.~\ref{s:numerics}, however, the numerical solution of the RHP in each of the regions $|\xi|<q_o$ and $|\xi|>q_o$ actually extends well beyond the boundaries of the domain, and the potentials reconstructed in each region show excellent agreement with each other, at least for bounded times. This seems to indicate that for the purpose of the numerical IST a separate study of the collisionless shock region, though desirable, is not required to compute solution profiles.
\end{remark}

In the following, we are going to assume that the IC is of box-type, and specifically of the form \eqref{eq:IC}.
As shown in \cite{BP2014}, if the asymptotic phase difference $\theta=0$ and $0<h<q_o$, the scattering problem always has at least one discrete eigenvalue. On the other hand, if $\theta=0$ and $h>q_o$, any choice of the box width $L$, and any $\alpha\in[0,\pi] $ satisfying: 
\begin{equation}
\label{e:nosolion_theta0}
0\le \alpha <\arccos{(q_o/h)}
\end{equation}
yields an IC for which no discrete eigenvalue is present. Conversely, if we assume $\alpha=0$,  for any choice of $L$ and $h>q_o$, any sufficiently small asymptotic phase difference $\theta$, satisfying the condition
\begin{equation}
\label{e:no_solitons_theta}
\sin \theta<\left( 1-\frac{q_o}{h}\right)\tanh{\left(2q_oL\sqrt{\frac{h^2}{q_o^2}-1}\right)}
\end{equation}
(cf. Eq.~(45) in \cite{BP2014}) also produces ICs with no discrete eigenvalues.

For definiteness, in most examples we will  consider Eq.~\eqref{eq:dNLS} with NZBCs \eqref{eq:NZBCs}, IC \eqref{eq:IC}, and the values
\begin{gather}\label{eq:values}
L=1, \quad q_o=1, \quad \theta =0, \quad \alpha=0, \quad h = 1.5.
\end{gather}
For these box parameter values, the expressions \eqref{e:ab} for the scattering coefficients and the reflection coefficient simplify to
\begin{subequations}\label{eq:eq318}
\begin{equation}\label{eq:eq318a}
a(z) = e^{2 i \lambda(z)} \left( \cos(2 \mu(z)) + i \frac{h-k^2(z)}{\lambda(z) \mu(z)} \sin(2 \mu(z)) \right),
\end{equation}
\begin{equation}\label{eq:eq318b}
b(z) = \frac{k(z)(h-1)}{\lambda(z) \mu (z)}\sin(2 \mu(z)),
\end{equation}
\begin{equation}\label{eq:eq318c}
\rho(z) = e^{-2 i \lambda(z)} \left( \frac{i k(z)(h-1)}{k^2(z) - h + i \lambda(z) \mu(z) \cot(2 \mu(z))} \right),
\end{equation}
\end{subequations}
with:
\begin{equation}
\label{e:klamu}
k(z)=\frac{1}{2}(z+1/z), \qquad \lambda(z)=\frac{1}{2}(z-1/z), 
\qquad \mu(z)=\sqrt{k(z)^2-h^2}\,.
\end{equation}
The function $\mu(z)$ has branching, with real branch points $k(z)=\pm h$, i.e. at $z=h\pm \sqrt{h^2+q_o^2}$ and $z=-h\pm \sqrt{h^2+q_o^2}$. However, as mentioned in Sec.~\ref{e:eqscdata}, the values of the scattering coefficients are independent of the choice of the branch for $\mu(z)$ since the dependence on $\mu$ is only via $\sin(2\mu(z))/\mu(z)$ or $\mu(z)\cot (2\mu(z))$, i.e., even functions of $\mu$. The box parameters \eqref{eq:values} are chosen so that $a(z)$ has no zeros in $\mathbb{C}^{+}$ (and $a^*(z^*)$ has no zeros in $\mathbb{C}^{-}$), indicating no solitons are present. Other choices of parameters which yield a purely radiative  (i.e., solitonless) solution,  e.g., satisfying \eqref{e:nosolion_theta0} or \eqref{e:no_solitons_theta}, will also be considered for illustrative purposes.

It is worth mentioning that since we are currently restricting ourselves to the solitonless case, $\tilde{q}(x)+H^{1,1}(\Real)$ provides an appropriate space for the ICs. Although our box-type ICs do not lie in this space, due to their piecewise-smooth, compactly supported (in relation to the background) nature, the scattering theory can nonetheless be implemented, analytically and numerically, without a hitch, giving an appropriately analytic reflection  coefficient.

As pointed out in Sec.~\ref{s:IST}, the scattering coefficients generically have simple poles at $z=\pm q_o$ unless the columns of $M_{\pm}$ are linearly dependent at $z=q_o$ or $z=-q_o$ or both, in which case $\alpha_{\pm}=0$. However, in this case it is straightforward to verify that $\alpha_{\pm} \ne 0$.

Moreover, the reflection coefficient can be analytically continued off the real $z$-axis, and it is in fact analytic in the entire complex plane. In this case, RHP \ref{eq:RHP1} simplifies into the following one.
\begin{RHP}[RHP for $m$ without solitons]\label{eq:RHP100} Find a $2 \times 2$ matrix-valued function $m(z)$ such that:
\begin{itemize}
    \item[1.] $m(z)$ is sectionally analytic in $\mathbb{C} \setminus \mathbb{R}$ taking continuous boundary values.
    \item[2.] The boundary values $m^{\pm}(z) = \lim_{\substack{\zeta \to z \\ \zeta \in \mathbb{C}^{\pm}}} m(\zeta)$ for any $z \in \mathbb{R} \setminus \{ 0 \}$ satisfy the jump relation across the real axis
\begin{equation}\label{eq:eq410}
m^{+}(z) = m^{-}(z) G(\xi,z), \quad G(\xi,z) = \begin{pmatrix}
1-\rho(z)\bar{\rho}(z) & -\bar{\rho}(z) e^{-2t \, i \Theta(\xi,z)}\\
\rho(z) e^{2t \, i \Theta(\xi,z)} & 1
\end{pmatrix}.
\end{equation}
\item[3.] $m(z)$ has the following asymptotic behavior as $z\to \infty$ and as $z\to 0$:  
 
\begin{gather}\label{eq:eqzerobeh}
m(z) = I_2 + \mathcal{O}(1/z), \quad z \to \infty, \\
m(z) = \frac{q_o}{z} \sigma_2 e^{-i \theta \sigma_3} + \mathcal{O}(1), \quad z \to 0, \quad z \in \mathbb{C}^{\pm}.
\end{gather}
\item[4.] $m(z)$ satisfies the symmetries:  
\begin{gather}\label{eq:eq66nw}
m(z) = \sigma_1 m^*(z^*) \sigma_1, \quad m (q_o^2/z) = \frac{z}{q_o} \, m(z) \, \sigma_2 e^{-i \theta \sigma_3} .\end{gather}
\end{itemize}
\end{RHP} 
In the following sections, we provide the necessary deformations that lead to a numerical method that is accurate for arbitrary $x\in \Real$, $t>0$ in both the solitonic and the solitonless regions.

\subsection{Jump matrix factorizations}
\label{s:jumpfact}

Direct calculations show that the jump matrix has the following two factorizations:
\begin{gather}
G(\xi,z) = M(\xi,z)P(\xi,z) \equiv L(\xi,z) D(z) U(\xi,z),
\end{gather}
where
\begin{subequations}\label{eq:eq321}
\begin{equation}\label{eq:eq321a}
M(\xi,z)=\begin{pmatrix}
1 & - \bar{\rho}(z) e^{-2t \, i \Theta(\xi,z)}\\
0 & 1
\end{pmatrix}, \quad P(\xi,z) = \begin{pmatrix}
1 & 0\\
\rho(z) e^{2t \, i \Theta(\xi,z)} & 1
\end{pmatrix},
\end{equation}
\begin{equation}\label{eq:eq321b}
L(\xi,z)=\begin{pmatrix}
1 & 0\\
\frac{\rho(z)}{1-\rho(z)\bar{\rho}(z)} e^{2t \, i \Theta(\xi,z)} & 1
        \end{pmatrix}, \quad U(\xi,z)=\begin{pmatrix}
1 & -\frac{\bar{\rho}(z)}{1-\rho(z)\bar{\rho}(z)}e^{-2t \, i \Theta(\xi,z)}\\
0 & 1     \end{pmatrix},
\end{equation}
\begin{equation}\label{eq:eq321c}
D(z)=\begin{pmatrix}
1-\rho(z)\bar{\rho}(z) & 0\\
0 & \frac{1}{1-\rho(z)\bar{\rho}(z)}
        \end{pmatrix}.
\end{equation}
\end{subequations}
Note that each of the matrices $P, M, L$, and $U$ involves only one of the exponents $e^{\pm 2 t \, i \Theta(\xi,z)}$, and therefore depending on the sign of $\mathrm{Re}(i \Theta)$, we use the appropriate factorizations of $G$ in the appropriate regions off the real axis. Next, we present the deformations in the solitonic and solitonless region, separately.

\subsection{Solitonic region}
\label{s:solitonregion}

In this subsection, we provide the required contour deformations in the solitonic region, corresponding to $\xi \in (-1,1)$. First, we choose a sufficiently small positive angle $\phi$ to ensure the deformed contours remain inside regions with the same sign in the sign chart and open lenses around $z=0$. While $\phi=\pi/4$ is the optimal choice for the direction of the fastest decay in the nonlinear steepest descent method, the primary requirement is simply to select a small positive angle $\phi$ that maintains the consistency of the sign in the sign chart of $\mathrm{Re}(i \Theta)$.

\begin{figure}[t!]
\centering
\includegraphics[width=\linewidth]{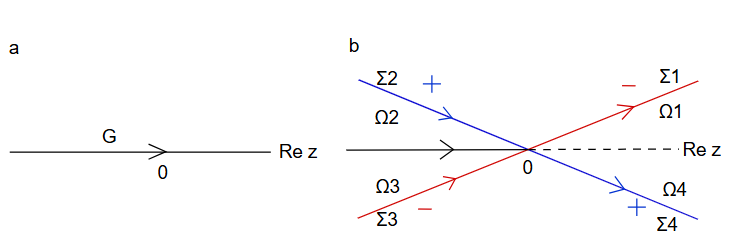}
\caption{Panel a: the initial jump for the function $m$. Panel b: the new jump contours after we open lenses in the solitonic region, and the sign of $\Re(i \Theta)$ when $-1 <\xi < 1$. $+$ indicates that $\Re(i \Theta) >0$ in the corresponding regions, and $-$ stands for $\Re(i \Theta) <0$.}

\label{fig:figgure1}
\end{figure}

Four new regions are therefore created, which we denote by $\Omega_k$, $k=1,\cdots,4$, given by
\begin{subequations}
\begin{gather}
\Omega_1 = \{z: \mathrm{arg}\, z \in (0,\phi)\}, \quad \Omega_2 = \{z: \mathrm{arg}\, z \in (\pi - \phi,\pi)\},\\
\Omega_3 = \{z: \mathrm{arg}\, z \in (-\pi, -\pi + \phi)\}, \quad \Omega_4 = \{z: \mathrm{arg}\, z \in (-\phi,0)\}.
\end{gather}
\end{subequations}
\vspace{-0.3cm}
Finally, denote by
\begin{subequations}
\begin{gather}
\Sigma_1 = e^{i \phi} \mathbb{R}_{+}, \quad \Sigma_2 = e^{i(\pi - \phi)}\mathbb{R}_{+}, \quad \Sigma_3 = e^{-i(\pi - \phi)} \mathbb{R}_{+}, \quad \Sigma_4 = e^{- i \phi} \mathbb{R}_{+},
\end{gather}
\end{subequations}
the left-to-right oriented boundaries of $\Omega = \bigcup_{k=1}^{4} \Omega_{k}$, see Fig.~\ref{fig:figgure1}.
Panels $(a)$ and $(d)$ in Fig.~\ref{e:figgure2} suggest to use the LDU-factorization in the regions $\Omega_2 \cup \Omega_3$, and the PM-factorization in the regions $\Omega_1 \cup \Omega_4$.

Next, we define the function $\check{m}(z)$:
\begin{gather}\label{eq:m_{1,d}*}
\check{m}(z) = \begin{cases}
m(z) P^{-1}(\xi,z), & z \in \Omega_1\\[5pt]
m(z) U^{-1}(\xi,z), & z \in \Omega_2\\[5pt]
m(z) L(\xi,z), & z \in \Omega_3\\[5pt]
m(z) M(\xi,z), & z \in \Omega_4\\[5pt]
m(z), & \text{elsewhere}
\end{cases}
\end{gather}
which satisfies the following conditions:
\begin{itemize}
    \item[1.] $\check{m}(z)$ is analytic for $z\in \mathbb{C} \setminus \Sigma$, where $\Sigma = (-\infty,0) \cup \bigcup_{j=1}^{4} \Sigma_j$, taking continuous boundary values.
    \item[2.] The boundary values of $\check{m}(z)$ satisfy the following jump relation across $\Sigma$:
\begin{subequations}
\begin{equation}
\check{m}^+(z) = \check{m}^-(z) \check{G}(\xi, z), \quad z \in \Sigma
\end{equation}
with jump matrix given by
\begin{equation}
\check{G}(\xi,z) = \begin{cases}
P(\xi,z), & z \in \Sigma_1\\[5pt]
U(\xi,z), & z \in \Sigma_2\\[5pt]
L(\xi,z), & z \in \Sigma_3\\[5pt]
M(\xi,z), & z \in \Sigma_4\\[5pt]
D(z), & z \in (-\infty,0)\\[5pt]
\end{cases}
\end{equation}
\end{subequations}
Here, $+/-$ are the limiting values of the function $\check{m}(z)$ as $z$ approaches the oriented contours of non-analyticity from the left and right, respectively.
\item[3.] $\check{m}(z)$ has the following asymptotic behavior:
\begin{equation}\label{eq:eq73nw}
\check{m}(z) = I_2 + \mathcal{O}(1/z), \quad z \to \infty
\end{equation}
\item[4.] For $t > 0$, $\check{m}(z)$ satisfies:
\begin{equation}
\label{e:mcheck0}
\check{m}(z) = \frac{q_o}{z} \sigma_2 e^{-i \theta \sigma_3} + \mathcal{O}(1), \quad z \to 0, \quad z \in \mathbb{C}^{\pm}.
\end{equation}
\item[5.] $\check{m}(z)$ satisfies the symmetries:  
\begin{gather}\label{eq:eq75nw}
\check{m}(z) = \sigma_1 \check{m}^*(z^*) \sigma_1, \quad \check{m} (q_o^2/z) = \frac{z}{q_o} \, \check{m}(z) \, \sigma_2 e^{-i \theta \sigma_3}.
\end{gather}
Note that since all the matrices in \eqref{eq:m_{1,d}*} have determinant equal to 1, according to Lemma~3 we also have $\det \check{m}(z)=1-q_o^2/z^2$.
\begin{proof}
Most of the properties above follow in a straightforward way from the definition \eqref{eq:m_{1,d}*} of $\check{m}(z)$ in terms of $m(z)$, and the corresponding properties of $m(z)$. It is worth discussing in some detail the behavior of $\check{m}(z)$ as $z\to 0$. Direct calculations show that: 
\begin{align*}
\check{m}_{11}(z) \sim iq_+\frac{\rho(z)}{z}e^{2it\Theta(z)}  & \qquad z\to 0,\ z\in \Omega_1, \\
\check{m}_{22}(z) \sim iq_-\frac{\rho^*(z^*)}{z(1-\rho(z)\rho^*(z^*))}e^{-2it\Theta(z)} & \qquad z\to 0, \ z\in \Omega_2, \\
\check{m}_{11}(z) \sim -iq_+\frac{\rho(z)}{z(1-\rho(z)\rho^*(z^*))}e^{2it\Theta(z)} & \qquad z\to 0, \ z\in \Omega_3, \\
\check{m}_{22}(z) \sim -iq_-\frac{\rho^*(z^*)}{z} e^{-2it\Theta(z)} & \qquad z\to 0, \ z\in \Omega_4,
\end{align*}
where we recall that $q_\pm =q_oe^{\pm i \theta}$, and the asymptotic behavior of all the remaining entries of $\check{m}(z)$ is given by \eqref{eq:eq73nw}. 
In order to verify \eqref{eq:eq73nw}, one then has to show that the non-tangential limits of the above diagonal entries in each sector $\Omega_j$ are zero, which is the case provided $t > 0$. This can be shown for instance by letting $z=\zeta e^{i\alpha}$ with $\alpha$ such that $z\in \Omega_j$, and taking the limit $\zeta\to 0^+$ so that $z\to 0$ in $\Omega_j$ along a ray, while using the explicit asymptotics of $\rho(z)=b(z)/a(z)$ as $z\to 0$ which can be obtained from the formulas in Appendix~\ref{e:appA}. For instance, for $z=\zeta e^{i\alpha}\in \Omega_1$ one has:
$$
|\rho(z)/z|\sim (h-1) e^{\sin \alpha/\zeta}, \qquad 0<\alpha<\phi<\pi/4,
$$
[with box parameters as in \eqref{eq:values}] which grows exponentially as $\zeta\to 0^+$, but as long as $t>0$
the necessary decay is provided by $e^{2it\Theta(z)}$, since
$$
|e^{2it\Theta(z)}|\sim e^{-t\sin(2\alpha)/\zeta^2}.
$$
Similarly, for $z=\zeta e^{i\alpha}\in \Omega_3$
$$
\left| \frac{\rho(z)}{z(1-\rho(z)\rho^*(z^*))} \right| \sim \frac{1}{h-1} \frac{1}{\zeta^2} e^{-\sin\alpha/\zeta}, \qquad
-\pi<\alpha<-\pi+\phi
$$ 
which grows exponentially as $\zeta\to 0^+$, but again for $t>0$ the necessary decay is provided by $e^{2it\Theta(z)}$.

As to the symmetries, taking into account that for $z \in \mathbb{C}^{+}$:
\[\rho(q_o^2/z)= - e^{-2i \theta} \rho^*(z^*), \quad \bar{\rho}(z) = \rho^*(z^*)\]
and $\Theta(q_o^2/z) = - \Theta(z)$, one can derive the following symmetries: 
\begin{subequations}\label{eq:eq71}
\begin{equation}\label{eq:eq71a}
\sigma_1 M^*(\xi,z^*) \sigma_1 = P^{-1}(\xi,z), \quad \sigma_1 L^*(\xi,z^*) \sigma_1 = U^{-1}(\xi,z)
\end{equation}
\begin{equation}\label{eq:eq71b}
e^{i \theta \sigma_3} \, \sigma_2 \, M(\xi,q_o^2/z) \, \sigma_2 \, e^{-i \theta \sigma_3} = P^{-1}(\xi,z), \quad e^{i \theta \sigma_3} \, \sigma_2 \, L(\xi,q_o^2/z) \, \sigma_2 \, e^{-i \theta \sigma_3} = U^{-1}(\xi,z).
\end{equation}
\end{subequations}
These symmetries, combined with the known symmetries of $m(z)$, then yield the desired symmetry relations for $\check{m}(z)$.
\end{proof}
\end{itemize}

\begin{remark}
We emphasize that in all RHPs beginning from the lens-opening step onward, it is essential to assume $t>0$. This assumption is crucial because, as demonstrated in the preceding proof, the existence of the non-tangential limit of $\check{m}(z)$ as $z \to 0$ relies on the exponential decay provided by the phase function. Without this decay, which fails to hold when $t=0$, even the non-tangential limit as $z \to 0$ fails to exist. This is not a crucial restriction, however, since at $t=0$, we do not need to solve the RHP, as the explicit solution is provided by the given IC. It is also worth pointing out that for smooth ICs one does not need to assume $t>0$, since in this case $\rho(z)$ decays at least like $z^2$ as $z\to 0$, and therefore one does not need the decay from the phase function.
\end{remark}

Next, we discuss the behavior of $\check{m}(z)$ at $z=\pm q_o$.

\begin{remark}\label{eq:remark5}
Since in $\Omega_1$ and $\Omega_4$, $\check{m}(z)$ is related to $m(z)$ via the matrices $P^{-1}$ and $M$, which are bounded at $z=q_o$, $\check{m}^\pm(q_o)$ are bounded, but, unlike the corresponding columns of $m(z)$, $\check{m}^+_1(q_o)\ne0$, $\check{m}^-_2(q_o)\ne 0$, although one still has $\det \check{m}^\pm(q_o)=0$ because the two columns are now proportional. Specifically:
\begin{equation}
\label{e:mcheck_qo}
\check{m}_1^\pm (q_o)=ie^{-i\theta}\check{m}_2^\pm(q_o),
\quad \check{m}_2^\pm(q_o)=m_2^+(q_o),
\end{equation}
where we have used \eqref{eq:eq222}.
Conversely, for $z \in \Omega_2$, $\check{m}(z) = m(z) U^{-1}(\xi, z)$, and for $z \in \Omega_3$, $\check{m}(z) = m(z) L(\xi,z)$. Proposition \ref{eq:prop3} shows that the $(1,2)$-entry of $U(\xi,z)$ has a double pole at $-q_o$, and the $(2,1)$-entry of $L(\xi,z)$ has a double pole at $-q_o$. Using Proposition~\ref{eq:prop1new} and Remark~\ref{eq:remark2}, one can then show that:
\begin{subequations}
\begin{gather}
    \check{m}^+_1(z) = \mathcal{O}(z+q_o) \quad \text{as } z\to -q_o, \ z\in \Complex^+ \\
    (z+q_o)\check{m}_2^+(z) = 
    \mathcal{O}(1) \quad \text{as } z\to -q_o, \ z\in \Complex^+ \\ 
    (z+q_o)\check{m}_1^-(z) = \mathcal{O}(1) \quad \text{as } z\to -q_o, \ z\in \Complex^- \\
    \check{m}^-_2(z) = \mathcal{O}(z+q_o) \quad \text{as } z\to -q_o, \ z\in \Complex^- 
\end{gather}
\end{subequations}
where $\check{m}_j(z)$ denotes the $j$-th column of $\check{m}(z)$. Therefore, $\check{m}(z)$ is singular at $z=-q_o$. However, as we will see below, we can formulate an RHP for a new function $\hat{m}(z)$ which is continuous across $\Real^-$, and which we can prove is non-singular at $-q_o$.
\end{remark}

It is worth noticing that for large $t$ the jumps across $\Sigma_j$, for $j=1,\cdots,4$ are exponentially close to the identity matrix, as the exponents $e^{\pm 2 t \, i \Theta(\xi,z)}$ decay rapidly. This leads to increased efficiency in the numerical scheme,  
since fewer terms need to be computed. However, this does not happen with the jump matrix $D(z)$ across $\Real^-$. Next, we define the matrix function $\Delta(z)$ explicitly as follows:
\begin{equation}\label{eq:eq326}
\Delta(z) = \mathrm{diag} \Big( \delta(z),1/\delta(z) \Big), \quad \delta(z) = \mathrm{exp} \left[ \frac{1}{2 \pi i} \int_{-\infty}^{0} \log \left(1-|\rho(s)|^2 \right) \left(  \frac{1}{s-z} - \frac{1}{2s} \right)\, ds \right],
\end{equation}
where $\delta(z)$ satisfies the following scalar RHP.
\begin{RHP}[RHP for $\delta$]\label{eq:RHP4} Find a scalar function $\delta(z)$ such that:
\begin{itemize}
    \item[1.] $\delta(z)$ is analytic for $\in \mathbb{C} \setminus (-\infty,0)$, taking continuous boundary values away from $-q_o$.
    \item[2.] $\delta(z)$ satisfies the jump relation
\begin{equation}
 \delta^{+}(z) = \delta^{-}(z) \left(1-|\rho(z)|^2 \right) , \quad z \in (-\infty,0).
\end{equation}
\item[3.] $\delta(z) = \delta_{\infty} + \mathcal{O}(1/z)$ as $z \to \infty$, where 
\begin{equation}
\delta_{\infty} = \mathrm{exp} \left[ - \frac{1}{4 \pi i} \int_{-\infty}^{0} \frac{\log \left(1-|\rho(s)|^2 \right) }{s} \, ds \right].
\end{equation}
\item[4.] $\delta(z)$ has the following asymptotic behavior as $z\to - q_o$:
\begin{subequations}
\begin{equation}
\delta(z) (z+q_o)^{-1} = \mathcal{O}(1), \quad z \to -q_o, \quad z \in \mathbb{C}^{+}
\end{equation}
\begin{equation}
\delta(z) (z+q_o) = \mathcal{O}(1), \quad z \to -q_o, \quad z \in \mathbb{C}^{-}.
\end{equation}
\end{subequations}
\item[5.] $\delta(z)$ satisfies the symmetries:
\begin{equation}
\label{e:symm_delta}
\delta^*(z^*) = \delta^{-1}(z) = \delta(q_o^2/z).
\end{equation}
\end{itemize}
\end{RHP}
In Sec.~\ref{s:numerics} and in Appendix~\ref{appendix:a}, we provide details on the numerical computation and the analytical derivation of the function $\delta(z)$, respectively. 
Note that in the numerical code we use:
\[\tilde{\delta}(z) = \mathrm{exp} \left[ \frac{1}{2 \pi i} \int_{-\infty}^{0} \frac{\log \left(1-|\rho(s)|^2 \right)}{s-z} \, ds \right], \]
which is simply proportional to $\delta(z)$:
\begin{equation}
\label{e:delta_to_hatdelta}
\tilde{\delta}(z) = \delta_{\infty}^{-1}\, \delta(z),
\end{equation}
and such that $\tilde{\delta}(z)\to 1$ as $z\to \infty$.
This definition preserves the first symmetry of $\delta(z)$ ($z \mapsto z^*$) but not the second one ($z \mapsto q_o^2/z$), since
$\tilde{\delta}^*(z^*)=\tilde{\delta}^{-1}(z)$, while $\tilde{\delta}(q_o^2/z)=\delta_\infty^{-2}\tilde{\delta}^{-1}(z)$.
However, one can easily go from $\delta(z)$ to $\tilde{\delta}(z)$
via \eqref{e:delta_to_hatdelta}, and the use of one versus the other amounts to the conjugation of the solution of the RHP by a constant diagonal matrix.

Using $\Delta(z)$, we can remove the jump across $\Real^-$, by letting:
\begin{equation}\label{e:m2d}
\hat{m}(z) = \Delta_{\infty} \, \check{m}(z) \Delta^{-1}(z), \quad \Delta_{\infty} = \mathrm{diag} \left(\delta_{\infty}, 1/\delta_{\infty} \right) 
\end{equation}
which satisfies the following RHP.
\begin{RHP}[RHP for $\hat{m}$]\label{eq:RHP3} Find a $2 \times 2$ matrix-valued function $\hat{m}(z)$ such that:
\begin{itemize}
    \item[1.] $\hat{m}(z)$ is analytic for $z\in \mathbb{C} \setminus \Sigma'$, where $\Sigma' = \bigcup_{j=1}^{4} \Sigma_j$, taking continuous boundary values.
    \item[2.] The boundary values of $\hat{m}(z)$ satisfy the jump relation across $\Sigma'$
\begin{subequations}
\begin{equation}
\hat{m}^+(z) = \hat{m}^-(z) \hat{G}(\xi,z), \quad z \in \Sigma'
\end{equation}
where the jump matrix is given by
\begin{equation}\label{eq:eq329b}
\hat{G}(\xi,z) = \begin{cases}
\Delta(z)  P(\xi,z)  \Delta^{-1}(z), & z \in \Sigma_1\\[5pt]
\Delta(z) U(\xi,z) \Delta^{-1}(z), & z \in \Sigma_2\\[5pt]
\Delta(z) L(\xi,z) \Delta^{-1}(z), & z \in \Sigma_3\\[5pt]
\Delta(z) M(\xi,z) \Delta^{-1}(z), & z \in \Sigma_4
\end{cases}
\end{equation}
\end{subequations}
\item[3.] $\hat{m}(z)$ has the asymptotic behavior:
\begin{equation}
\hat{m}(z) = I_2 + \mathcal{O}(1/z), \quad z \to \infty
\end{equation}
\item[4.] For $t > 0$, $\hat{m}(z)$ is such that:
\begin{equation}\label{eq:eq86new}
\hat{m}(z) = \frac{q_o \sigma_2 e^{-i \theta \sigma_3}}{z} + \mathcal{O}(1), \quad z \to 0, \quad z \in \mathbb{C}^{\pm}
\end{equation}
where we have used that $\delta(0)=1/\delta_\infty$ (consistently with the second symmetry for $\delta(z)$), and the simplification $\Delta_{\infty} \sigma_2 e^{-i \theta \sigma_3} \Delta_{\infty} \equiv \sigma_2 e^{-i \theta \sigma_3}$.
\item[5.] $\hat{m}(z)$ satisfies the symmetries:  
\begin{gather}\label{eq:eq87new}
\hat{m}(z) = \sigma_1 \hat{m}^*(z^*) \sigma_1, \quad \hat{m} (q_o^2/z) = \frac{z}{q_o} \, \hat{m}(z) \, \sigma_2 e^{-i \theta \sigma_3}.
\end{gather}
\begin{proof}
Most of the above properties follow in a straightforward way from the definition of $\hat{m}(z)$ and the corresponding properties of $m(z)$. As to the symmetries, they can be derived using the known symmetries for $\check{m}(z)$, along with the following symmetries for $\Delta$:
\begin{equation}\label{eq:eq81delta}
 \Delta(z) = \sigma_1 \Delta^*(z^*) \sigma_1, \quad \Delta(q_o^2/z) \equiv \Delta^{-1}(z) = \sigma_2 \Delta(z) \sigma_2, \quad \Delta_{\infty} = \sigma_1 \Delta_{\infty}^* \sigma_1.   
\end{equation}
\end{proof}
\end{itemize}
Note that since both $\Delta_\infty$ and $\Delta(z)$ have determinant equal to 1, it follows from \eqref{e:m2d} that 
 $\det \hat{m}(z)=\det \check{m}(z)=\det m(z)=1-q_o^2/z^2$.
\end{RHP}

\begin{remark}
We again observe that Eq.~\eqref{eq:eq86new} is, in a sense, redundant: once the second symmetry in \eqref{eq:eq87new} and the asymptotic behavior of $\hat{m}(z)$ as $z \to \infty$ are established, then \eqref{eq:eq86new} follows. The same holds also for Eqs.~\eqref{eq:eq73nw}--\eqref{eq:eq75nw}.
\end{remark}

\begin{lemma}
\label{lemma5}
If a solution to RHP~\ref{eq:RHP3} exists, then it is unique.
\end{lemma}
\begin{proof}
We apply Liouville's theorem to the matrix $\hat{m}_1 \, \hat{m}_2^{-1}$, where $\hat{m}_1$ and $\hat{m}_2$ are two arbitrary solutions to RHP \ref{eq:RHP3}. $\hat{m}_1 \, \hat{m}_2^{-1}$ is analytic everywhere, since $\hat{m}_1$ and $\hat{m}_2$ have the same jump condition. Moreover, note that at the origin, we have:
\[\lim_{z \to 0} \hat{m}_1 (z) \, \hat{m}_2^{-1}(z) = \lim_{z \to 0} z^2 (z^2 - q_o^2)^{-1} \hat{m}_1(z) \sigma_2 \hat{m}_2^{T}(z) \sigma_2 = I_2\]
where we used the property $\hat{m}_2^{-1}(z) = \left( \det \hat{m}_2 (z) \right)^{-1} \sigma_2 \hat{m}_2^{T}(z) \sigma_2$ for any invertible $2 \times 2$ matrix. Uniqueness then follows from the large-$z$ asymptotics of $\hat{m}_1 \, \hat{m}_2^{-1}$, and the fact that $\det \hat{m}_2(z)=1-q_o^2/z^2$.
\end{proof}

Next, we discuss the behavior of $\hat{m}(z)$ at $z=\pm q_o$.

\begin{remark}
Regarding the boundary values of $\hat{m}(z)$ as $z\to q_o$ from $\Complex^\pm$, we note that since $\check{m}(z)$ is bounded as $z\to q_o$ from $\Complex^\pm$, it follows that $\hat{m}(z)$, defined via \eqref{e:m2d}, is also bounded in this limit. Specifically, one has:
\[
\hat{m}^\pm (q_o)=
\begin{pmatrix}
\frac{\delta_\infty}{\delta(q_o)}\check{m}_{11}^\pm (q_o) &
\delta_\infty \delta(q_o) \check{m}_{12}^\pm (q_o) \\
\frac{1}{\delta_\infty \delta(q_o)}\check{m}_{21}^\pm (q_o) &
\frac{\delta(q_o)}{\delta_\infty} \check{m}_{22}^\pm (q_o)
\end{pmatrix},
\]
and all the entries are non-zero, because all entries of $\check{m}^\pm (q_o)$ are non-zero. On the other hand, $\det \check{m}^\pm(q_o)=0$ on account of \eqref{e:mcheck_qo}, which shows that
\begin{equation}
\label{e:mhat_qo}
\hat{m}_1^\pm (q_o)=\frac{ie^{-i\theta}}{(\delta(q_o))^2}\hat{m}_2^\pm(q_o), \quad
\hat{m}_2^\pm (q_o)=\Delta_\infty \delta(q_o)m_2^+(q_o).
\end{equation}
\end{remark}

\begin{proposition}\label{proposition1}
The boundary values of $\hat{m}(z)$ as $z\to - q_o$ from $\Complex^\pm$ are bounded.
\end{proposition}
\begin{proof}
A direct computation from the definition \eqref{e:m2d} of 
$\hat{m}(z)$   gives:
\begin{subequations}
\label{e:prop4}
\begin{equation}\label{eq:eq81n}
\lim_{\substack{z \to - q_o\  \\ z \in \mathbb{C}^{+}}} \hat{m}(z) = \Delta_{\infty} \lim_{z \to - q_o} \begin{pmatrix}
\frac{m^{+}_{11}(z)}{\delta^{+}(z)} & \quad m^{+}_{11}(z) \delta^{+}(z) \frac{\bar{\rho}(z)}{1-\rho(z) \bar{\rho}(z)} e^{-2t i \Theta(z)} + m^{+}_{12}(z)\delta^{+}(z)\\
\frac{m^{+}_{21}(z)}{\delta^{+}(z)} & \quad m^{+}_{21}(z) \delta^{+}(z) \frac{\bar{\rho}(z)}{1-\rho(z) \bar{\rho}(z)} e^{-2t i \Theta(z)} + m^{+}_{22}(z)\delta^{+}(z)
 \end{pmatrix}
\end{equation}
 and
 \begin{equation}\label{eq:eq82n}
\lim_{\substack{z \to - q_o \\ z \in \mathbb{C}^{-}}} \hat{m}(z) = \Delta_{\infty}\lim_{z \to - q_o }   \begin{pmatrix}
\frac{m^{-}_{11}(z)}{\delta^{-}(z)} + \frac{m^{-}_{12}(z)}{\delta^{-}(z)} \frac{\rho(z)}{1-\rho(z) \bar{\rho}(z)} e^{2t i \Theta(z)} & \quad m^{-}_{12}(z)\delta^{-}(z)\\
\frac{m^{-}_{21}(z)}{\delta^{-}(z)} + \frac{m^{-}_{22}(z)}{\delta^{-}(z)} \frac{\rho(z)}{1-\rho(z) \bar{\rho}(z)} e^{2t i \Theta(z)} & \quad m^{-}_{22}(z)\delta^{-}(z)
 \end{pmatrix}.
 \end{equation}
 \end{subequations}
Using Propositions \ref{eq:prop1new} and \ref{eq:prop3}, along with the known behavior of $\delta(z)$ as $z \to -q_o$ from $\mathbb{C}^{\pm}$, the fact that $\Delta_{\infty}$ is a constant matrix, and $\Theta(-q_o)=0$, we conclude that $\hat{m}^{\pm}(-q_o)$ exist and are finite. 
\newline
Moreover, we can express both columns of $\hat{m}^\pm (-q_o)$ as follows: 
\begin{subequations}
\label{e:mhat_pmqo}
\begin{gather}
\hat{m}_2^+(-q_o)=-ie^{i\theta}\left( \lim_{\substack{z \to -q_o \\ z \in \mathbb{C}^{+}}}
\frac{\delta^2(z)}{1-\rho(z)\bar{\rho}(z)}
\right)\hat{m}_1^+(-q_o), \qquad \hat{m}_1^+(-q_o)=\Delta_\infty \lim_{\substack{z \to -q_o \\ z \in \mathbb{C}^{+}}} \frac{m_1(z)}{\delta(z)},\\
\hat{m}_1^-(-q_o)=ie^{-i\theta}\left( \lim_{\substack{z \to -q_o \\ z \in \mathbb{C}^{-}}}
\frac{1}{\delta^2(z)(1-\rho(z)\bar{\rho}(z))}
\right)\hat{m}_2^-(-q_o), \quad
\hat{m}_2^-(-q_o)=\Delta_\infty \lim_{\substack{z \to -q_o \\ z \in \mathbb{C}^{-}}} \delta(z) m_2(z),
\end{gather}
\end{subequations}
which shows that both columns are non-zero and proportional to each other. This is consistent with $\det \hat{m}^\pm(-q_o)=0$.
The jump condition can be used to show that $\hat{m}(z)$ has no jump across $\Real^-$, so $\hat{m}_j^+(-q_o)=\hat{m}_j^-(-q_o)$ for $j=1,2$.
\end{proof}

\begin{remark}
It follows directly from Eq.~\eqref{e:mhat_qo} that $\hat{m}_1^+ (q_o) = \hat{m}_1^- (q_o)$ and $\hat{m}_2^+ (q_o) = \hat{m}_2^- (q_o)$, for all $t > 0$, which is consistent with the fact that $\hat{m}(z)$ has no jump across $\mathbb{R}^{+}$. Similarly, since $\hat{m}(z)$ has no jump across $\mathbb{R}^{-}$, it follows that $\hat{m}_1^+ (-q_o) = \hat{m}_1^- (-q_o)$ and $\hat{m}_2^+ (-q_o) = \hat{m}_2^- (-q_o)$, for all $t > 0$.  
As a consistency check, one can show that the two limits coincide at $t=0$ using the explicit expressions of $m(x,0;-q_o)$ computed from \eqref{e:mt0pmqo}.
\end{remark} 

\subsubsection{Removing the singularity at $z=0$}\label{subs34}

The contour deformations described in the previous subsection shift the original jump away from the real axis, transforming the oscillatory behavior into exponential decay. As a result, the jump matrices are now along rays where, for large $t$, they approach the identity exponentially fast. However, the matrix function $\hat{m}(z)$ in the last (so far) deformed RHP is still singular at $z=0$, which is numerically challenging. To circumvent this problem, we introduce a function $\widetilde{m}(z)$ that satisfies the same RHP as $\hat{m}(z)$ but is non-singular at $z=0$. Specifically, we define $\widetilde{m}(z)$ via the relation:
\begin{equation}
\label{e:def_mtilde}
\hat{m}(z)=E(z)\widetilde{m}(z), \qquad E(z)= I_2+\frac{q_o\sigma_2e^{-i\theta \sigma_3}}{z}\widetilde{m}^{-1}(0).
\end{equation}
As shown in Appendix~\ref{eq:appendixD}, if $\hat{m}(z)$ satisfies the solvability condition
\begin{equation}\label{eq:solve}
W\left(\hat{m}_2(q_o),\hat{m}_1(-q_o)\right)\ne 0,
\end{equation}
that is, if $\hat{m}_2(q_o)$ and $\hat{m}_1(-q_o)$ are linearly independent, then the function $\widetilde{m}(z)$ defined via relation \eqref{e:def_mtilde} is bounded at $z=\pm q_o$ (for further details, see Remark~\ref{eq:remark12}). Moreover, it satisfies the following RHP.
\begin{RHP}[RHP for $\widetilde{m}$]\label{eq:RHP4a} Find a $2 \times 2$ matrix-valued function $\widetilde{m}(z)$ such that:
\begin{itemize}
    \item[1.] $\widetilde{m}(z)$ is analytic for $z\in \mathbb{C} \setminus \Sigma' $, where $\Sigma' = \bigcup_{j=1}^{4} \Sigma_j$, taking continuous boundary values onto $\Sigma^\prime$.
    \item[2.] The boundary values of $\widetilde{m}(z)$ satisfy the jump relation across $\Sigma'$
\begin{subequations}
\begin{equation}
\widetilde{m}^+(z) = \widetilde{m}^-(z) \widetilde{G}(\xi,z), \quad z \in \Sigma'
\end{equation}
where $\widetilde{G}(\xi,z)$ is the same jump matrix as in Eq.~\eqref{eq:eq329b} (i.e., $\widetilde{G}(\xi,z) = \hat{G}(\xi,z)$).
\end{subequations}
\item[3.] $\widetilde{m}(z)$ has the following asymptotic behavior:
\begin{subequations}
\begin{equation}
\widetilde{m}(z) = I_2 + \mathcal{O}(1/z), \quad z \to \infty.
\end{equation}
\end{subequations}
\item[4.] The non-tangential limits of $\widetilde{m}(z)$ as $z \to 0$ from $\mathbb{C}^{\pm}$ exist and are equal, and we let 
\[\widetilde{m}(x,t;0) \coloneqq \lim_{\substack{\zeta \to 0 \\ \zeta \in \mathbb{C}^{\pm}}} \widetilde{m}(x,t;\zeta).\]
\item[5.] $\widetilde{m}(z)$ satisfies the symmetries 
\begin{gather}\label{eq:eqsym}
\widetilde{m}(z) = 
\sigma_1 \widetilde{m}^*(z^*) \sigma_1, \quad \widetilde{m}(q_o^2/z)=\widetilde{m}(0)\sigma_2 e^{-i\theta \sigma_3}\widetilde{m}(z)\sigma_2 e^{-i\theta \sigma_3}.
\end{gather}
\begin{proof}
This follows directly from symmetries \eqref{eq:eq87new}, the definition of $E(z)$, and relation \eqref{e:def_mtilde}.     
\end{proof}
\end{itemize}
\end{RHP}
\begin{lemma}\label{eq:lemma5nw}
Any solution $\widetilde{m}(z)$ of RHP \ref{eq:RHP4a} is such that $\det \widetilde{m}(z) = 1$, for all $z \in \mathbb{C}$. Consequently, standard RHP theory implies that if RHP~\ref{eq:RHP4a} has a solution, then the solution is unique.
\end{lemma}
\begin{proof}
Let $\widetilde{m}(z)$ be a solution to RHP \ref{eq:RHP4a}. Then, for all $z \in \Sigma'$, $\det \widetilde{m}^{+}(z) = \det \widetilde{m}^{-}(z)$, which implies that $\det \widetilde{m}(z)$ is entire. Moreover, $\det \widetilde{m}(z)=1+\mathcal{O}(1/z)$ as $z \to \infty$, which completes the proof.
\end{proof}
\begin{remark}\label{eq:remark12}
Here, we want to emphasize that although $\widetilde{m}(z)$ defined in Eq.~\eqref{e:def_mtilde} as $\widetilde{m}(z) = E^{-1}(z) \hat{m}(z)$ may appear to have poles at the points where $\det E(z)=0$, namely at $z = \pm q_o$, this is not actually the case. As shown in Appendix~\ref{eq:appendixD}, for any $x,t$ for which $\hat{m}(x,t;z)$ satisfies the solvability condition \eqref{eq:solve},  there exists a unique $\widetilde{m}(x,t;0)$, consistent with the symmetries, that ensures absence of singularities for $\widetilde{m}(x,t;z)$ at $z=\pm q_o$. 
\end{remark}

The question of existence of solution of RHP~\ref{eq:RHP4a} remains an open problem.  We note that by analytic Fredholm theory a solution can only fail on a set of measure zero in the $(x,t)$-plane.  This fact does not appear to adversely affect our numerical scheme.

\subsubsection{Reconstructing the solution of RHP \ref{eq:RHP3}}

The main idea is that we solve numerically the RHP for $\widetilde{m}$ and we recover $\hat{m}$ for any $z$ using relation \eqref{e:def_mtilde}. The following proposition holds.

\begin{proposition}\label{eq:proposition6nw}
If $\widetilde{m}(z)$ solves RHP \ref{eq:RHP4a}, then $\hat{m}(z)$ defined in \eqref{e:def_mtilde}, solves RHP \ref{eq:RHP3}.
\end{proposition}

\begin{proof}
\begin{enumerate}
    \item $\hat{m}(z)$ and $\widetilde{m}(z)$ share the same jump across $\Sigma'$, and $E(z)$ is analytic there, so the jump condition in RHP \ref{eq:RHP3} follows immediately.
    \item The asymptotic behavior of $\hat{m}(z)$ as $z \to 0$ and $z \to \infty$ in RHP \ref{eq:RHP3} follows directly from \eqref{e:def_mtilde} and RHP \ref{eq:RHP4a}.
    \item The first symmetry in \eqref{eq:eq87new} holds similarly. For the second symmetry, observe:
    \[\hat{m}(q_o^2/z) = E(q_o^2/z) \widetilde{m}(0) \sigma_2 e^{-i \theta \sigma_3} E^{-1}(z) \hat{m}(z) \sigma_2 e^{-i \theta \sigma_3}.\]
This simplifies to the desired one after using the property:
\[
\frac{z}{q_o}E^{-1}(q_o^2/z)E(z)=\widetilde{m}(0)\sigma_2e^{-i\theta \sigma_3}
\]
which is obtained directly from the definition of $E(z)$, and Lemma~\ref{eq:lemma5nw}. 
\end{enumerate}
\end{proof}
\begin{remark}
Note that $\hat{m}(z)$ cannot be singular at $z=\pm q_o$, since $E(z)$ is regular at these points and $\widetilde{m}(z)$ is also non-singular there.
\end{remark}
The question of which constraint/constraints are needed on the IC (or, correspondingly, on the scattering data) so that the solvability condition \eqref{eq:eq169n} is satisfied, and whether/how these constraints depend on $x$ and or $t$ is currently an open one. Remark~\ref{e:solv_t0} shows that at $t=0$ the solvability condition is satisfied for any $x\in \Real$ by any box-type IC provided $\theta\ne \pi/2$, but a proof that this is the case also for $t>0$ is currently not available. It is reasonable to conjecture that this constraint be related to the existence of solutions of RHP~\ref{eq:RHP4a}. Indeed, if such a solution $\widetilde{m}(z)$ exists, then, on one hand, it is necessarily non-singular at $z=\pm q_o$, and, on the other hand, by virtue of Lemma~\ref{eq:lemma5nw} this solution is unique. In turn, this unique solution, regular at $z=\pm q_o$, allows one to obtain via \eqref{e:def_mtilde} the (unique, by Lemma~\ref{lemma5}) solution $\hat{m}(z)$ of RHP~\ref{eq:RHP3}. And, based on the calculations in Appendix~D, this implies that: (i) this $\hat{m}(z)$ satisfies the solvability condition (otherwise, the corresponding $\widetilde{m}(z)$ defined via \eqref{e:def_mtilde} would not be bounded at both $z=\pm q_o$); and (ii) $\widetilde{m}(0)$ obtained from RHP~\ref{eq:RHP4a} must be related to $\hat{m}(\pm q_o)$ via Eqs.~\eqref{e:mtilde0}, because this is the unique value of $\widetilde{m}(0)$ which guarantees absence of singularities of $\widetilde{m}(z)$ at $z=\pm q_o$. 
Arguably, the situations when the solvability condition fails to be satisfied are the same as those in which RHP~\ref{eq:RHP4a} does not admit a solution, and this is consistent with the fact that in this case one would have to modify RHP~\ref{eq:RHP4a} so as to include a singularity at either $z=q_o$ or $z=-q_o$.

In our approach, once we choose the IC, we directly assess, at any given $x\in\Real$ and $t>0$, whether a numerical solution $\widetilde{m}(x,t;z)$ to RHP~\ref{eq:RHP4a} exists, and build from it the solution $\hat{m}(x,t;z)$ of RHP~\ref{eq:RHP3}, and, in turn, the potential $q(x,t)$ (see below). The solvability condition, as well as the fact that Eqs.~\eqref{e:mtilde0} hold, can then be checked a posteriori. This is a strong indication that in all these cases one can expect to be able to rigorously prove that indeed  RHP~\ref{eq:RHP4a} has a solution, but this is left for future work.

Now, Proposition~\ref{eq:proposition6nw} implies that the function $\hat{m}(z)$ we recover from relation \eqref{e:def_mtilde}, which is the one obtained after solving numerically for $\widetilde{m}(z)$, is the unique solution to RHP \ref{eq:RHP3}. Moreover, by tracking back the sequence of transformations leading to the original function $m(z)$, we obtain the following reconstruction formula for $m(z)$ in terms of $\hat{m}(z)$:
\begin{equation}\label{eq:eq87}
m(z) = \begin{cases}
\Delta_{\infty}^{-1} \hat{m}(z) \Delta(z) P(z), & z \in \Omega_1\\
\Delta_{\infty}^{-1} \hat{m}(z) \Delta(z) U(z), & z \in \Omega_2\\
\Delta_{\infty}^{-1} \hat{m}(z) \Delta(z) L^{-1}(z), & z \in \Omega_3\\
\Delta_{\infty}^{-1} \hat{m}(z) \Delta(z) M^{-1}(z),& z \in \Omega_4\\
\Delta_{\infty}^{-1} \hat{m}(z) \Delta(z), & z \in \mathbb{C} \setminus \Omega_j, \quad j=1,\cdots,4
\end{cases}.
\end{equation}
Using the conditions of RHP \ref{eq:RHP3}, we can verify that $m(z)$, defined as in \eqref{eq:eq87}, satisfies the conditions of RHP \ref{eq:RHP100}. Hence, by uniqueness, the solution constructed through \eqref{eq:eq87}, is the unique solution to RHP \ref{eq:RHP100}.

\subsubsection{Reconstruction formula for the potential}

Using the above deformations, one can compute $\widetilde{m}(z)$ numerically for any $z$. However, the potential $q(x,t)$ is reconstructed via \eqref{eq:recformula} in terms of the solution of the initial RHP formulated for $m(z)$. Therefore, we need to relate $m(z)$ to $\widetilde{m}(z)$. 
Tracking the subsequent transformations $m(z) \mapsto 
\hat{m}(z) \mapsto \widetilde{m}(z)$, we have:
\begin{equation}\label{eq:eq88n}
m(z) = \Delta_{\infty}^{-1} \left( I_2 + \frac{q_o  \sigma_2 \, e^{-i \theta \sigma_3}}{z} \widetilde{m}^{-1}(0)\right) \widetilde{m}(z) \Delta(z), \quad z \in \mathbb{C} \setminus \bigcup_{j=1}^4 \Omega_j.
\end{equation}
For $z \in \Omega_j$, one can still write the equivalent expression of $m$ in terms of $\widetilde{m}$ using the respective definition of $\check{m}$ in each region. However, the potential $q(x,t)$ is determined by the large-$z$ behavior of $m(x,t,z)$, and $\check{m}$ coincides with $m$ as $z \to \infty$. Thus, Eq.~\eqref{eq:eq88n} is sufficient for this purpose. Using Eqs.~\eqref{eq:recformula} and \eqref{eq:eq88n}, one can recover the potential $q(x,t)$ for any $(x,t)$ via the relation
\begin{equation}\label{eq:potentialq}
q(x,t) = i \left[ \Delta_{\infty}^{-1} \left( \widetilde{m}^{(-1)}(x,t) + q_o \sigma_2 \, e^{-i \theta \sigma_3}\widetilde{m}^{-1}(0) \right) \Delta_{\infty} \right]_{12}
\end{equation}
where $\widetilde{m}^{(-1)}(x,t)$ is the $\mathcal{O}(1/z)$-order term in the asymptotics of the function $\widetilde{m}$ as $z \to \infty$, and the subscript denotes the $(1,2)$ entry of the prescribed matrix.

In Sec.~\ref{s:numerics}, we will discuss how to implement a numerical scheme to
compute the solution to RHP \ref{eq:RHP3} in the solitonic region numerically by solving the corresponding integral equations for $\widetilde{m}$ using the {\tt RHPackage} \cite{rhpackage} and {\tt ISTPackage} packages \cite{istpackage}. 

\subsection{Solitonless region}
\label{s:solitonlessregion}

In this section, we describe the required contour deformations in the solitonless region.  

\subsubsection{The case $\xi < -q_o$}

\begin{figure}[t!]
\centering
\includegraphics[width=1.0\linewidth]{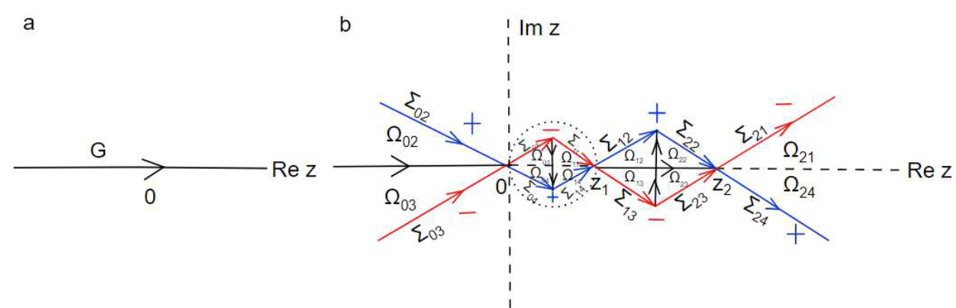}
\caption{Panel a: the initial jump for the function $m$. Panel b: the new jump contours after we open lenses in the solitonless region when $\xi < -q_o$, and the sign of $\Re(i \Theta)$, where $+$ indicates the real part $\Re(i \Theta) >0$ in the corresponding regions, and $-$ stands for $\Re(i \Theta) <0$.}
\label{e:solitonless}
\end{figure}

We first consider the case $\xi < -q_o$, see panel $(f)$ in Fig.~\ref{e:figgure2}. When $\xi > q_o$ (panel $(c)$ in Fig.~\ref{e:figgure2}), the sign chart changes, but the approach for the deformations is similar, and we will discuss it briefly in the next section. In the solitonless region, unlike the solitonic region, the phase function $\Theta$ has two real stationary phase points. Thus, we fix a sufficiently small positive angle $\phi$ to ensure the deformed contours remain inside regions with the same sign in the sign chart, and open lenses around the origin and the two stationary phase points, $z_1$ and $z_2$. Let us consider the real intervals
\begin{gather}\label{e:n154}
l_1 = \left( 0, \frac{|z_1|}{2} \, \sec \phi \right), \quad l_2 =  \left( 0, \frac{|z_2 - z_1|}{2} \, \sec \phi \right)
\end{gather}
and denote the boundaries produced after opening lenses by $\Sigma_{kj}$: 
\begin{subequations}\label{e:n155}
\begin{gather}
\Sigma_{01}=e^{i \phi} l_1, \quad \Sigma_{02} = e^{i (\pi - \phi)} \mathbb{R}^{+}, \quad \Sigma_{11} = z_1 + e^{i (\pi - \phi)} l_1, \quad\Sigma_{12} = z_1 + e^{i \phi} l_2
\\
\Sigma_{21} = z_2 + e^{i \phi} \mathbb{R}^{+}, \quad \Sigma_{22} = z_2 + e^{i (\pi - \phi)} l_2 \\
\Sigma_{k4} = \Sigma_{k1}^*, \quad \Sigma_{k3} = \Sigma_{k2}^*, \quad k=0,1,2.
\end{gather}
\end{subequations}
We then denote by $\Omega_{kj}$, $k=0,1,2$ and $j=1,\cdots,4$ the 12 new regions enclosed by the above boundaries, see Fig.~\ref{e:solitonless}. 
Based on panel $(f)$ in Fig.~\ref{e:figgure2}, we have 
\begin{subequations}\label{eq:eq343}
\begin{gather}
\mathrm{Re}(i \Theta(z)) > 0, \quad \text{for} \quad z \in \Omega_{kj}, \quad k=0,1,2, \quad j=2,4\\
\mathrm{Re}(i \Theta(z)) < 0, \quad
\text{for} \quad z \in \Omega_{kj}, \quad k=0,1,2, \quad j=1,3,
\end{gather}
\end{subequations}
which suggests to use the $\mathrm{LDU}$ factorization for the jump matrix $G$ in the regions $\Omega_{kj}$, for $k=0,1,2$ and $j=2,3$, and the $\mathrm{PM}$ factorization in the regions $\Omega_{kj}$, for $k=0,1,2$ and $j=1,4$.
Accordingly, we define the function $\check{m}$ as follows:
\begin{gather}\label{eq:m_{1,d}}
\check{m}(z) \coloneqq \check{m}(\xi,z) = \begin{cases}
m(z) P^{-1}(\xi,z), & z \in \Omega_{01} \cup \Omega_{11} \cup \Omega_{21}\\[5pt]
m(z) U^{-1}(\xi,z), & z \in \Omega_{02} \cup \Omega_{12} \cup \Omega_{22}\\[5pt]
m(z) L(\xi,z), & z \in \Omega_{03} \cup \Omega_{13} \cup \Omega_{23}\\[5pt]
m(z) M(\xi,z), & z \in \Omega_{04} \cup \Omega_{14} \cup \Omega_{24}\\[5pt]
m(z), & \text{elsewhere}
\end{cases}
\end{gather}

which satisfies the following conditions:

\begin{itemize}
    \item[1.] $\check{m}(z)$ is analytic for $z\in \mathbb{C} \setminus \Sigma$, where $\Sigma = (z_1,z_2) \cup (-\infty,0) \cup \bigcup_{k=0}^{2} \bigcup_{j=1}^{4} \Sigma_{k j}$
    \item[2.] $\check{m}(z)$ satisfies the jump relation across $\Sigma$
\begin{subequations}
\begin{equation}
\check{m}^+(z) = \check{m}^-(z) \check{G}(\xi,z), \quad z \in \Sigma
\end{equation}
with $\check{G}$ given by
\begin{equation}
\check{G}(\xi,z) = \begin{cases}
P(\xi,z), & z \in \Sigma_{01} \cup \Sigma_{11} \cup \Sigma_{21}\\[5pt]
U(\xi,z), & z \in \Sigma_{02} \cup \Sigma_{12} \cup \Sigma_{22}\\[5pt]
L(\xi,z), & z \in \Sigma_{03} \cup \Sigma_{13} \cup \Sigma_{23}\\[5pt]
M(\xi,z), & z \in \Sigma_{04} \cup \Sigma_{14} \cup \Sigma_{24}\\[5pt]
D(z), & z \in (-\infty,0) \cup (z_1,z_2)\\[5pt]
\end{cases}
\end{equation}
\end{subequations}
\item[3.] $\check{m}(z)$ has the following asymptotic behavior:
\begin{equation}
\check{m}(z) = I_2 + \mathcal{O}(1/z), \quad z \to \infty
\end{equation}
\item[4.] For $t>0$, $\check{m}(z)$ satisfies:
\begin{equation}
\label{e:mcheck0}
\check{m}(z) = \frac{q_o}{z} \sigma_2 e^{-i \theta \sigma_3} + \mathcal{O}(1), \quad z \to 0, \quad z \in \mathbb{C}^{\pm}
\end{equation}
\item[5.] $\check{m}(z)$ satisfies the symmetries:  
\begin{gather}
\check{m}(z) = \sigma_1 \check{m}^*(z^*) \sigma_1, \quad \check{m} (q_o^2/z) = \frac{z}{q_o} \, \check{m}(z) \, \sigma_2 e^{-i \theta \sigma_3}.
\end{gather}
\end{itemize}
Note that in order for $\check{m}(z)$ to satisfy the second symmetry for all $z$, the contour deformations have to be chosen appropriately for $|z|<q_o$ based on the contours for $|z|>q_o$. While we believe this can in general be done, it is not practical to enforce it numerically. For this reason, we will not include the symmetries explicitly in some of the deformed RHPs considered below.

\begin{remark}
Since $\Sigma_{02}$ and $\Sigma_{03}$ open around $(-\infty,0)$, and $\Sigma_{12} \cup \Sigma_{22}$ and $\Sigma_{13} \cup \Sigma_{23}$ open around $(z_1,z_2)$, and given that $q_o \in (z_1,z_2)$ and $-q_o \in (-\infty,0)$ (see Eq.~\eqref{eq:eq57new}), together with the fact that the definition of $\check{m}(z)$ in the respective domains involves the matrices $U(\xi,z)$ and $L(\xi,z)$ (whose entries have double poles at $z=\pm q_o$), Remark~\ref{eq:remark5} applies here as well. The key difference compared to the solitonic region is that we must now examine $\check{m}(z)$ as $z \to \pm q_o$, as it may exhibit singular behavior at both points. In fact, from the definition of $\check{m}(z)$ and Proposition~\ref{eq:prop1new}, one can show that:
\begin{itemize}
    \item $\check{m}^+_1(z) = \mathcal{O}(z \mp q_o)$, as $z\to \pm q_o$ from $\Complex^+$    
    \item $\check{m}^+_2(z)$ is singular at $\pm q_o$, with $(z \mp q_o)\check{m}_2^+(z) = \mathcal{O}(1)$, as $z\to \pm q_o$ from $\Complex^+$ 
    \item $\check{m}^-_1(z)$ is singular at $\pm q_o$, with $(z \mp q_o)\check{m}_1^-(z) = \mathcal{O}(1)$, as $z\to \pm q_o$ from $\Complex^-$
    \item  $\check{m}^-_2(z) = \mathcal{O}(z \mp q_o)$, as $z\to \pm q_o$ from $\Complex^-$.
\end{itemize}
\end{remark}

\begin{figure}[th!]
\centering
\includegraphics[width=0.8\linewidth]{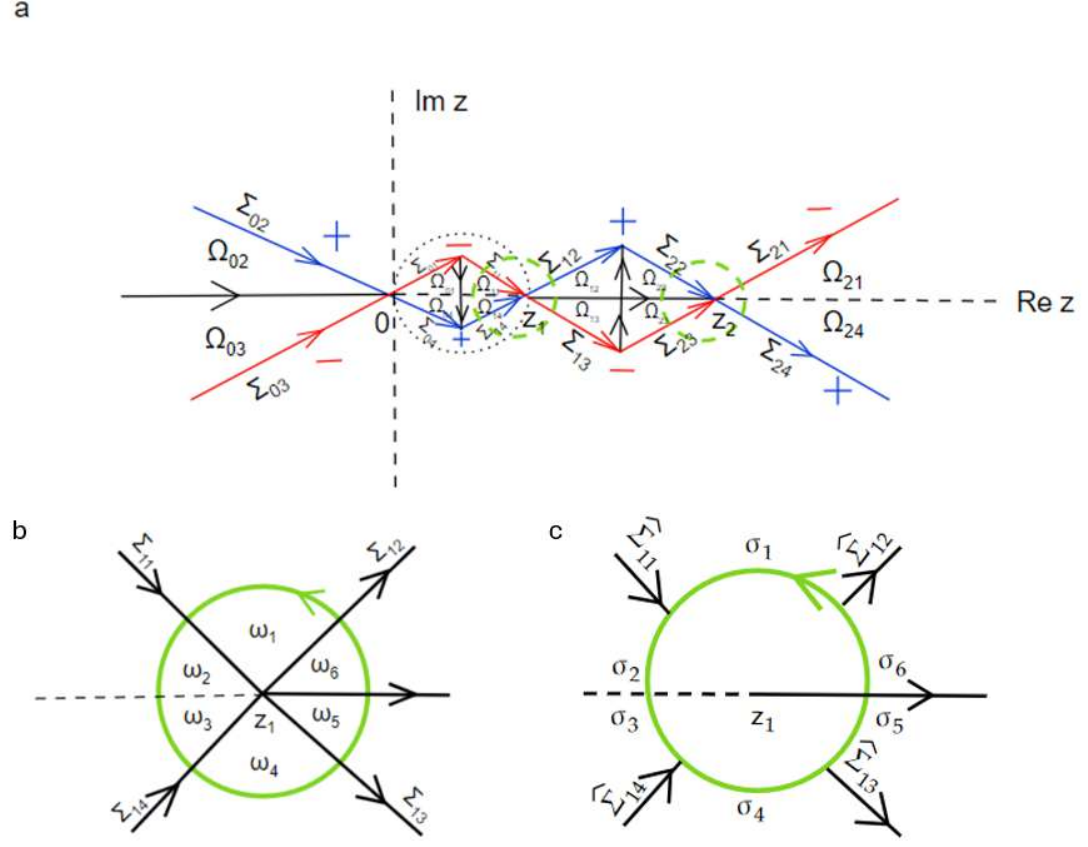}
\caption{Panel a: circles introduced to avoid the singularities of $\hat{\Delta}$ at the two stationary phase points. Panel b: new regions of non-analyticity for the function $\hat{m}$ near $z_1$. Panel c: jump contours for $\hat{m}$ near $z_1$.}
\label{e:figure7n}
\end{figure}

\begin{figure}[ht]
\centering
\includegraphics[width=0.8\linewidth]{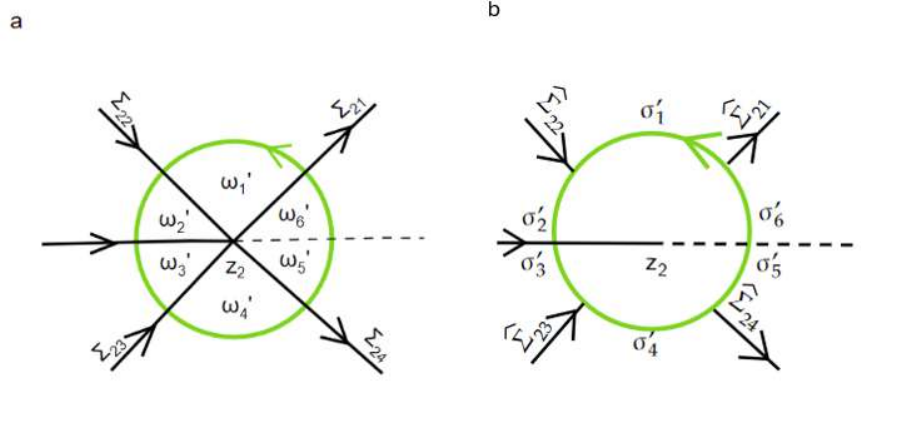}
\caption{Panel a: new regions of non-analyticity for the function $\hat{m}$ near $z_2$. Panel b: jump contours for $\hat{m}$ near $z_2$.}
\label{e:figure8n}
\end{figure}

Like in the solitoni region, we need to introduce an appropriate function $\hat{\Delta}$ to remove the jump matrix $D(z)$ across $(-\infty,0) \cup (z_1,z_2)$. Note that in general, $\hat{\Delta}$ is singular at the stationary phase points. To properly handle the singularities, we introduce circles around both $z_1$ and $z_2$, see Fig.~\ref{e:figure7n}. 
This introduces new regions, denoted by $\omega_j, \omega_j'$, for $j=1,\cdots,6$, see Figs.~\ref{e:figure7n} and \ref{e:figure8n}. 

We define the function $\hat{m}$ as follows:
\begin{gather}\label{eq:m2dnew}
\hat{m}(z) \coloneqq \hat{m}(\xi,z) = \begin{cases}
\check{m}(z) U^{-1}(\xi,z) \hat{\Delta}(z), & z \in \omega_1 \cup \omega_1'\\[5pt]
\check{m}(z) P(\xi,z) U^{-1}(\xi,z) \hat{\Delta}(z), & z \in \omega_2 \cup \omega_3 \cup \omega_5' \cup \omega_6'\\[5pt]
\check{m}(z) M(\xi,z) P(\xi,z) U^{-1}(\xi,z) \hat{\Delta}(z), & z \in \omega_4 \cup \omega_4'\\[5pt]
\check{m}(z) D(z) \hat{\Delta}(z), & z \in \omega_5 \cup \omega_3'\\[5pt]
\check{m}(z) \hat{\Delta}(z), & z \in \omega_6 \cup \omega_2'\\[5pt]
\check{m}(z), & \text{elsewhere}
\end{cases}
\end{gather}
which satisfies the following conditions:
\begin{itemize}
    \item[1.] $\hat{m}(z)$ is analytic for $z\in\mathbb{C} \setminus \Sigma'$, taking continuous boundary values, where
    \begin{gather}\label{eq:eq113nw}
    \Sigma' = \Sigma \cup \bigcup_{j=1}^{6} \sigma_{j} \cup \bigcup_{j=1}^{6} \sigma'_{j}, \quad \Sigma = (-\infty,0) \cup (z_1,z_2) \cup \bigcup_{j=1}^{4} \Sigma_{0 j} \cup \bigcup_{j=1}^4 \hat{\Sigma}_{1 j} \cup \bigcup_{j=1}^{4} \hat{\Sigma}_{2 j}
    \end{gather}
    \item[2.] $\hat{m}(z)$ satisfies the jump relation across $\Sigma'$
\begin{subequations}
\begin{equation}
\hat{m}^+(z) = \hat{m}^-(z) \hat{G}(\xi,z), \quad z \in \Sigma'
\end{equation}
where the jump matrix $\hat{G}$ is given by
\begin{equation}\label{eq:eq348n}
\hat{G}(\xi,z) = \begin{cases}
U^{-1}(\xi,z) \hat{\Delta}(z), & z \in \sigma_1 \cup \sigma_1'\\[5pt]
P(\xi,z) U^{-1}(\xi,z) \hat{\Delta}(z), & z \in \sigma_2 \cup \sigma_3 \cup \sigma_5' \cup \sigma_6'\\[5pt]
M(\xi,z) P(\xi,z) U^{-1}(\xi,z) \hat{\Delta}(z), & z \in \sigma_4 \cup \sigma_4'\\[5pt]
D(z) \hat{\Delta}(z), & z \in \sigma_5 \cup \sigma_3'\\[5pt]
\hat{\Delta}(z), & z \in \sigma_6 \cup \sigma_2'\\[5pt]
D(z), & z \in (-\infty,0) \cup (z_1,z_2)
\end{cases}
\end{equation}
\end{subequations}
and the jumps across the segments $\Sigma_{0j}$, for $j=1,\cdots,4$, and $\hat{\Sigma}_{kj}$, for $k=1,2$ and $j=1,\dots,4$ are the same as the jumps for the function $\check{m}$. Here, $\hat{\Sigma}_{kj}$ denote the portions of the segments $\Sigma_{kj}$ which lie outside the circular regions $\omega_j$ and $\omega_j'$. Notice that $\hat{m}(z)$ is analytic inside the circular regions.
\item[3.] $\hat{m}(z) = I_2 + \mathcal{O}(1/z), \quad z \to \infty$.
\item[4.] Since for $z \in \mathbb{C} \setminus \bigcup_{j=1}^{6} (\omega_j \cup \omega'_j)$, $\hat{m}(z)$ and $\check{m}(z)$ coincide, their limiting behavior at $z=0$ must also coincide, i.e., for $t>0$
\begin{equation}\label{eq:eq115nw}
\hat{m}(z) = \frac{q_o}{z} \sigma_2 e^{-i \theta \sigma_3} + \mathcal{O}(1), \quad z \to 0, \quad z \in \mathbb{C}^{\pm}
\end{equation}
\item[5.] For all $z \in \mathbb{C} \setminus \bigcup_{j=1}^{6}( \omega_j \cup \omega'_j)$, $\hat{m}(z)$ satisfies the same symmetries as $\check{m}(z)$ supposing that the contours have been chosen such that they remain invariant under the symmetry $z \mapsto q_o^2/z^2$.
\end{itemize}

\begin{remark}
Note that since $\hat{m}(z)$ coincides with $\check{m}(z)$ outside $\bigcup_{j=1}^{6} \left(\omega_j \cup \omega'_j\right)$, $\hat{m}(z)$ and $\check{m}(z)$ have the same behavior at $z = \pm q_o$, i.e.,
\begin{itemize}
    \item $\hat{m}^+_1(z) = \mathcal{O}(z \mp q_o)$, as $z\to \pm q_o$ from $\Complex^+$  
    \item $\hat{m}^+_2(z)$ is singular at $\pm q_o$, with $(z \mp q_o)\hat{m}_2^+(z) = \mathcal{O}(1)$, as $z\to \pm q_o$ from $\Complex^+$ 
    \item $\hat{m}^-_1(z)$ is singular at $\pm q_o$, with $(z \mp q_o)\hat{m}_1^-(z) = \mathcal{O}(1)$, as $z\to \pm q_o$ from $\Complex^-$
    \item  $\hat{m}^-_2(z) = \mathcal{O}(z \mp q_o)$, as $z\to \pm q_o$ from $\Complex^-$.
\end{itemize}
\end{remark}

Next, we define the matrix function $\hat{\Delta}(z)$ explicitly as follows: 
\begin{equation}
\hat{\Delta}(z) = \mathrm{diag} \Big( \hat{\delta}(z),1/\hat{\delta}(z) \Big), \quad  \hat{\delta}(z) = \mathrm{exp} \left[ \frac{1}{2 \pi i} \int_{(-\infty,0) \cup (z_1,z_2)} \log \left(1-|\rho(s)|^2 \right) \left(  \frac{1}{s-z} - \frac{1}{2s} \right)\, ds \right]
\end{equation}
and $\hat{\delta}(z)$ satisfies the following scalar RHP.
\begin{RHP}[RHP for $\hat{\delta}$]\label{eq:RHP66} Find a scalar function $\hat{\delta}(z)$ such that:
\begin{itemize}
    \item[1.] $\hat{\delta}(z)$ is analytic in $\mathbb{C} \setminus (-\infty,0) \cup (z_1,z_2)$, taking continuous boundary values away from $\pm q_o$
    \item[2.] $\hat{\delta}(z)$ satisfies the jump relation
\begin{equation}
 \hat{\delta}^{+}(z) = \hat{\delta}^{-}(z) \left(1-|\rho(z)|^2 \right) , \quad z \in (-\infty,0) \cup (z_1,z_2)
\end{equation}
\item[3.] $\hat{\delta}(z) = \hat{\delta}_{\infty} + \mathcal{O}(1/z)$, as $z \to \infty$, where $\hat{\delta}_{\infty} = \mathrm{exp} \left[ - \frac{1}{4 \pi i}  \int_{(-\infty,0) \cup (z_1,z_2)} \frac{\log \left(1-|\rho(s)|^2 \right) }{s} \, ds \right]$.
\item[4.] $\hat{\delta}(z)$ has the following asymptotic behavior as $z\to \pm q_o$:
\begin{subequations}
\begin{gather}
\hat{\delta}(z) (z \mp q_o)^{-1} = \mathcal{O}(1), \quad z \to \pm q_o, \quad z \in \mathbb{C}^{+}
\\
\hat{\delta}(z) (z \mp q_o) = \mathcal{O}(1), \quad z \to \pm q_o, \quad z \in \mathbb{C}^{-}
\end{gather}
\end{subequations}
\item[5.] $\hat{\delta}(z)$ satisfies the symmetries:
\[\hat{\delta}^*(z^*) = \hat{\delta}^{-1}(z) = \hat{\delta}(q_o^2/z).\]
\begin{proof}
Using the definition of $\hat{\delta}(z)$ and observing that the two stationary phase points satisfy the symmetry $\frac{1}{z_1} = \frac{z_2}{q_o^2}$, we can establish the second symmetry above.
\end{proof}
\end{itemize}
\end{RHP}
Using $\hat{\Delta}(z)$, we can remove the jump across $(-\infty,0) \cup (z_1,z_2)$ by letting:
\begin{equation}
\label{e:m3d}
\breve{m}(z) = \hat{\Delta}_{\infty} \, \hat{m}(z) \hat{\Delta}^{-1}(z), \quad \hat{\Delta}_{\infty} = \mathrm{diag} \left(\hat{\delta}_{\infty}, 1/\hat{\delta}_{\infty} \right)
\end{equation}
which satisfies the following RHP.
\begin{RHP}[RHP for $\breve{m}$]\label{eq:RHPnew3} Find a $2 \times 2$ matrix-valued function $\breve{m}(z)$ such that:
\begin{itemize}
   \item[1.] $\breve{m}(z)$ is analytic for $z\in \mathbb{C} \setminus \Sigma''$, where
\begin{equation}\label{eq:equ109}
\Sigma'' = \bigcup_{j=1}^{6} \sigma_{j} \cup \bigcup_{j=1}^{6} \sigma'_{j} \cup \bigcup_{j=1}^{4} \Sigma_{0 j} \cup \bigcup_{j=1}^4 \hat{\Sigma}_{1 j} \cup \bigcup_{j=1}^{4} \hat{\Sigma}_{2 j}
\end{equation}
   \item[2.] $\breve{m}(z)$ satisfies the jump relation across $\Sigma''$
\begin{subequations}
\begin{equation}
\breve{m}^+(z) = \breve{m}^-(z) \breve{G}(\xi,z), \quad z \in \Sigma''
\end{equation}
where the jump matrix $\breve{G}$ is given by
\begin{equation}\label{eq:eq357b}
\breve{G}(\xi,z) = \begin{cases}
\hat{\Delta}(z) P(\xi,z) \hat{\Delta}^{-1}(z), & z \in \Sigma_{01} \cup \hat{\Sigma}_{11} \cup \hat{\Sigma}_{21}\\[5pt]
\hat{\Delta}(z) U(\xi,z) \hat{\Delta}^{-1}(z), & z \in \Sigma_{02} \cup \hat{\Sigma}_{12} \cup \hat{\Sigma}_{22}\\[5pt]
\hat{\Delta}(z) L(\xi,z) \hat{\Delta}^{-1}(z), & z \in \Sigma_{03} \cup \hat{\Sigma}_{13} \cup \hat{\Sigma}_{23}\\[5pt]
\hat{\Delta}(z) M(\xi,z) \hat{\Delta}^{-1}(z), & z \in \Sigma_{04} \cup \hat{\Sigma}_{14} \cup \hat{\Sigma}_{24}\\[5pt]
\hat{\Delta}(z) U^{-1}(\xi,z), & z \in \sigma_{1} \cup \sigma_{1}'\\[5pt]
\hat{\Delta}(z) P(\xi,z) U^{-1}(\xi,z), & z \in \sigma_{2} \cup \sigma_{3} \cup \sigma_{5}' \cup \sigma_{6}'\\[5pt]
\hat{\Delta}(z) M(\xi,z) P(\xi,z) U^{-1}(\xi,z), & z \in \sigma_{4} \cup \sigma_{4}'\\[5pt]
\hat{\Delta}(z) D(z), & z \in \sigma_{5} \cup \sigma_{3}'\\[5pt]
\hat{\Delta}(z), & z \in \sigma_{6} \cup \sigma_{2}'
\end{cases}
\end{equation}
\end{subequations}
where again by $\hat{\Sigma}_{kj}$, for $k=1,2$ and $j=1,\cdots,4$ we denote the portions of the segments $\Sigma_{kj}$ which lie outside the circular regions $\omega_j$ and $\omega_j'$.
\item[3.] $\breve{m}(z)$ has the following asymptotic behavior:
\begin{equation}
\breve{m}(z) = I_2 + \mathcal{O}(1/z), \quad z \to \infty
\end{equation}
\item[4.] For $t>0$, $\breve{m}(z)$ satisfies
\begin{equation}\label{eq:eq123nw}
\breve{m}(z) = \frac{q_o \sigma_2 \, e^{-i \theta \sigma_3}}{z} + \mathcal{O}(1), \quad z \to 0, \quad z \in \mathbb{C}^{\pm}.
\end{equation}
\end{itemize}
\end{RHP}

Next, we discuss the behavior of $\breve{m}(z)$ as $z \to \pm q_o$.

\begin{proposition}\label{proposition1}
The boundary values of $\breve{m}(z)$ as $z \to \pm q_o$ from $\Complex^\pm$ are bounded.
\end{proposition}
\begin{proof}
For all $z \in \mathbb{C} \setminus \bigcup_{j=1}^6 (\omega_j \cup \omega_j')$, $\hat{m}(z)=\check{m}(z)$, and thus Eqs.~\eqref{e:m2d} and \eqref{e:m3d} are equivalent. Furthermore, the functions denoted by $\check{m}(z)$ in the two (solitonic and solitonless)  regions are defined identically in a neighborhood of $-q_o$. Consequently, we may use relations \eqref{e:prop4} for the limiting behavior of $\breve{m}(z)$ at $z=-q_o$. Moreover, from \eqref{eq:m_{1,d}}, observe that $\check{m}(z)$ is defined identically in a neighborhood of both $q_o$ and $-q_o$, and thus the behaviors of $\breve{m}(z)$ at $z=\pm q_o$ coincide. We then write: 
\begin{subequations}
\begin{equation}
\lim_{\substack{z \to \pm q_o\  \\ z \in \mathbb{C}^{+}}} \breve{m}(z) = \hat{\Delta}_{\infty} \lim_{z \to \pm q_o} \begin{pmatrix}
\frac{m^{+}_{11}(z)}{\hat{\delta}^{+}(z)} & \quad m^{+}_{11}(z) \hat{\delta}^{+}(z) \frac{\bar{\rho}(z)}{1-\rho(z) \bar{\rho}(z)} e^{-2t i \Theta(z)} + m^{+}_{12}(z) \hat{\delta}^{+}(z)\\
\frac{m^{+}_{21}(z)}{\hat{\delta}^{+}(z)} & \quad m^{+}_{21}(z) \hat{\delta}^{+}(z) \frac{\bar{\rho}(z)}{1-\rho(z) \bar{\rho}(z)} e^{-2t i \Theta(z)} + m^{+}_{22}(z) \hat{\delta}^{+}(z)
 \end{pmatrix}
\end{equation}
 and
 \begin{equation}
\lim_{\substack{z \to \pm q_o \\ z \in \mathbb{C}^{-}}} \breve{m}(z) = \hat{\Delta}_{\infty}\lim_{z \to \pm q_o }   \begin{pmatrix}
\frac{m^{-}_{11}(z)}{\hat{\delta}^{-}(z)} + \frac{m^{-}_{12}(z)}{\hat{\delta}^{-}(z)} \frac{\rho(z)}{1-\rho(z) \bar{\rho}(z)} e^{2t i \Theta(z)} & \quad m^{-}_{12}(z) \hat{\delta}^{-}(z)\\
\frac{m^{-}_{21}(z)}{\hat{\delta}^{-}(z)} + \frac{m^{-}_{22}(z)}{\hat{\delta}^{-}(z)} \frac{\rho(z)}{1-\rho(z) \bar{\rho}(z)} e^{2t i \Theta(z)} & \quad m^{-}_{22}(z) \hat{\delta}^{-}(z)
 \end{pmatrix}.
 \end{equation}
 \end{subequations}
Using Propositions \ref{eq:prop3} and \ref{eq:prop1new}, along with the known behavior of $\hat{\delta}(z)$ as $z \to \pm q_o$ from $\mathbb{C}^{\pm}$, the fact that $\hat{\Delta}_{\infty}$ is a constant matrix, and $\Theta(-q_o)=0$, we conclude that $\breve{m}^{\pm}(\pm q_o)$ exist and are finite.
\end{proof}

Notice that $\breve{m}(z)$ is singular at $z=0$. Proposition~\ref{proposition1} shows that in the region under consideration, opening lenses does not introduce singularities at $\pm q_o$. Therefore, we can use the same approach as in Sec.~\ref{subs34} to remove the singularity at $z=0$. In this case, we denote the analog of the function $\widetilde{m}(z)$ as $\tilde{m}(z)$, defined by $\breve{m}(z) = E(z) \tilde{m}(z)$, with $E(z)$ given as in \eqref{e:def_mtilde}. Like in the solitonic region, using the above deformations, we can compute $\tilde{m}(z)$ numerically for any $z \in \Complex$. Since in recovering the potential $q(x,t)$, we use the large-$z$ behavior of the respective transformations rather than their local properties, then $q(x,t)$ is still obtained by means of \eqref{eq:potentialq} with $\Delta_{\infty}$ replaced by $\hat{\Delta}_{\infty}$.

\begin{remark}\label{r:circ}
    In performing the deformations depicted in Fig.~\ref{e:figure8n} two things are important: (i) the angle at which the black contours leave each circle, and (ii) the radius of the circle.  For a generic second-order stationary phase point, the second derivative of the phase function $\Theta(z)$ is non-zero, and these contours should all be separated by angles of $\pi/2$, making angles of $\pi/4$ with the real axis, in order to match the local path of steepest descent. As far as (ii) is concerned, this means that the circle intersects the path of steepest ascent, and we may have growth on the circle. To avoid this issue, we choose the radius proportional to $1/\sqrt{\Theta''(z_j)}$ where $z_j$ is the stationary phase point under consideration.  This is feasible because while $\delta$ has a singularity inside the circle, it is a bounded singularity --it behaves as  $(z- z_j)^{\pm i \nu}$ for $\nu > 0$.  This scaling is discussed further in Sec.~\ref{s:numerics} below.  Additionally, in our numerical implementations, we find it more convenient to use squares instead of circles for the local deformations near the stationary phase points, which of course can be done without loss of generality.
\end{remark}

\subsubsection{The case of $\xi > q_o$}

\begin{figure}[t!]
\centering
\includegraphics[width=1.0\linewidth]{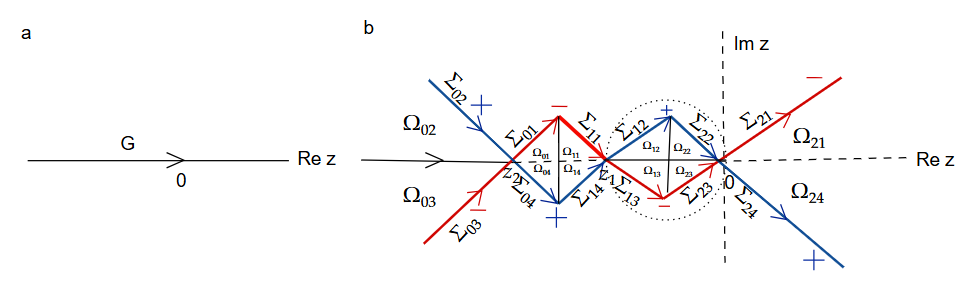}
\caption{Panel a: The initial jump for the function $m(z)$. Panel b: The new jump contours after we open lenses in the solitonless region when $\xi>q_o$, and the sign of $\Re(i \Theta)$, where $+$ indicates the real part $\Re(i \Theta) >0$ in the corresponding regions, and $-$ stands for $\Re(i \Theta) <0$.}
\label{e:solitonless2}
\end{figure}

We now briefly discuss the appropriate deformations for the solitonless region when $\xi > q_o$. Similarly to the case $\xi < -q_o$, we open lenses around the origin and the two stationary phase points, $\hat{z}_1$ and $\hat{z}_2$, creating 12 new regions $\Omega_{kj}$ enclosed by the boundaries $\Sigma_{kj}$, as shown in Fig.~\ref{e:solitonless2}. The boundaries $\Sigma_{kj}$ are now defined as follows:
\begin{subequations}\label{e:new155}
\begin{gather}
\Sigma_{01} = \hat{z}_2 + e^{i \phi} l_2, \quad \Sigma_{02} = \hat{z}_2 + e^{i (\pi - \phi)} \mathbb{R}^{+}, \quad \Sigma_{11} = \hat{z}_1 + e^{i (\pi - \phi)} l_2, \quad \Sigma_{12} = \hat{z}_1 + e^{i \phi} l_1
\\
\Sigma_{21} = e^{i \phi} \mathbb{R}^{+}, \quad \Sigma_{22} = e^{i (\pi - \phi)} l_1
\\
\Sigma_{k4} = \Sigma_{k1}^*, \quad \Sigma_{k3} = \Sigma_{k2}^*, \quad k=0,1,2
\end{gather}
\end{subequations}
and the real intervals $l_1$ and $l_2$ are still given by Eq.~ \eqref{e:n154}. Based on panel $(c)$ in Fig.~\ref{e:figgure2}, Eqs.~\eqref{eq:eq343} also hold in this case, suggesting the use of the $\mathrm{LDU}$ factorization for the jump matrix $G$ in the regions $\Omega_{kj}$, for $k=0,1,2$ and $j=2,3$, and the $\mathrm{PM}$ factorization in the regions $\Omega_{kj}$, for $k=0,1,2$ and $j=1,4$. Consequently, the definition of the first deformation $\check{m}(z)$ is identical to \eqref{eq:m_{1,d}}, and it satisfies the same conditions as in the solitonless region with $\xi < -q_o$, with the contour of non-analyticity now given by $\Sigma =  (-\infty,\hat{z}_2) \cup (\hat{z}_1,0) \cup \bigcup_{k=0}^{2} \bigcup_{j=1}^{4} \Sigma_{k j} $. Thus, the jump matrix $D(z)$ is across the contour $(-\infty,\hat{z}_2) \cup (\hat{z}_1,0)$. To eliminate this jump, we define 
\begin{gather}\label{eq:eq354}
\check{\Delta}(z) = \mathrm{diag} \Big( \check{\delta}(z),1/\check{\delta}(z) \Big), \quad \check{\delta}(z) = \mathrm{exp} \left[ \frac{1}{2 \pi i} \int_{(-\infty,\hat{z}_2) \cup (\hat{z}_1,0)}  \log \left(1-|\rho(s)|^2 \right) \left( \frac{1}{s-z} - \frac{1}{2s} \right) \, ds \right].
\end{gather}
The second deformation $\hat{m}(z)$ is defined as in Eq.~\eqref{eq:m2dnew}, except that in this case $\hat{\Delta}(z)$ is replaced by $\check{\Delta}(z)$, $\omega_j$'s are the new circular regions of non-analyticity of $\hat{m}(z)$ near $\hat{z}_1$ and $\omega_j'$'s are the new circular regions of non-analyticity of $\hat{m}(z)$ near $\hat{z}_2$, for $j=1 \cdots 4$. The function $\hat{m}(z)$ satisfies conditions analogous to those for $\hat{m}(z)$ in the solitonless region with $\xi < -q_o$, except that $\Sigma$ is now replaced by 
\[\Sigma = (-\infty,\hat{z}_2) \cup (\hat{z}_1,0) \cup \bigcup_{j=1}^{4} \hat{\Sigma}_{0j} \cup \bigcup_{j=1}^{4}  \hat{\Sigma}_{1j} \cup \bigcup_{j=1}^{4} \Sigma_{2j}\]
and again $\hat{\Sigma}_{kj}$ denote the portions of the segments $\Sigma_{kj}$, for $k=0,1$ and $j=1 \cdots 4$, which lie outside the circular regions $\omega_j$ and $\omega_j'$. The jump of $\hat{m}(z)$ across $\Sigma$ is given by Eq.~\eqref{eq:eq348n} where $\hat{\Delta}(z)$ is replaced by $\check{\Delta}(z)$, and $D(z)$ is now a jump across $(-\infty,\hat{z}_2) \cup (\hat{z}_1,0)$. Note that $\hat{m}(z)$ is analytic inside the circular regions. To eliminate the jump $D(z)$, we introduce the function $\breve{m}(z)$ given by
\[\breve{m}(z) = \check{\Delta}_{\infty} \, \hat{m}(z) \, \check{\Delta}^{-1}(z)\]
where $\check{\Delta}(z)$ is as in \eqref{eq:eq354}.

Notice that since $\Sigma_{01}$ and $\Sigma_{11}$ (and likewise $\Sigma_{04}$ and $\Sigma_{14}$) open around $(\hat{z}_2,\hat{z}_1)$, and $\Sigma_{21}$ and $\Sigma_{24}$ open around $(0,\infty)$, and given that $-q_o \in (\hat{z}_2,\hat{z}_1)$ and $q_o>0$ (see Eq.~\eqref{eq:eq57new}), together with the fact that $\check{m}(z)$ in the respective domains is defined via the matrices $P(\xi,z)$ and $M(\xi,z)$, which are bounded at $\pm q_o$, then $\check{m}(z)$ is also bounded at these points.

\section{Numerical implementation and {examples}}
\label{s:numerics}
In this section, we describe the numerical code, discuss the numerical challenges, and provide some plots to illustrate the time evolution of the IC for the parameters given in \eqref{eq:values}. For completeness, we also include an example of a box IC with $\theta \ne 0$. A {\tt Mathematica} implementation of the code will be made available as electronic supplementary material. 

To start, we initialize the values $q_o, h, \theta, \alpha, \phi$ as specified in \eqref{eq:values}. Next, we define the reflection coefficient \eqref{eq:eq318} as a function of the uniformization variable $z$ using \eqref{e:klamu}. 
Note that, although we must define numerically the branch cut arising from $\mu=\sqrt{k^2-h^2}$, as clarified earlier the expression of $\rho(z)$ is independent of the choice of the branch cut for $\mu$.

In both the solitonic and solitonless regions, the goal is to numerically solve the RHP associated to the final transformation, i.e., the RHP associated to the function $\widetilde{m}(z)$ in the solitonic region, and with the function $\tilde{m}(z)$ in the solitonless region. For instance, recall that in the solitonic region $\widetilde{m}$ satisfies the jump condition
\begin{gather}\label{eq:eq41}
\widetilde{m}^+(z) = \widetilde{m}^-(z) \, \widetilde{G}(\xi,z), \quad z \in \Sigma' = \bigcup_{j=1}^{4} \Sigma_j
\end{gather}
where $\widetilde{G}$ is given explicitly in Eq.~\eqref{eq:eq329b} (recall that $\widetilde{G}(\xi,z)=\hat{G}(\xi,z)$). For any $(x,t)$ with $|x|<2t$ (so that $|\xi|<q_o$), we write\footnote{This ansatz can be justified by additional estimates on the Jost solutions, showing that they are in an appropriate Hardy space.}
\begin{align*}
 \widetilde{m}(z)= I_2 + \mathcal C_{\Sigma'}u(z)
\end{align*}
for a new unknown matrix-valued function $u$, where we have defined the Cauchy operator for a general contour $\Sigma$
\begin{align*}
\mathcal C_\Sigma u(z) := \frac{1}{2 \pi i} \int_{\Sigma} \frac{u(s)}{s -z} ds.
\end{align*}
We then define the boundary operators
\begin{align*}
    \mathcal C_\Sigma^\pm u(z) := \lim_{\overset{z' \to z}{z' \in \Omega_{z,\pm}}} \mathcal C_\Sigma u(z'),
\end{align*}
where $\Omega_{z,+}$ (resp., $\Omega_{z,-}$) is a region just to the left (resp., right) of $\Sigma$. So, consider the prototypical problem:
\begin{RHP}
Find $m(z)$ satisfying
    \begin{itemize}
        \item[1.] $m(z) = I_2 + \mathcal C_{\Sigma} u(z)$, $u \in L^2(\Sigma)$, and
        \item[2.] $m^+(z) = m^-(z) G(z)$, $z \in \Gamma$.
    \end{itemize}
\end{RHP}
In substituting the expression $m(z) = I_2 + \mathcal C_{\Sigma} u(z)$ into the jump condition, using the Plemelj lemma, $\mathcal C^+_\Sigma - \mathcal C^-_\Sigma = I$, we find the singular integral equation for $u$
\begin{align}\label{eq:uSIE}
    u - (\mathcal C_{\Sigma}^- u) (G - I_2) = G - I_2.
\end{align}
We then write \eqref{eq:eq41} as an integral equation
\begin{gather}\label{eq:eq42}
\widetilde{m}(z) = I_2 + \mathcal C_{\Sigma'} \widetilde{u}(z),\qquad
\widetilde{u} - (\mathcal C_{\Sigma'}^- \widetilde{u}) (\widetilde{G} - I_2) = \widetilde{G} - I_2.
\end{gather}
Similarly, in the solitonless region $\widetilde{m}, \widetilde{u}$ in \eqref{eq:eq42} will be replaced by $\tilde{m}, \tilde{u}$ and $\Sigma'$ will be replaced by $\Sigma''$, where $\Sigma''$ are as in \eqref{eq:equ109}. 

Singular integral equations like \eqref{eq:eq42} are tractable numerically using the method of Olver \cite{SO2012} (see also \cite{TO2015}) provided that the contour $\Sigma'$ is a union of line segments and the jump matrix $\widetilde{G}$ satisfies some regularity conditions \cite{TO2015}. The method is a Chebyshev collocation method that makes use of an explicit representation of the Cauchy integrals, and their boundary values, of Chebyshev polynomials in terms of special functions, or alternatively, one can use Legendre polynomials and their associated three-term recurrence \cite{OST2020}, with special care taken at intersection points.

For the current work, as noted above,  we use the two packages {\tt RHPackage} and {\tt ISTPackage} which are implemented in {\tt Mathematica} \cite{rhpackage,istpackage}.  These packages provide tools for the convenient input of jump contours and jump matrices.  The resulting collocation linear system can then be reliably solved to the desired level of accuracy by monitoring the decay of the Chebyshev coefficients of the resulting approximate solution.  We refer the reader to \cite{TO2015} for more details.

The behavior of the solution $m(z)$ at infinity, or at the origin, can be (formally) computed via
\begin{align}
    m(z) &= I_2 - \frac{1}{2 \pi i z} \int_{\Sigma} u(s) ds + o(z^{-1}), \quad z \to \infty, \label{eq:resinf}\\
    m(z) &= I_2 + \frac{1}{2 \pi i} \int_{\Sigma} \frac{u(s)}{s} ds + o(1), \quad z \to 0.\label{eq:valzero}
\end{align}
To justify these expressions, we see that if $G - I_2 \in L^2(\Sigma) \cap L^1 (\Sigma)$ (recall that sufficient decay is still obtained from the exponential factors appearing in the jump $G$ after we deform the contour of non-analyticity $\Sigma$ off the real axis into regions where these factors decay according to the sign chart of $\Re(i \Theta)$), then \eqref{eq:uSIE} implies that $u \in L^1(\Sigma)$ and \eqref{eq:resinf} follows.  Similarly, when, for simplicity, $\Sigma$ is a union of line segments, if $G -I_2 \in H^1(\Sigma)$ and $G(0) = I_2$, we have that $m(z)$ takes continuous boundary values in a neighborhood of $z = 0$ and the boundary values are all given by \eqref{eq:valzero}.

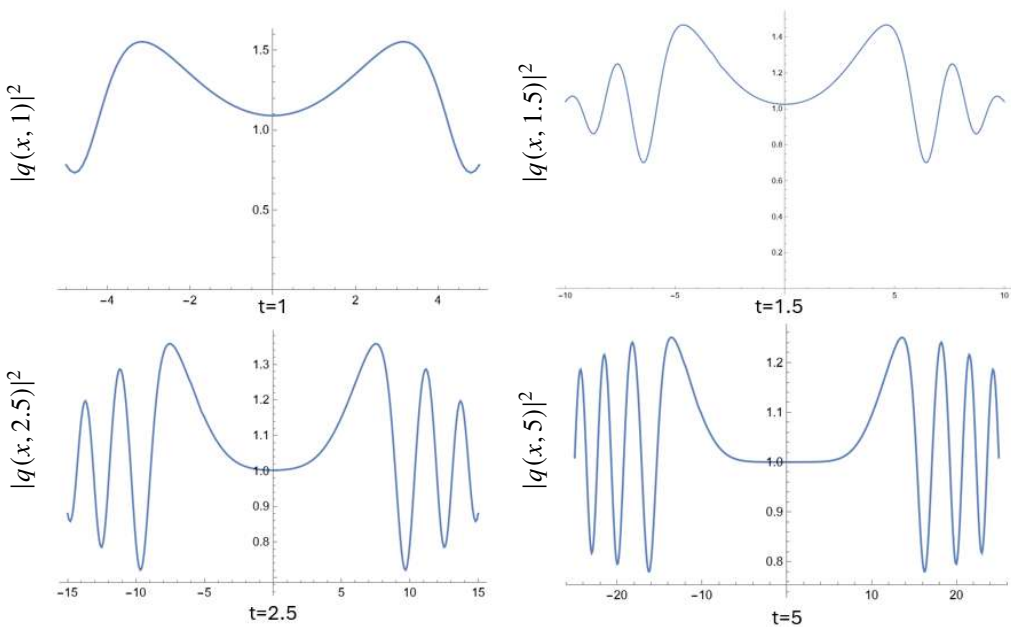
\begin{figure}[th!]
\centering
\begin{overpic}[width=\linewidth,unit=1mm]{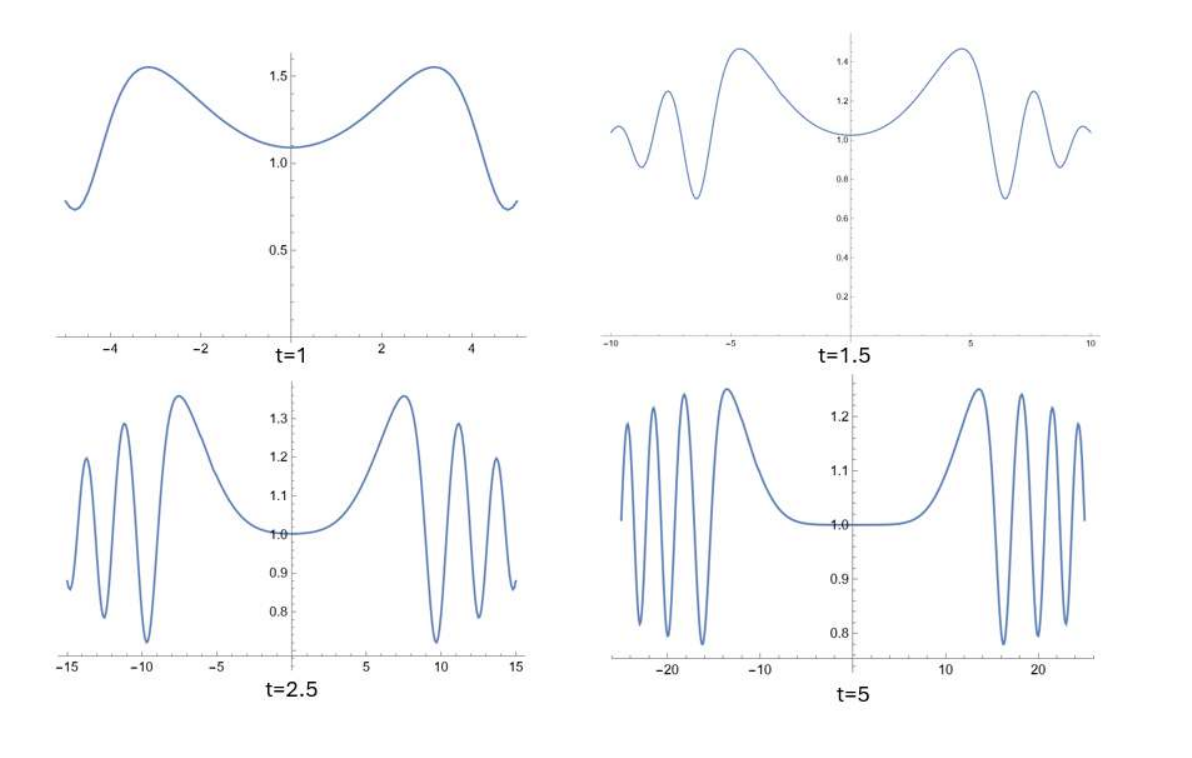}
\put(0,45){\rotatebox{90}{$|q(x,1)|^2$}}
\put(0,17){\rotatebox{90}{$|q(x,2.5)|^2$}}
\put(47,45){\rotatebox{90}{$|q(x,1.5)|^2$}}
\put(47,17){\rotatebox{90}{$|q(x,5)|^2$}}
\end{overpic}
\caption{Plot of $|q(x,t)|^2$ for fixed values of $t$ as a function of $x$ for IC as in \eqref{eq:IC}, with parameters as in \eqref{eq:values}.} 
\label{fig:timeevol}
\end{figure}

\begin{figure}[tbp]

\centering
\begin{overpic}[width=0.4\linewidth,units=1mm]{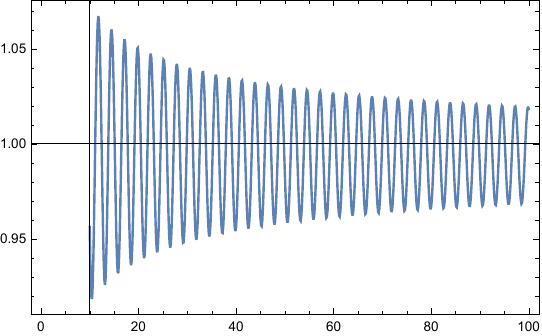}
\put(-5,20){\rotatebox{90}{$\mathrm{Re}\, q(-4t ,t)$}}
\put(50,-2){$t$}
\end{overpic}
\hspace{.1in}
\begin{overpic}[width=0.4\linewidth,units=1mm]{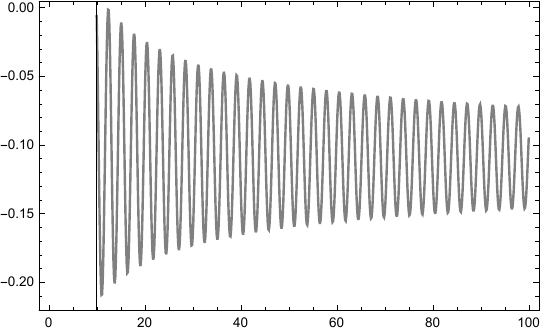}
\put(-5,20){\rotatebox{90}{$\mathrm{Im}\, q(-4t ,t)$}}
\put(50,-2){$t$}
\end{overpic} \vspace{.2in}
\begin{overpic}[width=0.4\linewidth,units=1mm]{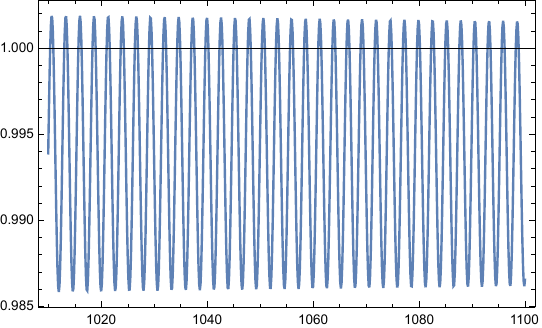}
\put(-5,20){\rotatebox{90}{$\mathrm{Re}\, q(-4t ,t)$}}
\put(50,-2){$t$}
\end{overpic}
\hspace{.1in}
\begin{overpic}[width=0.4\linewidth,units=1mm]{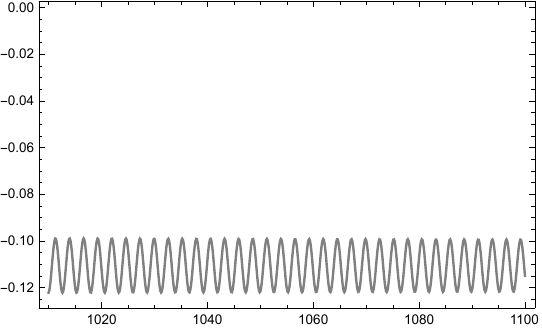}
\put(-5,20){\rotatebox{90}{$\mathrm{Im}\, q(-4t ,t)$}}
\put(50,-2){$t$}
\end{overpic}
\vspace{.1in}
\caption{The real (left) and imaginary parts (right) of the solution of the NLS equation with IC \eqref{eq:IC} and  parameters as in \eqref{eq:values} for large values of $(x,t)$ such that $x = -4 t$.} \label{f:asymp}
\end{figure}

\begin{figure}[tbp]
\centering
\includegraphics[width=.3\linewidth]{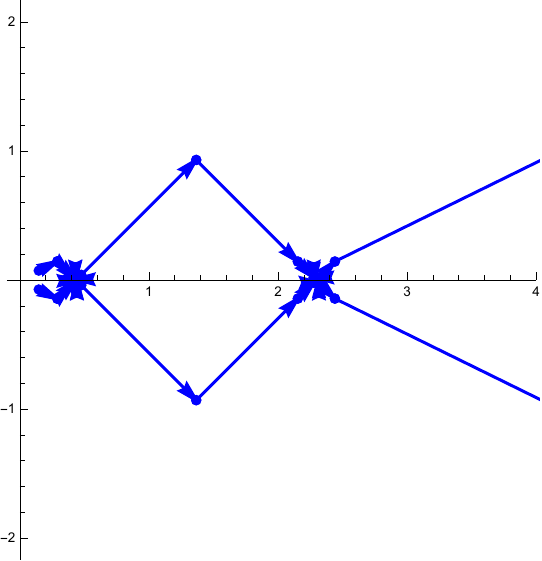}
\raisebox{.68\height}{\includegraphics[width=.3\linewidth]{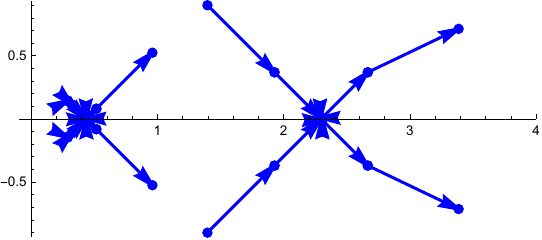}}
\raisebox{2.9\height}{\includegraphics[width=.3\linewidth]{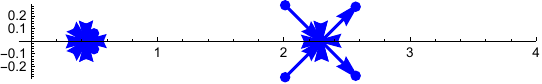}}
\caption{The contours used in the computation of $\tilde{m}(z)$ for $\xi = -2 < -1$ (right-half plane only) for $t = 1,10,100$ (left, center, right, respectively). As it is easily understood, here and in Figs.~\ref{f:allc} and \ref{f:z1} the horizontal and vertical axis are $\Re z$ and $\Im z$, respectively.}\label{f:allc10}
\end{figure}

\begin{figure}[tbp]
\centering
\includegraphics[width=.3\linewidth]{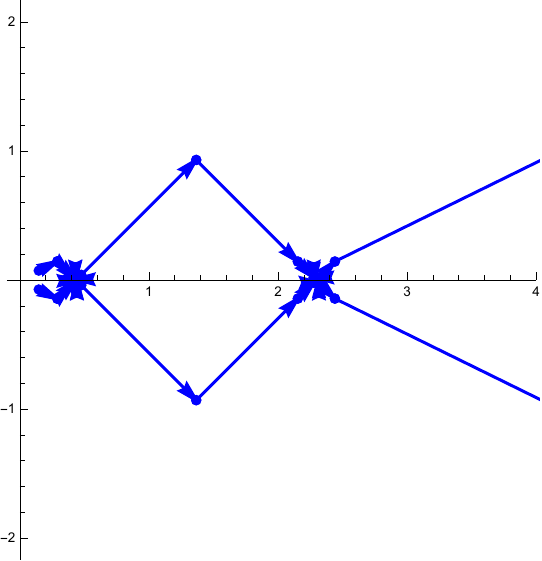}
\raisebox{.68\height}{\includegraphics[width=.3\linewidth]{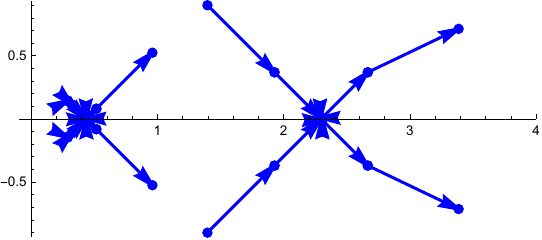}}
\raisebox{2.9\height}{\includegraphics[width=.3\linewidth]{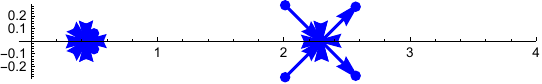}}
\caption{The contours used in the computation of $\tilde{m}$ for $\xi = -2 < -1$ (right-half plane only) for $t = 1,10,100$ (left, center, right, respectively).}\label{f:allc}
\end{figure}

\begin{figure}[tbp]
\centering
\includegraphics[width=.3\linewidth]{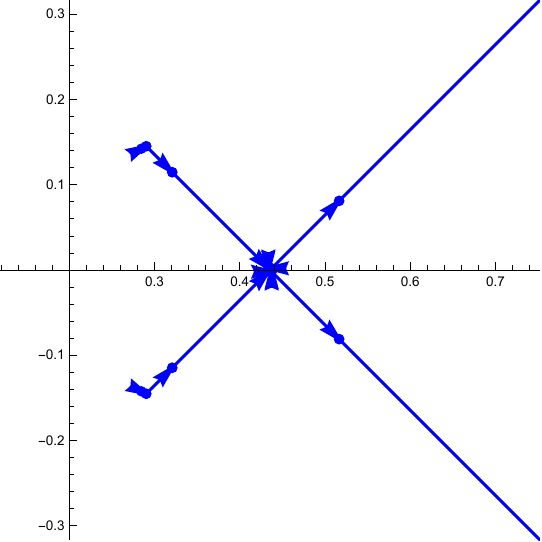}
\includegraphics[width=.3\linewidth]{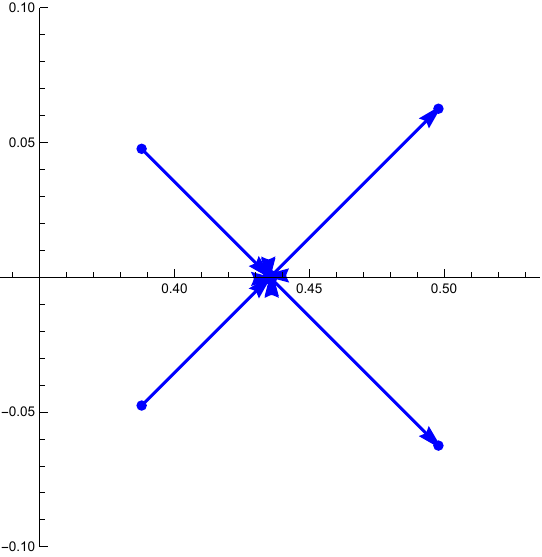}
\includegraphics[width=.3\linewidth]{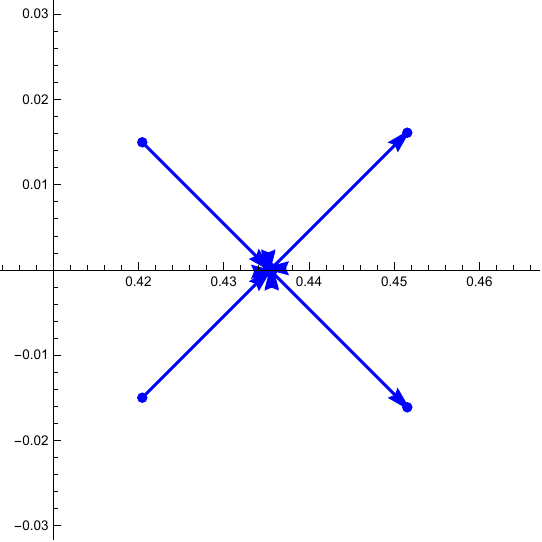}
\caption{The contours used in the computation of $\tilde{m}$ for $\xi = -2 < -1$ for $t = 10, 100, 1000$ (left, center, right, respectively) near the left-most stationary phase point $z_1$.  As $t$ increases, the contours are scaled and truncated following Remark~\ref{r:circ}.}
\label{f:z1}
\end{figure}

\begin{figure}[tbp]
\centering
\includegraphics[width=.3\linewidth]{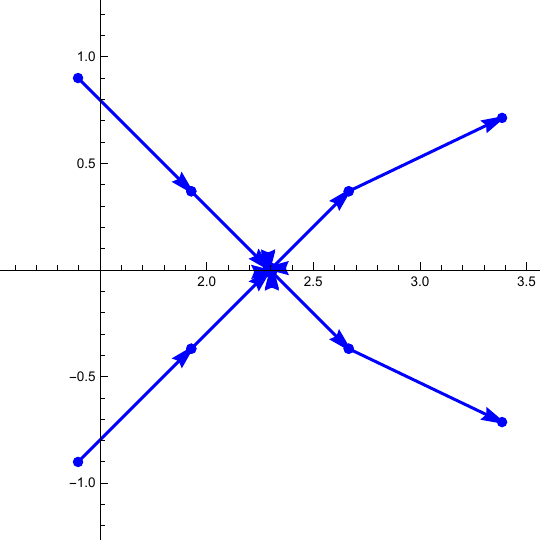}
\includegraphics[width=.3\linewidth]{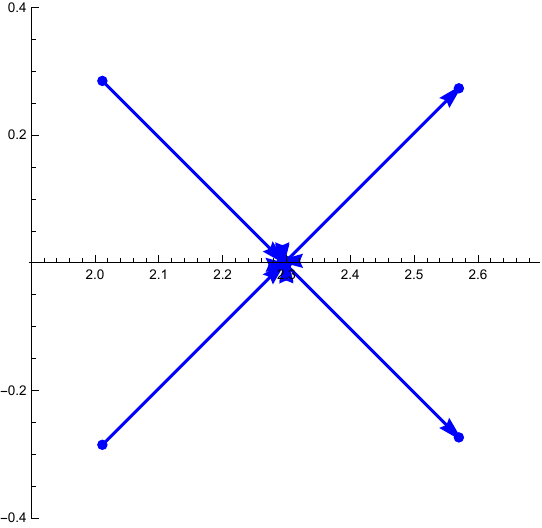}
\includegraphics[width=.3\linewidth]{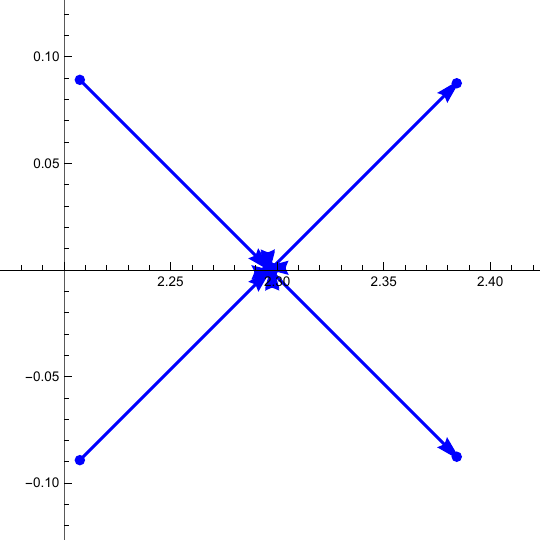}
\caption{The contours used in the computation of $\tilde{m}$ for $\xi = -2 < -1$ for $t = 10, 100, 1000$ (left, center, right, respectively) near the right-most stationary phase point $z_2$.  As $t$ increases, the contours are scaled and truncated following Remark~\ref{r:circ}.}
\label{f:z2}
\end{figure}

\begin{figure}[tbp]
    \centering
    \begin{overpic}[width=.9\linewidth,unit=1mm]{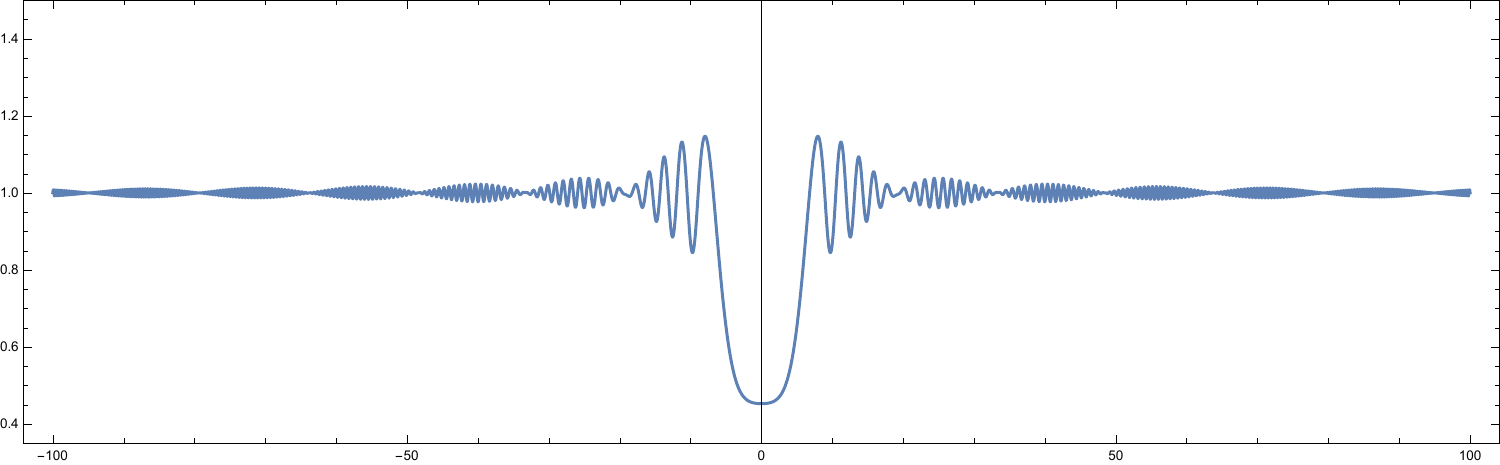}
    \put(-4,10){\rotatebox{90}{$\mathrm{Re}\,q(x,2.5)$}}
    \put(50,-1.5){$x$}
    \end{overpic}\vspace{.1in}

    \begin{overpic}[width=.9\linewidth,unit=1mm]{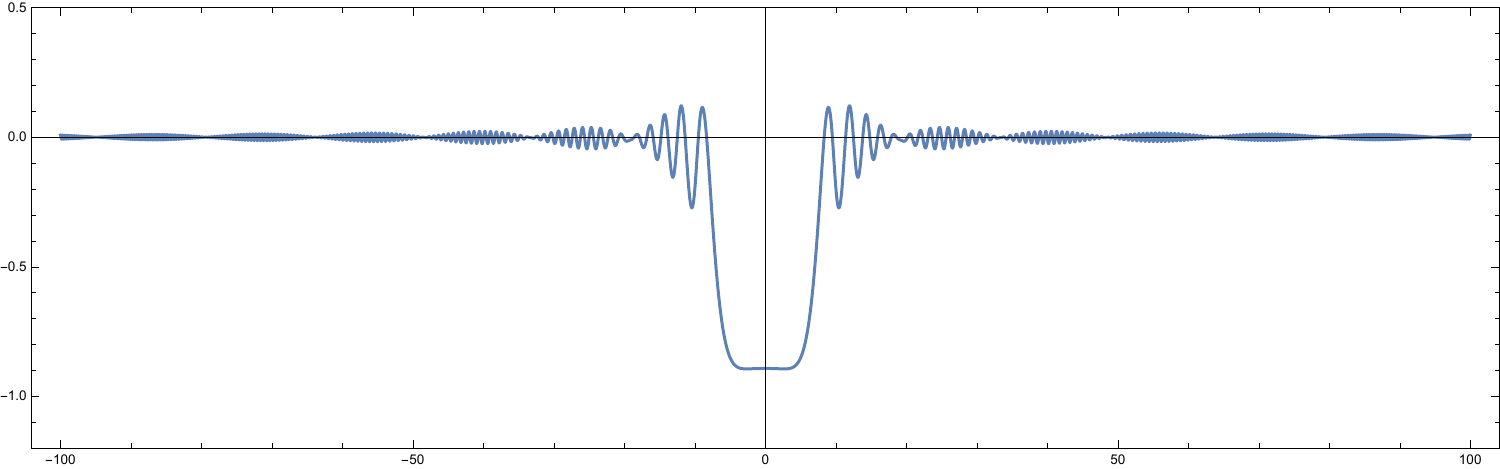}
    \put(-4,10){\rotatebox{90}{$\mathrm{Im}\,q(x,2.5)$}}
    \put(50,-1.5){$x$}
    \end{overpic}\vspace{.1in}
    \caption{The real (top) and imaginary parts (bottom) of the solution of the NLS equation with IC \eqref{eq:IC} at $t = 2.5$ with parameters as in \eqref{eq:values}.}
    \label{fig:t_2_5}
\end{figure}

\begin{figure}[tbp]
    \centering
    \centering
    \begin{overpic}[width=.9\linewidth,unit=1mm]{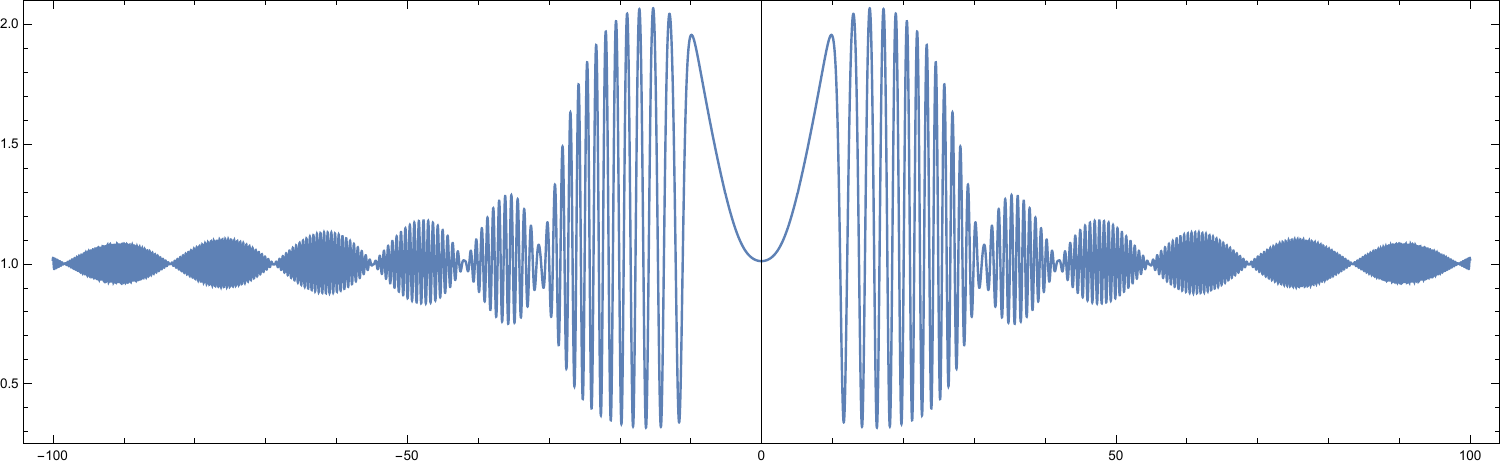}
    \put(-4,10){\rotatebox{90}{$|q(x,2.5)|^2$}}
    \put(50,-1.5){$x$}
    \end{overpic}\vspace{.1in}

    \begin{overpic}[width=.9\linewidth,unit=1mm]{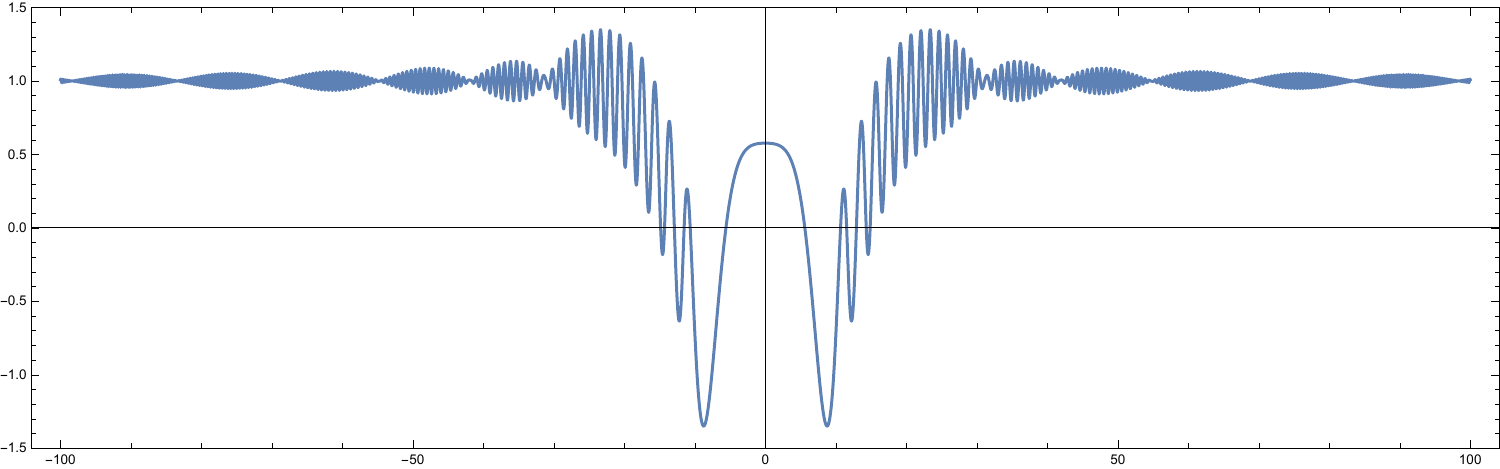}
    \put(-4,10){\rotatebox{90}{$\mathrm{Re}\,q(x,2.5)$}}
    \put(50,-1.5){$x$}
    \end{overpic}\vspace{.1in}

    \begin{overpic}[width=.9\linewidth,unit=1mm]{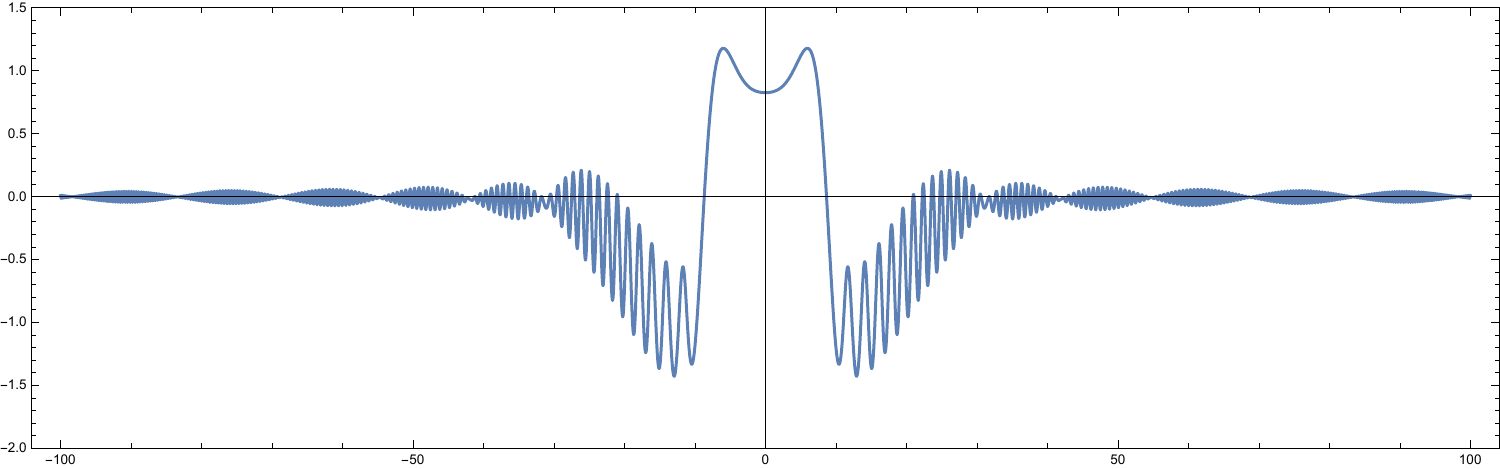}
    \put(-4,10){\rotatebox{90}{$\mathrm{Im}\,q(x,2.5)$}}
    \put(50,-1.5){$x$}
    \end{overpic}\vspace{.1in}
    \caption{The squared modulus (top), real part (middle) and imaginary part (bottom) of the solution of the NLS equation with IC \eqref{eq:IC} at $t = 2.5$ with parameters as in \eqref{eq:values} but with $h = 3.0$.}
    \label{fig:h_3}
\end{figure}

A visual representation of the time evolution of the IC with the parameters given in \eqref{eq:values} for small values of $x$ is provided in Fig.~\ref{fig:timeevol}. 
Plots of the solution over larger spatial domains are shown in Fig.~\ref{fig:mod_t_2_5}, \ref{fig:t_2_5}, and \ref{fig:h_3}.  In Fig.~\ref{f:asymp} we plot the solution off into asymptotic regimes to demonstrate the asymptotic effectiveness of the method.  In actuality, the numerical method becomes more accurate and efficient for larger values of parameters.  This is due to the scaling that is discussed in Remark~\ref{r:circ} and described in more detail in \cite{OT2012}.  Off into asymptotic regimes, one is left with essentially disconnected, fixed contours that scale as $1/\sqrt{t}$. We demonstrate this in Fig.~\ref{f:allc10} where we see that as $t$ increases, $x/t = -4$, the contours on which the Riemann--Hilbert problem is (numerically) posed, after scaling, limit to a fixed contour. Behind this limiting configuration is a contour truncation algorithm that prunes both entire contours when the jump matrix is sufficiently close to the identity and shrinks existing contours using the same metric.  For the specifics of this method, we refer the reader to \cite{istpackage}.  The exact implementation of the contours for $\xi = -2 < -1$ is demonstrated in Fig.~\ref{f:allc}, \ref{f:z1}, \ref{f:z2}. 

\begin{figure}[tbp]
    \centering
    \begin{overpic}[width=.9\linewidth,unit=1mm]{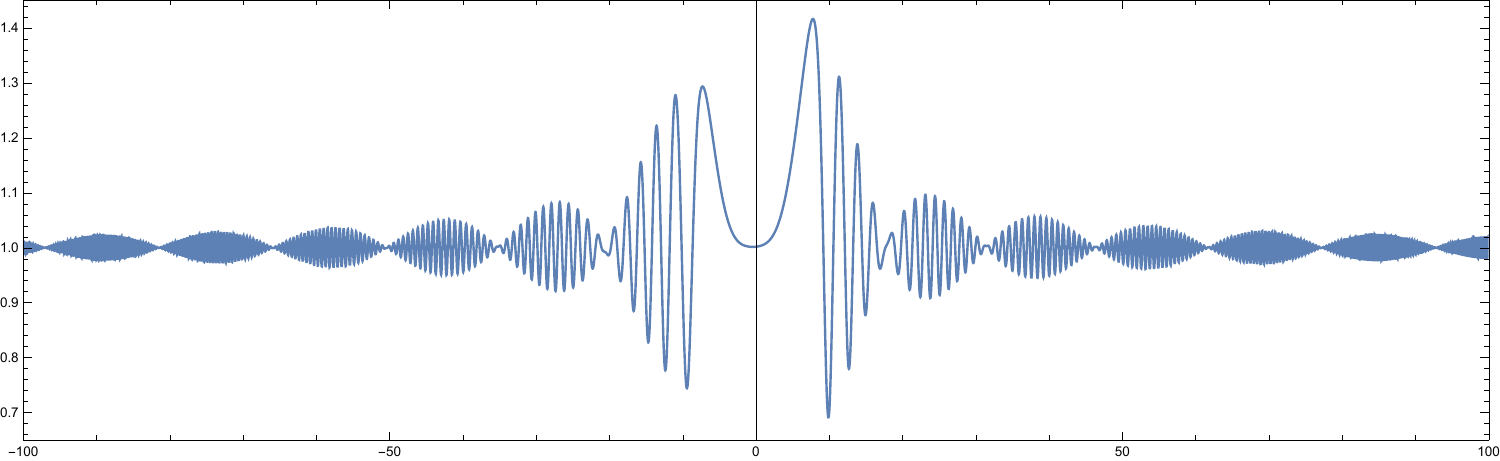}
    \put(-4,10){\rotatebox{90}{$|q(x,2.5)|^2$}}
    \put(50,-1.5){$x$}
    \end{overpic}\vspace{.1in}
    \caption{The squared modulus of the solution of the NLS equation with IC \eqref{eq:IC} at $t = 2.5$ and box parameters $q_o=1$, $L=1$, $h=1.5$, $\alpha=0$ and $\theta=0.15$.}
    \label{fig:mod1_t_2_5}
\end{figure}

\begin{figure}[tbp]
    \centering
    \begin{overpic}[width=.9\linewidth,unit=1mm]{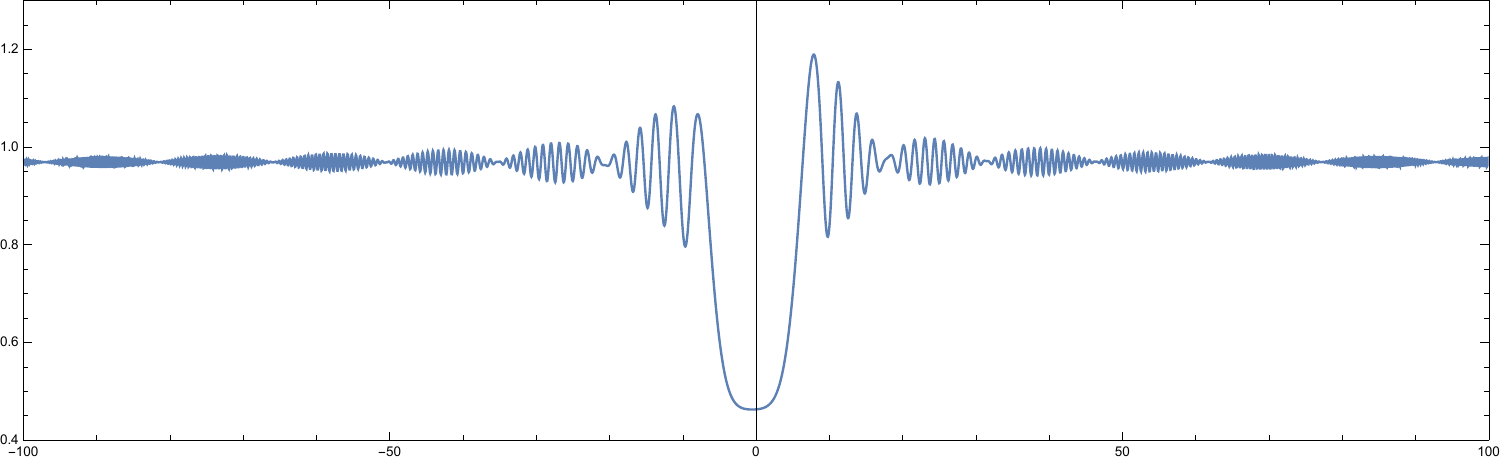}
    \put(-4,10){\rotatebox{90}{$\Re q (x,2.5)$}}
    \put(50,-1.5){$x$}
    \end{overpic}\vspace{.2in}
    \begin{overpic}[width=.9\linewidth,unit=1mm]{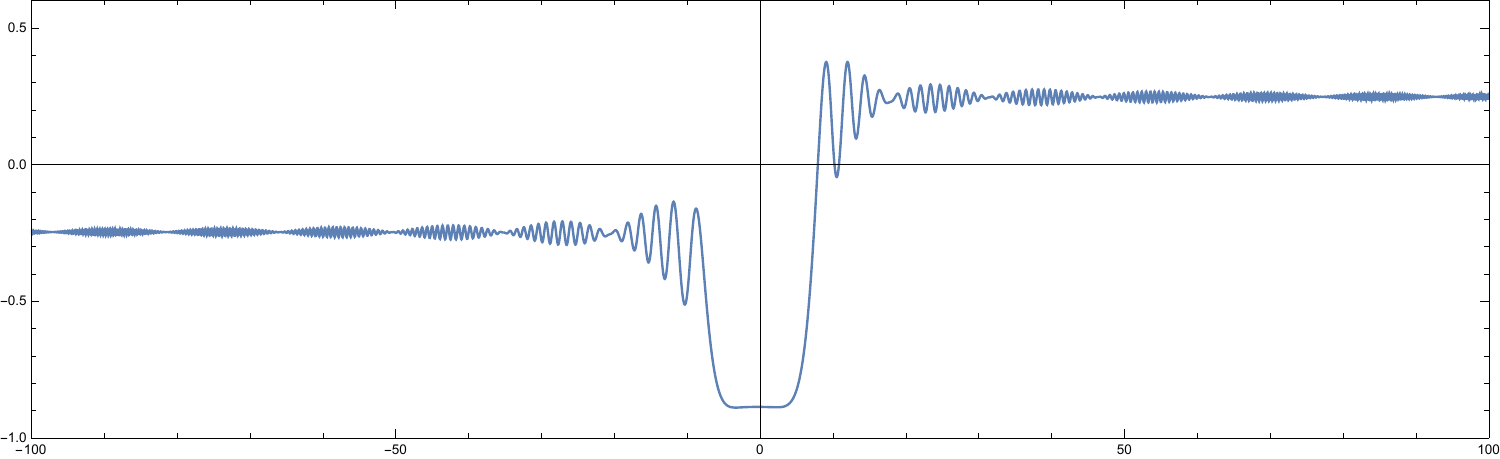}
    \put(-4,10){\rotatebox{90}{$\Im q(x,2.5)$}}
    \put(50,-1.5){$x$}
    \end{overpic}\vspace{.2in}
    \begin{overpic}[width=.9\linewidth,unit=1mm]{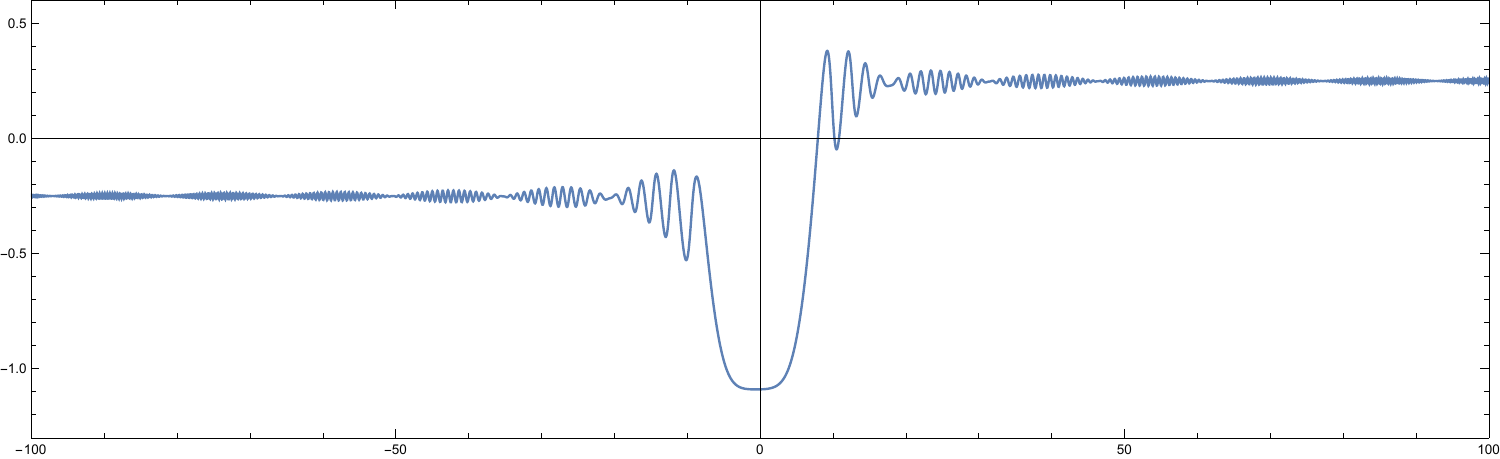}
    \put(-4,10){\rotatebox{90}{$\mathrm{arg}\, q(x,2.5)$}}
    \put(50,-1.5){$x$}
    \end{overpic}\vspace{.1in}
    \caption{Plot of the real part (top), imaginary part (middle) and argument (bottom) of the solution $q(x,2.5)$ for the values $q_o=1$, $L=1$, $\theta=0.15$, $\alpha=0, h=1.5$.}
    \label{fig:mod2_t_2_5}
\end{figure}

For completeness, we give below an example of the time evolution of a box-type IC with $\theta\ne 0$. Let us pick $\alpha=0$, $q_o=L=1$; then Eqs.~\eqref{e:ab} for $\rho(z)=b(z)/a(z)$ give:
\begin{subequations}
\begin{gather}
\rho(z)=e^{-i(2\lambda(z)+\theta)}\times\\
\frac{-\sin \theta \cos (2\mu(z))+\left(h-\cos \theta\right)k(z)\sin (2\mu(z))/\mu(z)}{\cos(2\mu(z))\left[\lambda(z) \cos \theta-ik(z) \sin \theta\right]+i
\left[h-k^2(z)\cos\theta+ik(z)\lambda(z)\sin \theta\right]\sin(2\mu(z))/\mu(z)} \notag \\
k(z)=\frac{1}{2}(z+1/z), \qquad \lambda(z)=\frac{1}{2}(z-1/z),
\qquad \mu(z)=\sqrt{k(z)^2-h^2}\,.
\end{gather}
\end{subequations}
For $h=3/2>q_o=1$, condition (62) to exclude discrete eigenvalues is $\sin \theta < \tanh(\sqrt{5})/3$, which gives $0<\theta<0.331862$. In Figs.~\ref{fig:mod1_t_2_5} and ~\ref{fig:mod2_t_2_5}, we present plots of the solution for values of $\theta \neq 0$ that satisfy the above constraint.

It is crucial to emphasize that the function $\Delta$, which eliminates the jump $D$ in both regions, plays a key role in the definition of the jump matrices for $\widetilde{m}(z)$ (or $\tilde{m}(z)$). Consequently, it is essential to define $\Delta(z)$ appropriately by computing numerically the relevant Cauchy integrals. We give these details in Appendix~\ref{a:deltacomp}.

To address the accuracy of the method described here, we discuss the distribution of collocation nodes and the number of basis functions use per contour.  Let $n$ be the degree of our approximation.  For example, on the contours shown in Fig.~\ref{f:allc} that are close to the stationary points, we use an $n$th-order Chebyshev approximation.  On the longer contours away from the stationary points, we use a $2n$th-order Chebyshev approximation.  We refer to this approximation of the solution of \eqref{eq:dNLS} as $q(x,t,n)$.

\begin{figure}[tbp]
    \centering

    \begin{overpic}[width=.5\linewidth,unit=1mm]{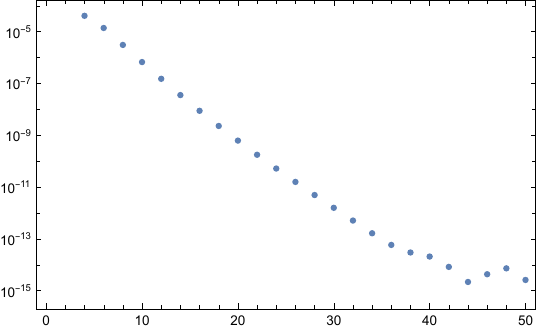}
    \put(-8,10){\rotatebox{90}{$|q(x,t,n)- q(x,t,100)|$}}
    \put(50,-1.5){$n$}
    \end{overpic}
    
    \caption{The Cauchy error in the convergence of $q(x,t,n)$ for $n = 4,\ldots,50$ when $x = -400$, $t = 100$ and the parameters are as in \eqref{eq:values}.}
    \label{fig:cauchyerr}
\end{figure}

We then recall the discussion in the introduction. Using second-order finite differences and $n = 20$, with step size $\Delta x = \Delta t= 0.001$, we find that the discretized NLS equation \eqref{eq:dNLS} is satisfied to an accuracy of $0.000025$ at $x = -11, t = 5$ (parameters as in \eqref{eq:values}).  Additionally, we have run the damped Fourier spectral method of \cite{TrogdonLiu} with over a million of Fourier modes to $t = 1$, to see an error of $O(10^{-3})$ in both regions (parameters as in \eqref{eq:values}, $n = 20$).  Lastly, again with the parameters \eqref{eq:values}, we can verify the accuracy of the Riemann--Hilbert solver by analyzing the \emph{Cauchy error.}  Specifically, in Fig.~\ref{fig:cauchyerr} we compute the difference $|q(x,t,n) - q(x,t,100)|$, for $n = 4,\ldots,50$.

\section{Concluding remarks}\label{s:conclusion}

In this work, we implemented the numerical IST for the defocusing NLS equation with constant nonzero boundary conditions and a box-type IC at $t=0$ in order to approximate the solution of the Cauchy problem for $t>0$.
So far, the numerical IST had been implemented in the literature for a number of nonlinear integrable equations always assuming ICs in Schwartz class (i.e., smooth and rapidly decaying to zero). To our knowledge, this is the first work in which the method is generalized to the NLS equation with ICs that do not decay rapidly at space infinity and also exhibit a jump discontinuity.
Several directions for further exploration remain open. One avenue for future research is the numerical study of the collisionless shock region, which is an open problem also as far as the long-time asymptotics is concerned. Another direction for future investigation is to consider an IC that admits solitons, and solve the associated RHP numerically. Including
one or more solitons can also be done following a similar approach as in the rapidly decaying case, which amounts to turning the residue conditions at each discrete eigenvalue into suitable jump conditions, and then implementing the additional required contour deformations.
Finally, a natural follow-up problem is the case of a smooth IC decaying sufficiently rapidly to the nonzero background. An advantage here is that the reflection coefficient will have faster decay as $z\to 0$ and as $z \to \infty$, thus simplifying the numerical evaluation of the associated Cauchy integrals. 
On the other hand, the numerical implementation of the direct problem will be required, which can be done in a similar way as to the case of rapidly
decaying ICs. A key challenge will be incorporating the $\overline{\partial}$-problem into the inverse problem of the IST, since in this case the reflection coefficient does not admit analytic continuation to the entire complex plane. This issue arises from the non-analyticity of the reflection coefficient and represents a novel challenge from a numerical perspective.

\appendix

\section{Asymptotics of the scattering coefficients for a box IC} \label{e:appA}
In this Appendix we derive the asymptotics as $z\to\infty$, $z\to 0$ and $z\to \pm q_o$ of the scattering coefficients \eqref{e:ab} for a generic box IC.
First, note that one can simplify \eqref{e:ab} and explicitly single out the $z$ dependence as follows:
{\small
\begin{subequations}
\label{e:ab_new}
\begin{gather}
 a(z) = \frac{1}{2\lambda(z)}  e^{i(2 L \lambda(z) + \theta)} \Bigg\{ \cos(2 L \mu(z)) \left(ze^{-i\theta}-\frac{q_o^2}{z}e^{i\theta} \right)
\nonumber \\
+ i \frac{\sin(2L \mu(z))}{2\mu(z)} \left[ -z^2 e^{-i\theta} +2q_o(2h\cos \alpha-q_o \cos \theta) -\frac{q_o^4}{z^2}e^{i\theta}
\right] \Bigg\}\\
b(z) = \frac{1}{\lambda(z)} \Bigg\{ -q_o \sin \theta \cos(2 L \mu(z)) 
+ \left[ z(h e^{-i\alpha}-q_o\cos \theta)+\frac{q_o^2}{z} (h e^{i\alpha}-q_o\cos \theta)\right] \frac{\sin(2L \mu(z))}{2\mu(z)} \Bigg\}.
\end{gather}
\end{subequations}}
We can then use the following asymptotics:
\begin{subequations}
\begin{gather}
\frac{1}{\lambda(z)}=\frac{2}{z}\left(\frac{1}{1-q_o^2/z^2}\right)\sim \frac{2}{z}+\frac{2q_o^2}{z^3}+O(1/z^5) \qquad z\to \infty   \\
\frac{1}{2\lambda(z)}\equiv -\frac{z}{q_o^2}\left(
\frac{1}{1-z^2/q_o^2}\right)\sim -\frac{z}{q_o^2} - \frac{z^3}{q_o^4}+ O(z^5)  \qquad z\to 0 \\
2\mu(z)=z+\frac{q_o^2-2h^2}{z}+O(1/z^3) \qquad z\to \infty \\
2\mu(z)=\frac{q_o^2}{z}+\left( 1-2h^2/q_o^2\right)z+O(z^3) \qquad z\to 0 \\
\frac{1}{2\mu(z)}=\frac{1}{z}+\frac{2h^2-q_o^2}{z^2}+O(1/z^3) \qquad z\to \infty \\
\frac{1}{2\mu(z)}=\frac{z}{q_o^2}+\frac{2h^2-q_o^2}{q_o^6}z^3+O(z^5) \qquad z\to 0.
\end{gather}
\end{subequations}
[The above expansions for $\mu(z)$ pick one of the branches, but as pointed out, only even functions of $\mu(z)$ appear, so the result would be the same even if the other branch were considered.]

For the asymptotic expansions of the $\sin$ and $\cos$ terms, one can use the asymptotics of $\mu(z)$ and appropriate trig identities to obtain:
\begin{subequations}
\begin{gather}
\cos (2L\mu(z))
\sim \cos (Lz)+\frac{L}{z}(2h-q_o^2)\sin(Lz)+O(1/z^2) \qquad z\to \infty, \\
\sin (2L\mu(z))\sim \sin(Lz)-\frac{L}{z} (2h^2-q_o^2)\cos (Lz)+O(1/z^2) \qquad z\to \infty, \\
\cos (2L\mu(z))\sim \cos (Lq_o^2/z)-z L(1-2h^2/q_o^2)\sin (Lq_o^2/z)+O(z^2) \qquad z\to 0, \\
\sin (2L\mu(z))\sim \sin (Lq_o^2/z)+z L(1-2h^2/q_o^2)\cos(Lq_o^2/z)+O(z^2) \qquad z\to 0.
\end{gather}
\end{subequations}
Substituting the above asymptotic expansions into the expressions \eqref{e:ab_new} of $a(z)$ and $b(z)$ one then obtains:
\begin{subequations}
\begin{gather}
b(z) 
=- \frac{2}{z}\left[-q_o\sin (Lz+\theta)+he^{-i\alpha}\sin (Lz)\right]+O(1/z^2) \qquad z\to \infty, \\
b(z)=\frac{2z}{q_o^2}\left[ q_o\sin (Lq_o^2/z+\theta)-he^{i\alpha}\sin (Lq_o^2/z)\right]+O(z^2) \qquad z\to 0, \\
a(z)=1+\frac{i}{z}L(q_o^2-2h^2)+O(1/z^2) \qquad z\to \infty, \\
a(z)=e^{2i\theta}+z e^{2i\theta}L (1-2h^2/q_o^2)+O(z^3) \qquad z\to 0.
\end{gather}
\end{subequations}

It is also useful to compute the behavior of $a(z)$ and $b(z)$ as $z\to \pm q_o$. Taking into account that $k(\pm q_o)=\pm q_o$, $\lambda(\pm q_o)\sim (z\mp q_o)$, $\mu(\pm q_o)=i\mu_o$ with $\mu_o=\sqrt{h^2-q_o^2}$ (the latter is real when $h>q_o$), the explicit expressions of $a$ and $b$ in Eqs.~\eqref{e:ab} yield:
\begin{gather}
a(\pm q_o)\sim \frac{iq_o}{z\mp q_o}e^{i\theta}\left[\mp \sin \theta \cosh (2L\mu_o)+(h\cos \alpha-q_o\cos \theta)\frac{\sinh(2L\mu_o)}{\mu_o} \right], \\
b(\pm q_o)\sim \frac{-q_o}{z\mp q_o}\left[\sin \theta \cosh (2L\mu_o)\mp (h\cos \alpha-q_o\cos \theta)\frac{\sinh(2L\mu_o)}{\mu_o} \right],
\end{gather}
giving
\begin{gather}
\alpha_\pm =iq_o e^{i\theta}\left[\mp \sin \theta \cosh (2L\mu_o)+(h\cos \alpha-q_o\cos \theta)\frac{\sinh(2L\mu_o)}{\mu_o} \right], \\
\beta_\pm=\left[-\sin \theta \cosh (2L\mu_o)\pm (h\cos \alpha-q_o\cos \theta)\frac{\sinh(2L\mu_o)}{\mu_o} \right].
\end{gather}
The expressions for $h<q_o$ can be obtained from $\cosh (2L\sqrt{h^2-q_o^2})\to \cos (2L\sqrt{q_o^2-h^2})$ and \newline
$\sinh (2L\sqrt{h^2-q_o^2})/\sqrt{h^2-q_o^2}\to \sin(2L\sqrt{q_o^2-h^2})/\sqrt{q_o^2-h^2}$.

This shows that $\beta_\pm=\mp i e^{-i\theta}\alpha_\pm$. Moreover, $\beta_\pm \in \Real$ for any choice of the box parameters. It is possible, however, to have $\beta_+ =0$ or $\beta_-=0$ or both (and, correspondingly, $\alpha_+ =0$ or $\alpha_-=0$ or both), as this would happen for
$$
\tanh (2L\sqrt{h^2-q_o^2})=\pm \sqrt{h^2-q_o^2}\frac{\sin \theta}{h\cos \alpha -q_o \cos \theta} \qquad  \text{if } h>q_o
$$
and
$$
\tan (2L\sqrt{q_o^2-h^2})=\pm \sqrt{q_o^2-h^2}\frac{\sin \theta}{h\cos \alpha -q_o \cos \theta} \qquad \text{if } h<q_o.
$$
If $\theta =0$ there are no non-trivial (i.e., with $h\ne q_o$ and $L\ne 0$) solutions for either equation. But both equations could have solutions (for sufficiently large $L$, the first one always does, the second one does if the RHS is between -1 and 1), and one would need to evaluate the coefficients for the specific parameter choices to ensure the coefficients are non-zero. 

Note that this does not (a priori) contradict the result that $a(\pm q_o)\ne 0$ for IC in $L^{1,4}$, since one can have $\alpha_\pm =0$ but the $O(1)$ term non-zero (which is what the result claims).

Moreover, notice that $\mu(z)$ has branch points at $k(z)=\pm h$, which in terms of $z$ actually gives 4 branch points. Specifically, if $h>q_o$ all branch points are real, and located at
$
-h\pm \sqrt{h^2-q_o^2}$ and $h\pm \sqrt{h^2-q_o^2}$ 
which are such that: 
$$
-h-\sqrt{h^2-q_o^2}<-h+\sqrt{h^2-q_o^2}<-q_o<0<q_o<h-\sqrt{h^2-q_o^2}<h+\sqrt{h^2-q_o^2}.
$$
On the other hand, if $h<q_o$ the branch points are located on the circle of radius $h$ at $-h\pm i\sqrt{q_o^2-h^2}$ and  $h\pm i\sqrt{q_o^2-h^2}$.

\section{$\delta$ in the solitonic region}
\label{appendix:a}

In this section, we are providing the solution to RHP \ref{eq:RHP4}. Recall that the function $\delta$ satisfies the following jump condition
\begin{gather}
 \delta_{+}(z) = \delta_{-}(z) \left(1-|\rho(z)|^2 \right) , \quad z \in (-\infty,0).
\end{gather}
Applying the Sokhotski–Plemelj formula to this condition yields 
\begin{align}\label{e:eq138}
\delta(z) &=\delta_{\infty} \, \mathrm{exp} \Bigg( \frac{1}{2 \pi i} \int_{-\infty}^{0} \frac{\log \big(1 - |\rho(s)|^2\big)}{s-z} ds \Bigg), 
\end{align}
where $\delta_{\infty}$ is as follows:
\[\delta_{\infty} = \mathrm{exp} \left[ - \frac{1}{4 \pi i} \int_{-\infty}^{0} \frac{\log \left(1-|\rho(s)|^2 \right) }{s} \, ds \right].\]
We can rewrite \eqref{e:eq138} as follows:
\begin{gather}\label{e:eq_delat}
\delta(z)  
= \delta_{\infty} \, \mathrm{exp} \left( \frac{1}{2 \pi i} \int_{-\infty}^{\gamma_{b}} \frac{\log \big(1 - |\rho(s)|^2\big)}{s-z} ds  + \frac{1}{2 \pi i} \int_{\gamma_{b}}^{\gamma} \frac{\log \big(1 - |\rho(s)|^2\big)}{s-z} ds \right. \nonumber \\ 
\left. +  \frac{1}{2 \pi i} \int_{\gamma}^{0} \frac{\log \big(1 - |\rho(s)|^2\big)}{s-z} ds \right),
\end{gather}
where $\gamma,\gamma_{b}\in(-\infty,0)$ with $\gamma_{b} < -q_o < \gamma$, for some fixed amplitude $q_o>0$. The first and third integrals in the above relation can be evaluated numerically, with details provided in Appendix~\ref{a:deltacomp}. The second integral can likewise be computed numerically using {\tt{Mathematica}}-computable special functions. Additionally, we show below how the singular part of that integral can be dealt with analytically, following a similar strategy as the one used for the Toda lattice in \cite{BT2017}. Let us define
\begin{equation}
\mathcal{T}(z) = \frac{1}{2 \pi i} \int_{\gamma_{b}}^{\gamma} \frac{\log \big(1 - |\rho(s)|^2\big)}{s-z} ds.
\end{equation}
Note that the function $\log \big(1 - |\rho(z)|^2 \big)$ is singular at $z=-q_o \in (\gamma_{b},\gamma)$ since $\rho(-q_o)=i e^{-i \theta}$. We regularize the integrand by introducing the function
\begin{equation}
\tilde{\tau}(z) = \frac{1 - |\rho(z)|^2}{\left(z - q_o^2/z\right)^2} > 0, \quad z \in (\gamma_{b},\gamma)
\end{equation}
and rewriting $\mathcal{T}(z)$ as
\begin{equation}
\mathcal{T}(z) = \frac{1}{2 \pi i} \int_{\gamma_{b}}^{\gamma} \frac{\log \big(\tilde{\tau}(s)\big)}{s-z} ds + \frac{1}{2 \pi i} \int_{\gamma_{b}}^{\gamma} \frac{\log \left(s-q_o^2/s \right)^2}{s-z} ds
\end{equation}
where the function $\log \big(\tilde{\tau}(z)\big)$ is smooth over $(\gamma_{b},\gamma)$. The function $\log \left(z-q_o^2/z \right)^2$ is singular at $z=-q_o$, but its Cauchy integral
\begin{equation}
\mathcal{T}_{1}(z) = \frac{1}{2 \pi i} \int_{\gamma_{b}}^{\gamma} \frac{\log \left(s-q_o^2/s^2 \right)^2}{s-z} ds
\end{equation}
can be computed analytically. Indeed, for $z \in (\gamma_{b},\gamma)$ we have
\begin{equation}
\log \left(z-q_o^2/z \right)^2 = 2 \log \left( z+q_o \right) + 2 \log \left(1-q_o/z \right),
\end{equation}
which yields
\begin{equation}
\mathcal{T}_{1}(z) = \frac{1}{2 \pi i} \int_{\gamma_{b}}^{\gamma} \frac{2 \log \left(s+q_o \right)}{s-z} ds + \frac{1}{2 \pi i} \int_{\gamma_{b}}^{\gamma} \frac{2 \log \left(1- q_o/z \right)}{s-z} ds,
\end{equation}
where the function $\log \left(1-q_o/z \right)$ is smooth over $(\gamma_{b},\gamma)$. It remains to compute the integral
\begin{equation}
\mathcal{T}_{2}(z) = \frac{1}{2 \pi i} \int_{\gamma_{b}}^{\gamma} \frac{2 \log \left( s+ q_o \right)}{s-z} ds.
\end{equation}
To deal with the singularity of the function $\log \left(z+q_o\right)$ at $z=-q_o$, let us consider a closed curve $\sigma$ that encircles $z=-q_o$. We denote by $D_{\sigma}$ the region enclosed by $\sigma$, and $D_{\sigma}^{c} = \mathbb{C}\setminus D_{\sigma}$. We also choose the branch cut of the function $\log \left( z+ q_o \right)$ to be the line $z+q_o = r e^{i \theta}$ with $-\pi/2 \leq \theta \leq \pi/2$. From Cauchy’s integral formula (integrating over the closed curve $\sigma$), we get
\begin{gather}
\mathcal{T}_{2}(z) \equiv \lim_{\epsilon \to 0} \Big( \frac{1}{2 \pi i} \int_{ \Gamma_{\epsilon}} f(s,z) ds \Big) = - \frac{1}{2 \pi i} \int_{ \big( \gamma,\gamma_{b}\big)_{\mathrm{arc}}} f(s,z) ds - \lim_{\epsilon \to 0} \Big( \frac{1}{2 \pi i} \int_{ C_{\epsilon}} f(s,z) ds \Big) + \nonumber \\ 
+\begin{cases}
2 \log \left( z+ q_o \right), \quad z \in D_{\sigma}\\
0, \quad z \in D_{\sigma}^{c}\label{app:eqA10}
\end{cases}
\end{gather}
where $\big( \gamma,\gamma_{b}\big)_{\mathrm{arc}}$ denotes the semicircle centered at $z=-q_o$ with radius $r=\frac{|\gamma - \gamma_{b}|}{2}$ and counterclockwise orientation, $C_{\epsilon}$ denotes the semicircle centered at $z=-q_o$ with radius $\epsilon$ and counterclockwise orientation, $\Gamma_{\epsilon} = [\gamma_{b},\gamma] \setminus C_{\epsilon}$, and we introduced the notation $f(s,z)$ for the function $2 \log \left( s+ q_o \right)/(s-z)$ for brevity. One can show that there exists $K_{\epsilon}$ such that $|(s+q_o)f(s,z)| \leq K_{\epsilon}$, where $K_{\epsilon}$ depends on $\epsilon$ but not on the argument $\mathrm{arg}(s+q_o)$, and therefore $(s+q_o)f(s,z)$ approaches zero uniformly as $\epsilon \to 0$, which implies that
\begin{equation}
\lim_{\epsilon \to 0} \Big( \frac{1}{2 \pi i} \int_{ C_{\epsilon}} f(s,z) ds \Big) = 0.
\end{equation}
Moreover, it turns out that
\begin{equation}
\frac{1}{2 \pi i} \int_{ \big( \gamma,\gamma_{b}\big)_{\mathrm{arc}}}
\hspace*{-5mm}f(s,z) ds = \frac{1}{\pi i} \Bigg( \log r \log \frac{z+q_o-r}{q_o+z} - \log (-r) \log \frac{z+q_o+r}{q_o+z} - \mathrm{L}_{i2} \Big( \frac{-r}{q_o+z} \Big) + \mathrm{L}_{i2} \Big( \frac{r}{q_o+z} \Big) \Bigg),
\end{equation}
where $\mathrm{L}_{i2}$ is the dilogarithm function. Combining all the above into \eqref{app:eqA10} we get
\begin{gather}
\mathcal{T}_{2}(z) = - \frac{1}{\pi i} \Bigg( \log r \log \frac{z+q_o-r}{q_o+z} - \log (-r) \log \frac{z+q_o+r}{q_o+z} - \mathrm{L}_{i2} \Big( \frac{-r}{q_o+z} \Big) + \mathrm{L}_{i2} \Big( \frac{r}{q_o+z} \Big) \Bigg) \nonumber + \\  +\begin{cases}
2 \log \big( z+q_o \big), \quad z \in D_{\sigma}\\
0, \quad z \in D_{\sigma}^{c}
\end{cases}.
\end{gather}

\section{Computing $\delta$}\label{a:deltacomp}

We give below some details regarding the numerical evaluation of the integrals appearing in the definition of $\delta$.
Recall that in the solitonic region $\Delta(z) = \mathrm{diag} \Big( \delta(z),1/\delta(z) \Big)$, where $\delta$ is given by expression \eqref{e:eq138}.
Moreover, recall that the reflection coefficient is such that $\rho(\pm q_o) = \mp i e^{-i \theta}$, which makes the function $\log \big(1 - |\rho(s)|^2\big)$ singular at $s=-q_o$. We then rewrite $\delta$ as follows
\begin{gather}\label{eq:eqn44}
\delta(z) = \delta_{\infty} \, \mathrm{exp} \Bigg[ \frac{1}{2 \pi i} \Bigg( \int_{-\infty}^{\gamma_{b}} \frac{\log \big(1 - |\rho(s)|^2\big)}{s-z} ds + \int_{\gamma_{b}}^{\gamma} \frac{\log \big(1 - |\rho(s)|^2\big)}{s-z} ds + \int_{\gamma}^{0} \frac{\log \big(1 - |\rho(s)|^2\big)}{s-z} ds \Bigg) \Bigg]
\end{gather}
where $\gamma,\gamma_{b}\in (-\infty,0)$ with $\gamma_{b} < -q_o < \gamma$. Notice that the function $\log \big(1 - |\rho(s)|^2\big)$ is regular inside $(-\infty,\gamma_{b})$ and $(\gamma,0)$, and is singular inside the interval $(\gamma_{b},\gamma)$. The main challenge in numerically evaluating the first Cauchy integral in Eq.~\eqref{eq:eqn44} lies in the slow decay of the reflection coefficient as $z \to -\infty$. Conversely, the difficulty with the third Cauchy integral in \eqref{eq:eqn44} is given by the fast oscillations of the reflection coefficient near zero. To address these issues, we truncated the Cauchy integrals to the left of $z=0$ and to the right of $z \to - \infty$ enabling us to achieve high accuracy while managing computational costs effectively. We then use a Chebyshev approximation for the function $\log \big(1 - |\rho(s)|^2\big)$ as it remains smooth within the integration interval. The Cauchy integrals are then computed using {\tt Mathematica}’s special functions that define numerically the Cauchy transform. For the second integral in \eqref{eq:eqn44}, the function $\log \big(1 - |\rho(s)|^2\big)$ is singular at $s=-q_o$. We therefore rewrite this integral as follows
\begin{gather}
\int_{\gamma_{b}}^{\gamma} \frac{\log \big(1 - |\rho(s)|^2\big)}{s-z} ds = \int_{\gamma_{b}}^{\gamma} \log \Bigg( \frac{1 - |\rho(s)|^2}{(s+q_o)^2}\Bigg)\frac{1}{s-z} ds + \int_{\gamma_{b}}^{\gamma} \frac{\log\big( (s+q_o)^2 \big)}{s-z} ds
\end{gather}
where the function $\log \Bigg( \frac{1 - |\rho(s)|^2}{(s+q_o)^2} \Bigg)$ is smooth along $(\gamma_{b},\gamma)$ and its Cauchy integral can be computed using a Chebyshev approximation. The function $\log\big( (s+q_o)^2 \big)$ is still singular at $s=-q_o$. However, we can derive a closed-form expression for its Cauchy integral using special function routines in {\tt Mathematica}. On the other hand, in the solitonless region when $\xi < -q_o$, $\hat{\Delta}(z) = \mathrm{diag} \Big( \hat{\delta}(z),1/\hat{\delta}(z) \Big)$ where $\hat{\delta}$ is given by the expression
\begin{gather}\label{eq:eqn45}
\hat{\delta}(z) =  \hat{\delta}_{\infty} \, \mathrm{exp} \left[ \frac{1}{2 \pi i} \int_{-\infty}^{0}  \frac{\log \left(1-|\rho(s)|^2 \right)}{s-z} \, ds +  \frac{1}{2 \pi i} \int_{z_1}^{z_2}  \frac{\log \left(1-|\rho(s)|^2 \right)}{s-z} \, ds \right]
\end{gather}
where $\hat{\delta}_{\infty} = \mathrm{exp} \left[ - \frac{1}{4 \pi i}  \int_{(-\infty,0) \cup (z_1,z_2)} \frac{\log \left(1-|\rho(s)|^2 \right) }{s} \, ds \right]$ and $z_1 < q_o < z_2$. The function $\log \big(1 - |\rho(s)|^2\big)$ now becomes singular at $s=\pm q_o$. $\hat{\delta}$ is defined exactly the same way as in the solitonic region when $s \in (-\infty,0)$. When $s \in (z_1,z_2)$, we rewrite the second integral in Eq.~\eqref{eq:eqn45} as follows
\begin{gather}
 \frac{1}{2 \pi i} \int_{z_1}^{\gamma_{b}'}  \frac{\log \left(1-|\rho(s)|^2 \right)}{s-z} \, ds + \frac{1}{2 \pi i} \int_{\gamma_{b}'}^{\gamma'}  \frac{\log \left(1-|\rho(s)|^2 \right)}{s-z} \, ds + \frac{1}{2 \pi i} \int_{\gamma'}^{z_2}  \frac{\log \left(1-|\rho(s)|^2 \right)}{s-z} \, ds 
\end{gather}
with $z_1<\gamma'_b<\gamma'<z_2$. The function $\log \big(1 - |\rho(s)|^2 \big)$ is smooth along $(z_1,\gamma_{b}') \cup (\gamma',z_2)$, and the corresponding Cauchy integrals can be computed numerically using a Chebyshev approximation. To deal with the singularity at $s=q_o$, we write
\begin{gather}
\frac{1}{2 \pi i} \int_{\gamma_{b}'}^{\gamma'}  \frac{\log \left(1-|\rho(s)|^2 \right)}{s-z} \, ds = \int_{\gamma_{b}'}^{\gamma'} \log \Bigg( \frac{1 - |\rho(s)|^2}{(s-q_o)^2}\Bigg)\frac{1}{s-z} ds + \int_{\gamma_{b}'}^{\gamma'} \frac{\log\big( (s-q_o)^2 \big)}{s-z} ds
\end{gather}
where the function $\log \Bigg( \frac{1 - |\rho(s)|^2}{(s-q_o})^2 \Bigg)$ is smooth along $(\gamma_{b}',\gamma')$ and its Cauchy integral can be computed using a Chebyshev approximation. The function $\log\big( (s-q_o)^2 \big)$ is still singular at $s=q_o$, but its Cauchy integral can be computed using special function routines in {\tt Mathematica}. Moreover, in the solitonless region when $\xi > q_o$, we define $\check{\Delta}(z) = \mathrm{diag} \left( \check{\delta}(z),1/\check{\delta}(z) \right)$ where $\check{\delta}$ is given by the expression
\begin{gather}\label{eq:eq49}
\check{\delta}(z) = \check{\delta}_{\infty} \, \mathrm{exp} \left[ \frac{1}{2 \pi i} \int_{-\infty}^{\hat{z}_2} \frac{\log \left(1-|\rho(s)|^2 \right)}{s-z} \, ds + \frac{1}{2 \pi i} \int_{\hat{z}_1}^{0} \frac{\log \left(1-|\rho(s)|^2 \right)}{s-z} \, ds \right].
\end{gather}
where $\check{\delta}_{\infty} = \mathrm{exp} \left[ - \frac{1}{4 \pi i}  \int_{(-\infty,\hat{z}_2) \cup (\hat{z}_1,0)} \frac{\log \left(1-|\rho(s)|^2 \right) }{s} \, ds \right]$. Notice that since $\hat{z}_2 < -q_o < \hat{z}_1 < 0$, the values $s=\pm q_o$ do not lie on the contour of integration. Therefore, the function $\log \left(1-|\rho(s)|^2 \right)$ is smooth along $(-\infty,\hat{z}_2) \cup (\hat{z}_1,0)$, and its Cauchy integral can be computed numerically using a Chebyshev approximation. As before, we truncate the first Cauchy integral in \eqref{eq:eq49} to the right of $z \to - \infty$ and the second integral to the left of $z=0$, in order to account for the slow decay and fast oscillations of the reflection coefficient at $- \infty$ and at $0$, respectively. From a numerical perspective, the definition of $\check{\delta}$ is simpler in the solitonless region when $\xi > q_o$, compared to the definition of $\hat{\delta}$ in the same region when $\xi < -q_o$, where as we saw earlier, the singularities of the function $\log \left(1-|\rho(s)|^2 \right)$ at both values $s=\pm q_o$ must be accounted for. In practice, we see that it suffices to truncate these Cauchy integrals by integrating over $s \in [-10^{3},-10^{-3}] \cup [10^{-3},10^3]$. Truncating further can be used if less accuracy is desired.

\section{On the determinant, symmetries and singularities of $\widetilde{m}$} \label{eq:appendixD}

In this Appendix, we analyze the properties of the function $\widetilde{m}(z)$ defined via relation \eqref{e:def_mtilde} in terms of the solution $\hat{m}(z)$ of RHP~4.

\begin{proposition}\label{eq:proposition7}
If the function $\widetilde{m}(z)$ is defined via Eq.~\eqref{e:def_mtilde}, we have that $\det \widetilde{m}(z) = 1$, for any $z \in \mathbb{C}$, if and only if:
\begin{equation}
\label{e:m0_conds}
\widetilde{m}_{21}(0)=e^{-2i\theta}\widetilde{m}_{12}(0) \quad \text{and} \quad \det \widetilde{m}(0)=1.
\end{equation}
(Note that the second symmetry in Eq.~\eqref{eq:eqsym} implies \eqref{e:m0_conds}.)
\end{proposition}
\begin{proof}
From \eqref{e:def_mtilde}, we have that:
\begin{equation}
\label{e:det_cond}
1-\frac{q_o^2}{z^2}=\det \hat{m}(z)=\det E(z) \det \widetilde{m}(z)
\end{equation}
which implies that $\det E(z)=1-q_o^2/z^2$ is a necessary and sufficient condition to have the desired result. A direct computation on the definition of $E(z)$ shows that
$$
\det E(z)=1+\frac{iq_o}{z}\frac{1}{\det \widetilde{m}(0)}\left[ e^{i\theta} \widetilde{m}_{21}(0)-e^{-i\theta} \widetilde{m}_{12}(0)\right]-\frac{q_o^2}{z^2}\frac{1}{\det \widetilde{m}(0)}.
$$
Clearly, $\det E(z)=1-q_o^2/z^2$ if and only if Eq.~\eqref{e:m0_conds} holds. For $z = \pm q_o$: in principle, if $\det E(z)=1-q_o^2/z^2$, then $\det \widetilde{m}(z)$ could take any value at $\pm q_o$ and \eqref{e:det_cond} would still be satisfied. However, $\det \widetilde{m}(z)$ cannot be singular at $\pm q_o$, otherwise it would violate \eqref{e:det_cond}. If $\det \widetilde{m}(z)$ is to be continuous across $\Real$, then its value has to be 1 also at $z=\pm q_o$.
\end{proof}
Therefore, we can say $\det \widetilde{m}(z)=1$ if and only if $\widetilde{m}(0)$ satisfies conditions \eqref{e:m0_conds}, and the general form for $\widetilde{m}(0)$ then becomes
$$
\widetilde{m}(0)=\begin{pmatrix}
\widetilde{m}_{11}(0) & \widetilde{m}_{12}(0) \\
\widetilde{m}_{12}(0)e^{-2i\theta} & (1+\tilde{m}_{12}(0)^2e^{-2i\theta})/\widetilde{m}_{11}(0)
\end{pmatrix}
$$
where $\widetilde{m}_{11}(0)$ and $\widetilde{m}_{12}(0)$ are two arbitrary complex constants.

\begin{proposition}
Conditions \eqref{e:m0_conds} are deduced as a consequence of the second symmetry for $\widetilde{m}(z)$, i.e., if $\widetilde{m}(z)$ satisfies the second symmetry in \eqref{eq:eqsym}, then conditions \eqref{e:m0_conds} follow, and therefore $\det \widetilde{m}(z)=1$, for all $z\in \Complex$.
Conversely, if $\widetilde{m}(z)$ defined via Eq.~\eqref{e:def_mtilde} satisfies \eqref{e:m0_conds}, then the expected symmetry follows.
\end{proposition}
\begin{proof}
The first part of the result is straightforward. Conversely, direct computation on $E(z)$, assuming only conditions \eqref{e:m0_conds}, gives the following:
\begin{gather}
E^{-1}(z)=\frac{z^2}{z^2-q_o^2}\left[
I_2-\frac{q_o}{z}\tilde{m}(0)\sigma_2 e^{-i\theta \sigma_3}
\right], \quad
E^{-1}(q_o^2/z)=\frac{q_o^2}{q_o^2-z^2}\left[
I_2-\frac{z}{q_o}\tilde{m}(0)\sigma_2 e^{-i\theta \sigma_3}\right].
\end{gather}
Now we use that $\hat{m}(z)=E(z)\widetilde{m}(z)$, and the second symmetry for $\hat{m}(z)$, to obtain
\begin{equation*}
\widetilde{m}(q_o^2/z)=\frac{z}{q_o}E^{-1}(q_o^2/z)E(z)\widetilde{m}(z) \sigma_2 e^{-i\theta \sigma_3}.
\end{equation*}
Now the factor in front of $\widetilde{m}(z)$ on the RHS can be computed explicitly, and a direct computation shows that
\[
\frac{z}{q_o}E^{-1}(q_o^2/z)E(z)=\widetilde{m}(0)\sigma_2e^{-i\theta \sigma_3}
\]
from which we then deduce the second symmetry for $\widetilde{m}(z)$:
\begin{equation}
\label{e:2nd_symm_mtilde}
\widetilde{m}(q_o^2/z)=\widetilde{m}(0)\sigma_2 e^{-i\theta \sigma_3}\widetilde{m}(z)\sigma_2 e^{-i\theta \sigma_3}
\end{equation}
which is the `expected' second symmetry. Finally, note that if we evaluate the above symmetry as $z\to 0$, we find
$$
\widetilde{m}^{-1}(0)=\sigma_2 e^{-i\theta \sigma_3}\widetilde{m}(0)\sigma_2 e^{-i\theta \sigma_3}.
$$
A direct computation of both sides yields
\[
\frac{1}{\det \widetilde{m}(0)}
\begin{pmatrix}
\widetilde{m}_{22}(0) & -\widetilde{m}_{12}(0) \\
-\widetilde{m}_{21}(0) & \widetilde{m}_{11}(0)
\end{pmatrix}=\begin{pmatrix}
\widetilde{m}_{22}(0) &-e^{2i\theta}\widetilde{m}_{21}(0) \\
-e^{-2i\theta}\widetilde{m}_{12}(0) & \widetilde{m}_{11}(0)
\end{pmatrix}
\]
and then it is obvious that if the second symmetry is satisfied, even in the form of \eqref{e:2nd_symm_mtilde} with (an invertible) arbitrary $\widetilde{m}(0)$, it implies that $\widetilde{m}(0)$ must satisfy conditions \eqref{e:m0_conds} which guarantee that $\det E(z)=1-q_o^2/z^2$ or equivalently $\det \widetilde{m}(z)=1$ for all $z\in \Complex$.
\end{proof}

\begin{proposition}
As to the first symmetry of $\widetilde{m}(z)$, the following holds: 
\begin{equation}
\widetilde{m}^*(z)=\sigma_1 \widetilde{m}(z)\sigma_1 \quad \Leftrightarrow \quad E(z)=\sigma_1E^*(z^*)\sigma_1.
\end{equation}
\end{proposition}
\begin{proof}
For the respective symmetry of $E(z)$, we compute:
$$
\sigma_1 E^*(z^*)\sigma_1=I_2+\frac{q_o}{z}\sigma_2 e^{-i\theta \sigma_3} \sigma_1 \left( \widetilde{m}^{-1}(0)\right)^* \sigma_1
$$
which implies that 
\begin{equation}
\label{e:1stsymmE}
E(z)=\sigma_1E^*(z^*)\sigma_1 \quad \Leftrightarrow \quad
\widetilde{m}(0)=\sigma_1 \widetilde{m}^*(0)\sigma_1.
\end{equation}
If the above condition holds, then it should follow that the first symmetry holds for all $z$, i.e.
\begin{equation}
\widetilde{m}^*(z)=\sigma_1 \widetilde{m}(z)\sigma_1.
\end{equation}
Furthermore, the second of \eqref{e:1stsymmE} is equivalent to:
\begin{equation}
\widetilde{m}_{22}(0)=\widetilde{m}_{11}^*(0) \quad \text {and} \quad \widetilde{m}_{21}^*(0)=\widetilde{m}_{12}(0),
\end{equation}
and combining these conditions with \eqref{e:m0_conds} yields
\begin{equation}
\label{e:mtilde0+1st}
\widetilde{m}(0)=
\begin{pmatrix}
\widetilde{m}_{11}(0) & e^{i\theta}\sqrt{|\widetilde{m}_{11}(0)|^2-1} \\
e^{-i\theta} \sqrt{|\widetilde{m}_{11}(0)|^2-1} & \widetilde{m}_{11}^*(0)
\end{pmatrix}.
\end{equation}
This is the most general form of $\widetilde{m}(0)$, which is consistent with the conditions necessary to have $\det E(z)=1-q_o^2/z^2$ and which satisfies the first symmetry.
\end{proof}

\begin{lemma}
If for given $x,t$ one has that $\hat{m}(x,t;z)$, related to $\widetilde{m}(x,t;z)$ via \eqref{e:def_mtilde}, satisfies the solvability condition
\begin{equation}\label{eq:eq169n}
W\left(\hat{m}_2(x,t;q_o),\hat{m}_1(x,t;-q_o)\right)\ne 0,
\end{equation}
that is if $\hat{m}_2(x,t;q_o)$ and $\hat{m}_1(x,t;-q_o)$ are linearly independent, then there exists a unique $\widetilde{m}(x,t;0)$ (with entries given by Eqs.\eqref{e:mtilde0}), which guarantees that $\widetilde{m}(x,t;z)$ is bounded as $z\to \pm q_o$.
\end{lemma}
\begin{proof}
In the proof we omit the $x,t$-dependence for brevity. From Eq.~\eqref{e:def_mtilde}, we have:
\begin{equation}
\label{e:mtilde}
\widetilde{m}(z)=\frac{z^2}{z^2-q_o^2}\left[ I_2 -\frac{q_o}{z}\widetilde{m}(0) \sigma_2 e^{-i\theta \sigma_3}\right]\hat{m}(z)
\end{equation}
where the columns of $\hat{m}(z)$ as $z\to \pm q_o$ from $\Complex^\pm$ are given in Eqs.~\eqref{e:mhat_qo} and \eqref{e:mhat_pmqo}.
Explicitly, \eqref{e:mtilde} gives:
\begin{subequations}
\label{e:mtilde_cols}
\begin{gather}
\widetilde{m}_1(z)=\frac{z^2}{z^2-q_o^2}
\begin{pmatrix}
\left(1-\frac{iq_-}{z}\widetilde{m}_{12}(0) \right)\hat{m}_{11}(z)+\frac{iq_+}{z}\widetilde{m}_{11}(0)\hat{m}_{21}(z) \\ \\[1pt]
-\frac{iq_-}{z}\widetilde{m}_{22}(0)\hat{m}_{11}(z)+\left(1+\frac{iq_+}{z}\widetilde{m}_{21}(0) \right)\hat{m}_{21}(z)
\end{pmatrix}, \\
\widetilde{m}_2(z)=\frac{z^2}{z^2-q_o^2}
\begin{pmatrix}
\left(1-\frac{iq_-}{z}\widetilde{m}_{12}(0) \right)\hat{m}_{12}(z)+\frac{iq_+}{z}\widetilde{m}_{11}(0)\hat{m}_{22}(z)
\\ \\[1pt] -\frac{iq_-}{z}\widetilde{m}_{22}(0)\hat{m}_{12}(z)+\left(1+\frac{iq_+}{z}\widetilde{m}_{21}(0) \right)\hat{m}_{22}(z)
\end{pmatrix}.
\end{gather}
\end{subequations}
We seek for conditions on $\widetilde{m}(0)$ that guarantee that  all the entries of $\widetilde{m}_1(z)$ and 
 $\widetilde{m}_2(z)$ are bounded as $z\to \pm q_o$. Eqs.~\eqref{e:mhat_qo} and \eqref{e:mhat_pmqo} indicate that $\hat{m}_1(q_o)\propto \hat{m}_2(q_o)$ and $\hat{m}_2(-q_o)\propto \hat{m}_1(-q_o)$, which implies that it is sufficient to impose the conditions only on either $\widetilde{m}_1(z)$ or $\widetilde{m}_2(z)$ to ensure both are bounded. Specifically, $\widetilde{m}(z)$ is bounded at $z=q_o$ under the following conditions.
\begin{itemize}
\item If $\hat{m}_{22}(q_o)\hat{m}_{12}(q_o)\ne 0$, then
\begin{equation}
\widetilde{m}_{11}(0)=ie^{-i\theta}\frac{\hat{m}_{12}(q_o)}{\hat{m}_{22}(q_o)}\left[ 1-ie^{-i\theta}\widetilde{m}_{12}(0)\right], \quad
\widetilde{m}_{22}(0)=-ie^{i\theta}\frac{\hat{m}_{22}(q_o)}{\hat{m}_{12}(q_o)}\left[ 1+ie^{i\theta}\widetilde{m}_{21}(0)\right].
\end{equation}
\item If $\hat{m}_{22}(q_o)=0$, $\hat{m}_{12}(q_o)\ne 0$, then
\begin{equation}
\widetilde{m}_{22}(0)=0, \qquad \widetilde{m}_{12}(0)=-ie^{i\theta}.
\end{equation}
\item If $\hat{m}_{12}(q_o)=0$, $\hat{m}_{22}(q_o)\ne 0$, then
\begin{equation}
\widetilde{m}_{11}(0)=0, \qquad \widetilde{m}_{21}(0)=ie^{-i\theta}.
\end{equation}
\end{itemize}
 We disregard the case in which $\hat{m}_{22}(q_o)=\hat{m}_{12}(q_o)=0$, because one would have $\hat{m}(q_o)=0$ and $\det \hat{m}(z)$ would have a double zero at $z=q_o$.
Similarly, $\widetilde{m}(z)$ is bounded at $z=-q_o$ under the following conditions.
\begin{itemize}
\item If $\hat{m}_{11}(-q_o)\hat{m}_{21}(-q_o)\ne 0$, then
\begin{equation}
\widetilde{m}_{11}(0)=ie^{-i\theta}\frac{\hat{m}_{11}(-q_o)}{\hat{m}_{21}(-q_o)}\left[ 1+ie^{-i\theta}\widetilde{m}_{12}(0)\right], \quad
\widetilde{m}_{22}(0)=ie^{i\theta}\frac{\hat{m}_{21}(-q_o)}{\hat{m}_{11}(-q_o)}\left[ 1-ie^{i\theta}\widetilde{m}_{21}(0)\right].
\end{equation}
\item If $\hat{m}_{21}(-q_o)=0$, $\hat{m}_{11}(-q_o)\ne 0$, then
\begin{equation}
\widetilde{m}_{22}(0)=0, \qquad \widetilde{m}_{12}(0)=ie^{i\theta}.
\end{equation}
\item If $\hat{m}_{11}(-q_o)=0$, $\hat{m}_{21}(-q_o)\ne 0$, then
\begin{equation}
\widetilde{m}_{11}(0)=0, \qquad \widetilde{m}_{21}(0)=-ie^{-i\theta}.
\end{equation}
\end{itemize}
As before, we disregard the case in which $\hat{m}_{11}(-q_o)=\hat{m}_{21}(-q_o)=0$ because one would have $\hat{m}(-q_o)=0$ and $\det \hat{m}(z)$ would have a double zero at $z=-q_o$. 

For $\widetilde{m}(z)$ to be bounded at both $z=\pm q_o$, the above conditions have to be satisfied simultaneously, and this requires solvability condition \eqref{eq:eq169n}. 
Moreover, the followings hold.
\begin{itemize}
\item If $\hat{m}_{12}(q_o)\hat{m}_{22}(q_o)\ne 0$ and $\hat{m}_{11}(-q_o)\hat{m}_{21}(-q_o)\ne 0$, then $\widetilde{m}(z)$ is bounded at $z=\pm q_o$ provided that Eq.~\eqref{eq:eq169n} holds,
and the entries of $\widetilde{m}(0)$ are uniquely specified by: 
\begin{subequations}
\label{e:mtilde0}
\begin{gather}
\widetilde{m}_{12}(0)=ie^{i\theta}\frac{\left[ \hat{m}_{11}(-q_o)\hat{m}_{22}(q_o)+\hat{m}_{21}(-q_o)\hat{m}_{12}(q_o)\right]}{W\left(\hat{m}_2(q_o),\hat{m}_1(-q_o)\right)}, \\
\widetilde{m}_{21}(0)=ie^{-i\theta}\frac{\left[ \hat{m}_{11}(-q_o)\hat{m}_{22}(q_o)+\hat{m}_{21}(-q_o)\hat{m}_{12}(q_o)\right]}{W\left(\hat{m}_2(q_o),\hat{m}_1(-q_o)\right)}, \\
\widetilde{m}_{11}(0)=2ie^{-i\theta}\frac{\hat{m}_{11}(-q_o)\hat{m}_{12}(q_o)}{W\left(\hat{m}_2(q_o),\hat{m}_1(-q_o)\right)}, \qquad \widetilde{m}_{22}(0)=
2ie^{i\theta}\frac{\hat{m}_{21}(-q_o)\hat{m}_{22}(q_o)}{W\left(\hat{m}_2(q_o),\hat{m}_1(-q_o)\right)}.
\end{gather}
\end{subequations}
One can easily check that both conditions in Proposition~\ref{eq:proposition7} are satisfied. The first symmetry follows from the first symmetry for $\hat{m}_1(q_o)$ and $\hat{m}_2(-q_o)$.
\item If $\hat{m}_{22}(q_o)=0$, $\hat{m}_{12}(q_o)\ne 0$ and $\hat{m}_{21}(-q_o)\hat{m}_{11}(-q_o)\ne 0$, then  $\widetilde{m}(z)$ is bounded at $z=\pm q_o$ provided that Eq.~\eqref{eq:eq169n} holds,
and the entries of $\widetilde{m}(0)$ are uniquely specified by:
\begin{gather}
\widetilde{m}_{12}(0)=-ie^{i\theta}, \quad \widetilde{m}_{21}(0)=-ie^{-i\theta}, \quad
\widetilde{m}_{22}(0)=0, \quad \widetilde{m}_{11}(0)=-2ie^{-i\theta}
\frac{\hat{m}_{11}(-q_o)}{\hat{m}_{21}(-q_o)}.
\end{gather}
Again, both conditions in Proposition~\ref{eq:proposition7} are satisfied, consistently with the second symmetry, but the first symmetry is violated unless one also has $\hat{m}_{11}(-q_o)=0$, in which case $\widetilde{m}_{11}(0)=0$ as well.
\item If $\hat{m}_{12}(q_o)=0$, $\hat{m}_{22}(q_o)\ne 0$ and $\hat{m}_{21}(-q_o)\hat{m}_{11}(-q_o)\ne 0$, then  $\widetilde{m}(z)$ is bounded at $z=\pm q_o$ provided that Eq.~\eqref{eq:eq169n} holds,
and the entries of $\widetilde{m}(0)$ are uniquely specified by:
\begin{gather}
\widetilde{m}_{12}(0)=ie^{i\theta}, \quad \widetilde{m}_{21}(0)=-ie^{-i\theta}, \quad
\widetilde{m}_{11}(0)=0, \quad \widetilde{m}_{22}(0)=2ie^{i\theta}
\frac{\hat{m}_{21}(-q_o)}{\hat{m}_{11}(-q_o)}.
\end{gather}
Again, both conditions in Proposition~\ref{eq:proposition7} are satisfied, consistently with the second symmetry, but the first symmetry is violated unless one also has $\hat{m}_{21}(-q_o)=0$, in which case $\widetilde{m}_{22}(0)=0$ as well.
\item Finally, note that both $\hat{m}_{22}(q_o)=0$, $\hat{m}_{12}(q_o)\ne 0$, $\hat{m}_{21}(-q_o)=0$, $\hat{m}_{11}(-q_o)\ne 0$
and $\hat{m}_{22}(q_o)\ne0$, $\hat{m}_{12}(q_o)= 0$, $\hat{m}_{21}(-q_o)\ne0$, $\hat{m}_{11}(-q_o)= 0$ as well as the case in which all 4 are zero are incompatible with the solvability condition \eqref{eq:eq169n}.
\item From the above discussion, it also follows that if any of the entries of $\hat{m}(\pm q_o)$ vanishes, the values of $\widetilde{m}(0)$ that are compatible with both symmetries can be obtained as reductions of Eqs.~\eqref{e:mtilde0}.
\end{itemize}
In conclusion, for any $x,t$ for which the solvability condition \eqref{eq:eq169n} is satisfied, there exists a unique $\widetilde{m}(x,t;0)$ given by Eqs.~\eqref{e:mtilde0} for which $\widetilde{m}(x,t;z)$ defined via \eqref{e:def_mtilde} is bounded at both $z=\pm q_o$.
\end{proof}

\begin{remark}
\label{e:solv_t0}
As an example, one can easily check that at $t=0$, the solvability condition \eqref{eq:eq169n} is satisfied for any $x\in \Real$ if and only if $\theta = \pm \pi/2$. For $\theta \neq \pm \pi/2$, the unique $\widetilde{m}(x,0;0)$ that guarantees absence of singularities of $\widetilde{m}(x,0;z)$ at $z = \pm q_o$ is given by:
\begin{subequations}
\begin{gather}
\widetilde{m}_{12}(x,0;0)=-i\frac{e^{i\theta}-e^{-i\theta}}{1+e^{-2i\theta}}\equiv -ie^{i\theta}\tanh \theta, \quad
\widetilde{m}_{11}(x,0;0)=\frac{2e^{-2i\theta}}{1+e^{-2i\theta}}\delta_\infty^2\equiv e^{-i\theta}\delta_\infty\mathrm{sech}\, \theta, \\
\widetilde{m}_{21}(x,0;0)=-i\frac{e^{i\theta}-e^{-i\theta}}{1+e^{2i\theta}}\equiv -ie^{-i\theta}\tanh \theta, \quad
\widetilde{m}_{22}(x,0;0)=\frac{2e^{2i\theta}}{1+e^{2i\theta}}\frac{1}{\delta_\infty^2}\equiv e^{i\theta} \delta_\infty^{-2} \mathrm{sech}\, \theta.
\end{gather}
\end{subequations}
\end{remark}

\begin{Backmatter}

\paragraph{Acknowledgments}
We would like to thank K. McLaughlin, R. Jenkins and D. Bilman for many insightful discussions, as well as the anonymous reviewers whose comments helped us improve the manuscript. We also acknowledge the Isaac Newton Institute for Mathematical Sciences, Cambridge, for support and hospitality during the satellite programme ``Emergent phenomena in nonlinear dispersive waves'' (supported by EPSRC grant EP/R014604/1), where work on this paper was undertaken.

\paragraph{Funding Statement}
B.P. and T.T. acknowledge partial support for this research through grants from the National Science Foundation DMS-2406626 and DMS-2306438, respectively.

\paragraph{Competing Interests}
None.

\paragraph{Data Availability Statement}
The code used to produce the figures in this paper will be available in the electronic supplementary materials.

\paragraph{Ethical Standards}
The research meets all ethical guidelines.

\paragraph{Author Contributions}
Conceptualization: A.G.; B.P.; T.T. Methodology: A.G.; B.P.; T.T. Writing original draft: A.G.; B.P.; T.T. All authors approved the final submitted draft.

\end{Backmatter}

\end{document}